\documentclass[preprint]{aastex}
\usepackage{hyperref}
\usepackage{xcolor}
\pdfoutput=1

\begin{document}

\title{The Green Bank Ammonia Survey: Observations of Hierarchical Dense Gas Structures in Cepheus-L1251}

\author{Jared Keown\altaffilmark{1}, James Di Francesco\altaffilmark{1,2}, Helen Kirk\altaffilmark{2}, Rachel K. Friesen\altaffilmark{3}, Jaime E. Pineda\altaffilmark{4}, Erik Rosolowsky\altaffilmark{5}, Adam Ginsburg\altaffilmark{6}, Stella S. R. Offner\altaffilmark{7}, Paola Caselli\altaffilmark{4}, Felipe Alves\altaffilmark{4}, Ana Chac\'on-Tanarro\altaffilmark{4}, Anna Punanova\altaffilmark{4}, Elena Redaelli\altaffilmark{4}, Young Min Seo\altaffilmark{8}, Christopher D. Matzner\altaffilmark{9}, Michael Chun-Yuan Chen\altaffilmark{1}, Alyssa A. Goodman\altaffilmark{10}, How-Huan Chen\altaffilmark{10}, Yancy Shirley\altaffilmark{11}, Ayushi Singh\altaffilmark{9}, Hector G. Arce\altaffilmark{12}, Peter Martin\altaffilmark{13}, Philip C. Myers\altaffilmark{10}}

\email{jkeown@uvic.ca}

\altaffiltext{1}{Department of Physics and Astronomy, University of Victoria, Victoria, BC, V8P 5C2, Canada} 

\altaffiltext{2}{NRC Herzberg Astronomy and Astrophysics, 5071 West Saanich Road, Victoria, BC, V9E 2E7, Canada}

\altaffiltext{3}{Dunlap Institute for Astronomy and Astrophysics, University of Toronto, 50 St. George St., Toronto, ON, M5S 3H4, Canada}

\altaffiltext{4}{Max-Planck-Institut f\"{u}r extraterrestrische Physik, Giessenbachstrasse 1, D-85748, Garching, Germany}

\altaffiltext{5}{Department of Physics, University of Alberta, Edmonton, AB, Canada}

\altaffiltext{6}{National Radio Astronomy Observatory, Socorro, NM 87801, USA}

\altaffiltext{7}{Department of Astronomy, The University of Texas, Austin, TX 78712, USA}

\altaffiltext{8}{Jet Propulsion Laboratory, NASA, 4800 Oak Grove Dr, Pasadena, CA 91109, USA}

\altaffiltext{9}{Department of Astronomy \& Astrophysics, University of Toronto, 50 St. George Street, Toronto, Ontario, Canada M5S 3H4}

\altaffiltext{10}{Harvard-Smithsonian Center for Astrophysics, 60 Garden St., Cambridge, MA 02138, USA}

\altaffiltext{11}{Steward Observatory, 933 North Cherry Avenue, Tucson, AZ 85721, USA}

\altaffiltext{12}{Department of Astronomy, Yale University, P.O. Box 208101, New Haven, CT 06520-8101, USA}

\altaffiltext{13}{Canadian Institute for Theoretical Astrophysics, University of Toronto, 60 St. George St., Toronto, Ontario, Canada, M5S 3H8}


\begin{abstract}
We use Green Bank Ammonia Survey observations of NH$_3$ (1,1) and (2,2) emission with 32$\arcsec$ FWHM resolution from a $\sim$ 10 pc$^{2}$ portion of the Cepheus-L1251 molecular cloud to identify hierarchical dense gas structures.  Our dendrogram analysis of the NH$_3$ data results in 22 top-level structures, which reside within 13 lower-level, parent structures.  The structures are compact (0.01 pc $\lesssim R_{eff} \lesssim$ 0.1 pc) and are spatially correlated with the highest H$_2$ column density portions of the cloud.  We also compare the ammonia data to a catalog of dense cores identified by higher-resolution (18.2$\arcsec$ FWHM) \textit{Herschel Space Observatory} observations of dust continuum emission from Cepheus-L1251.  Maps of kinetic gas temperature, velocity dispersion, and NH$_3$ column density, derived from detailed modeling of the NH$_3$ data, are used to investigate the stability and chemistry of the ammonia-identified and \textit{Herschel}-identified structures.  We show that the dust and dense gas in the structures have similar temperatures, with median $T_{dust}$ and $T_K$ measurements of 11.7 $\pm$ 1.1 K and 10.3 $\pm$ 2.0 K, respectively.  Based on a virial analysis, we find that the ammonia-identified structures are gravitationally dominated, yet may be in or near a state of virial equilibrium.  Meanwhile, the majority of the \textit{Herschel}-identified dense cores appear to be not bound by their own gravity and instead confined by external pressure.  CCS $(2_0-1_0)$ and HC$_5$N $(9-8)$ emission from the region reveal broader line widths and centroid velocity offsets when compared to the NH$_3$ (1,1) emission in some cases, likely due to these carbon-based molecules tracing the turbulent outer layers of the dense cores.

\end{abstract}

\section{Introduction}
Recent large-scale surveys of dust continuum emission from nearby star-forming regions have provided unprecedented insights into the structure of molecular clouds.  The Gould Belt Legacy surveys on the \textit{Herschel Space Observatory} \citep[HGBS;][]{Andre_2010} and James Clerk Maxwell Telescope \citep[JGBS;][]{WT_2007} have fueled these advancements, providing photometric censuses of nearby ($<$ 500 pc) Galactic molecular clouds.  In particular, they have revealed that filaments pervade both active \citep{Andre_2010, Menshchikov_2010} and quiescent \citep{Miville_2010, Ward-Thompson_2010} molecular clouds.  Dense cores, the high density regions of molecular clouds where stars are born, also appear to be spatially correlated with filaments \citep{Konyves_2010, Konyves_2015}.  These results suggest mass flow onto or through filaments may be a requirement to gain the locally high density conditions necessary for core formation and subsequent gravitational collapse to form stars.  

The HGBS has shown that the mass distribution of prestellar dense cores, i.e., gravitationally-bound structures on the verge of forming new stars, bears a resemblance to the stellar initial mass function (IMF) across several star-forming environments (Aquila, \citealt{Konyves_2015}; Taurus, \citealt{Marsh_2016}; Cepheus, \citealt{DiFrancesco_prep}).  Assuming the observed prestellar cores will indeed form stars in the future, the relationship between the IMF and core mass function lends credence to the theory that stellar masses are set at the dense core stage.  Without adequate spectral line measurements for the prestellar cores, however, it is difficult to determine whether these structures are truly gravitationally bound objects or rather they are simply pressure-confined by the molecular cloud within which they reside.  

To provide a large-scale spectral counterpart to the photometry-based HGBS and JGBS, the Green Bank Ammonia Survey \cite[GAS;][]{Friesen_submitted} has mapped NH$_3$ emission across the highest H$_2$ column density portions of the northern Gould Belt molecular clouds.  NH$_3$ is an ideal tracer for such a survey because it is excited in the cold, dense gas ($n > 2 \times 10^3$ cm$^{-3}$ at 10 K) found in cores and filaments \citep{Shirley_2015}.  Observing both NH$_3$ (1,1) and (2,2) emission also provides a convenient way of measuring kinetic gas temperatures \citep{Ho_1983}, which, along with velocity dispersion measurements, can be used to determine the stability of filaments and dense cores.  

In this paper, we analyze the stability of dense gas structures identified in the Cepheus-L1251 molecular cloud by combining HGBS photometry with GAS NH$_3$ spectral data for the region.  Cepheus-L1251 is a prime candidate for such an analysis due to its large population of both starless and protostellar dense cores \citep{Lee_2007, Kun_2009, Kim_2015}, as well as its prominent network of parsec-length filaments \citep{Sato_1994}.  The cloud consists of three main submillimeter-bright regions - L1251A in the west, L1251C in the center, and L1251E/B in the east - all of which contain protostar-driven outflows (L1251A: \citealt{Lee_2010}; L1251C: \citealt{Kim_2015}, L1251E/B: \citealt{Lee_2007}).  These outflows are indicative of the active nature of the region and may produce noticeable chemo-dynamical effects upon the surrounding gas.  

Furthermore, the dense core population in Cepheus-L1251 has been catalogued by both the HGBS \citep{DiFrancesco_prep}, which identified 187 dense cores in a $\sim$ 3 square degree field in the region, and the JGBS \citep{Pattle_2017}, which found 51 sources in a $\sim$ 1.6 square degree field.  Newly obtained NH$_3$ observations of Cepheus-L1251 by GAS provide the link necessary to determine the dynamical states of dust continuum-identified structures.  Here, we analyze the stability of cores in the \cite{DiFrancesco_prep} catalog since the \cite{Pattle_2017} observations do not include the southwestern portion of Cepheus-L1251 observed by the HGBS and GAS.  We also use the NH$_3$ (1,1) data to identify hierarchical dense gas structures in Cepheus-L1251, which are cataloged here and compared to the continuum-identified structures.  We assume a distance of 300 pc to Cepheus-L1251 throughout this paper, which appears to be a consistent distance measurement found from a variety of methods \citep{Kun_1993, Kun_1998, Balazs_2004}.



In \S~2, we describe the observations and data sets used for our analysis.  In \S~3, we outline the methods used to identify structures in the NH$_3$ data and derive stability parameters for those structures.  In \S~4, we compare our results to a similar analysis performed by \cite{Friesen_2016} in Serpens South, discuss the role of external pressure within the observed structures, and investigate potential chemical effects within L1251 by analyzing CCS $(2_0-1_0)$ and HC$_5$N $(9-8)$ emission from the region.  We conclude, in \S~5, with a summary of the paper.


\section{Observations}

\subsection{GBT NH$_3$ Data}

Our ammonia data were obtained as part of the Green Bank Ammonia Survey \cite[GAS;][]{Friesen_submitted}, a large project on the Green Bank Telescope which mapped NH$_3$ (\textit{J,K}) = (1,1), (2,2), and (3,3), HC$_7$N \textit{J} = $(21-20)$ and $(22-21)$, HC$_5$N \textit{J} = $(9-8)$, and CCS \textit{J} = $(2_0-1_0)$ emission across much of the northern Gould Belt star-forming regions where $A_V$ $>$ 7 mag.  The Cepheus-L1251 region was observed for a total of $\sim$ 21 hours between October 2015 and February 2016.  Observations covered the highest H$_2$ column density portions of L1251 and totaled $\sim$ 0.36 square degrees ($\sim$ 10 pc$^{2}$ at a distance of 300 pc).  Figure \ref{obs} shows the outline of the region observed by GAS overlaid onto the \textit{Herschel}-derived H$_2$ column density map for L1251 \citep{DiFrancesco_prep}.  The angular resolution for all observed lines is 32$\arcsec$ (FWHM), which is equivalent to $\sim$ 0.05 pc at a distance of 300 pc.  The spectral resolution is 5.7 kHz, or $\sim$ 0.07 km s$^{-1}$ at 23.7 GHz.  The median rms noise in the off-line channels of both NH$_3$ (1,1) and (2,2) is 0.12 K.  If the noisy map edges are excluded, the median rms noise in the maps is 0.09 K.  Emission was detected from all observed transitions except NH$_3$ (3,3), HC$_7$N (21-20), and HC$_7$N (22-21).  All GAS-observed data and parameter maps used in this paper are publicly available at \url{https://dataverse.harvard.edu/dataverse/Cepheus-L1251}.

\subsection{\textit{Herschel} Dust Continuum Data}
As part of the \textit{Herschel} Gould Belt Survey \citep[HGBS;][]{Andre_2010}, five separate fields in the Cepheus star-forming region, including L1251, were mapped by the PACS and SPIRE instruments at $70-500 ~\mu$m.  After subtracting background sub-millimeter emission observed toward the region by \textit{Planck}, \cite{DiFrancesco_prep} fit SEDs to the \textit{Herschel} photometric measurements to derive H$_2$ column densities and dust temperatures across the observed fields.  In this paper, we adopt the ``high-resolution'' (18.2$\arcsec$) H$_2$ column density map produced by \citeauthor{DiFrancesco_prep}, which was created using a multi-scale decomposition method outlined in \cite{Palmeirim_2013}.  In addition, \cite{DiFrancesco_prep} used the multi-scale, multi-wavelength source extraction algorithm \textit{getsources} \citep{Menshchikov_2012} to identify dense cores and protostars throughout Cepheus.  They classified cores depending on the likelihood they were gravitationally bound based on the ratio of their critical Bonnor-Ebert mass \citep{Ebert_1955, Bonnor_1956} to their observed mass ($\alpha_{BE} = M_{BE,crit} / M_{obs}$), which provides an estimate for the virial state of a core in the absence of spectroscopically obtained velocity dispersion and gas temperature measurements.  Cores with $\alpha_{BE}$ $>$ 5 are deemed gravitationally unbound and termed ``starless,'' cores with $\alpha_{BE}$ $\leq$ 5 are classified as ``prestellar candidate'' cores that are likely to be gravitationally bound, while cores with $\alpha_{BE}$ $\leq$ 2 are coined ``robust prestellar'' cores that are most likely to be gravitationally bound.  Such a core classification scheme depends heavily on whether or not the dust-derived temperatures are representative of the \textit{gas} temperatures of structures and excludes any non-thermal motions that could support the structures against gravity.  Determining whether or not these prestellar-classified cores are truly gravitationally bound structures using data of lines from dense cores like those of NH$_3$ will lead to a more informed interpretation of the relationship between the IMF and core mass function.      

\section{Analysis and Results}

\subsection{NH$_3$ Line Fitting}
Although we detect NH$_3$ (1,1) and (2,2) emission throughout Cepheus-L1251, both of which are caused by the para-NH$_3$ species, we do not detect emission from the ortho-NH$_3$ species (e.g., the NH$_3$ (3,3) transition).  Thus, we are unable to determine an ortho-to-para NH$_3$ ratio and must focus our analysis on the (1,1) and (2,2) para-NH$_3$ transitions.  The GAS analysis pipeline\footnote{available at \url{http://gas.readthedocs.io/}} was used for data calibration, imaging, and NH$_3$ (1,1) and (2,2) line fitting.  This pipeline, and the methods used in the GAS line-fitting procedure, are discussed in detail by \cite{Friesen_submitted}.  The GAS line-fitting pipeline simultaneously fits the NH$_3$ (1,1) and (2,2) lines pixel-by-pixel, assuming LTE and a single velocity component along the line-of-sight, to produce a set of best-fit parameter maps which includes kinetic gas temperature ($T_{K}$), excitation temperature ($T_{ex}$), para-NH$_3$ column density ($N_{para-NH_3}$), velocity dispersion ($\sigma$), and centroid velocity ($V_{LSR}$).  

In this paper, we use the GAS analysis pipeline to fit all pixels in Cepheus-L1251 with signal-to-noise ratio (SNR) $> 3$ in the NH$_3$ (1,1) spectrum, where the SNR has been estimated from the peak emission channel and rms.  The resulting best-fit parameter maps were used to calculate the kinematic and chemical properties of the structures presented in our analysis.  All pixels with peak SNR $\leq 3$ in the NH$_3$ (1,1) spectrum are not included in our analysis.  We also exclude any pixels from our analysis that do not meet the following requirements: 5 K $< T_{K} <$ 30 K (outside this range, the NH$_3$ (1,1) and (2,2) lines cannot constrain $T_{K}$), 0.05 km s$^{-1}$ $< \sigma <$ 2.0 km s$^{-1}$, $-7$ km s$^{-1}$ $< V_{LSR} <$ $-1$ km s$^{-1}$,  $T_{K, err} <$ 5 K, $\sigma_{err} <$ 2.0 km s$^{-1}$, and $V_{LSR, err} <$ 1 km s$^{-1}$.  For the para-NH$_3$ abundance analysis presented in \S~3.6, we additionally require $T_{ex} <$ 30 K and $T_{ex, err} <$ 5 K since reliable NH$_3$ estimates are dependent upon accurate $T_{ex}$ measurements.  The final parameter maps used for the analyses are shown in Figures \ref{Params1} ($\sigma$ and $V_{LSR}$) and \ref{Params2} ($T_{K}$ and para-NH$_3$ abundance).  Detailed analyses and discussions of these maps are presented in the following sections.   



\subsection{Dendrogram Structure-Finding}
Although there exists a variety of methods for extracting structures from both 2D and 3D emission maps of molecular clouds, dendrograms capture the hierarchical nature of clouds better than most other extraction products.  This is due to the dendrogram's basis on tree diagrams, which display the relationship between the various structures found within a map.  The analysis begins by locating the emission peaks in a map, then proceeds by grouping the fainter surrounding pixels until two or more local maxima, i.e., \textit{leaves}, contain adjoining pixels, at which point they merge at a \textit{branch}.  Fainter pixels are continually added to this merged structure until it is either merged with another local maximum, or reaches a user-defined noise threshold, at which point it becomes a \textit{trunk}.  A detailed description of dendrograms as a source extraction method is discussed in \cite{Rosolowsky_2008}.  Throughout this paper, we use the term ``parent'' for structures that are either a branch or a trunk.  

Due to the 18 hyperfine components of the NH$_3$ (1,1) line, running a dendrogram extraction algorithm on our NH$_3$ (1,1) data cube produces spurious identifications, as well as inaccurate source sizes and velocity dispersions.  To circumvent this complication, we constructed a simulated Gaussian emission data cube based on the parameters obtained from the line-fitting procedure described in Section 3.1.  Namely, a Gaussian spectrum was created for each pixel with SNR $> 3$ in the observed NH$_3$ (1,1) map.  The Gaussian profile was scaled to each pixel's peak brightness temperature, centroid velocity, and velocity dispersion as measured in the NH$_3$ (1,1) map.  \cite{Friesen_2016} note that some error is introduced when using this method on spectra that contain multiple velocity components, for which our single component fit would produce skewed estimates for $V_{LSR}$ and $\sigma$.  Cepheus-L1251 is a relatively quiescent environment compared to the Serpens South region observed by \cite{Friesen_2016}, however, and our data show no signs of the multiple velocity components that they observed.  Furthermore, \cite{Friesen_2016} add the residual of their single velocity component fit of the NH$_3$ (1,1) emission back into their simulated Gaussian profile, which can lead to spurious sources being identified by the dendrogram when multiple velocity components are present.  To circumvent this complication, in our analysis, random noise was added back into each simulated Gaussian spectrum with an rms equivalent to the rms measured for that pixel in the observed NH$_3$ (1,1) cube.  We also mask out the edges of the map where rms noise levels were high in the original NH$_3$ (1,1) cube.  

In this paper, we use the \textit{astrodendro} package to identify hierarchical structures in our simulated Gaussian data cube.  The input parameters for the dendrogram algorithm are: (1) \texttt{min\char`_value}, the minimum threshold value to consider in the data cube (i.e., the lowest intensity a pixel can have to be joined to a structure).  We chose to set \texttt{min\char`_value} at 0.45 K, which is equivalent to 5 times the median rms noise.  (2) \texttt{min\char`_delta}, the minimum difference in brightness between two structures before they are merged into a single structure.  We set \texttt{min\char`_delta} to 0.18 K, or 2 times the median rms noise.  (3) \texttt{min\char`_npix}, the minimum number of pixels a structure must contain to remain independent.  We set \texttt{min\char`_npix} to 10, which reduces the number of noise spikes identified by the extraction.  Our choices for the \textit{astrodendro} input parameters follow the NH$_3$ structure-finding prescription outlined by \cite{Friesen_2016}, which based their parameter selections on standard recommendations for dendrogram-based source extraction \citep[e.g.,][]{Rosolowsky_2008}.  

The major and minor axes of sources identified in our dendrogram analysis are calculated by \textit{astrodendro} based on the intensity weighted second moment in the direction of greatest elongation in the position-position (PP) plane.  After completing source identification with \textit{astrodendro}, we also require that the major and minor axes of all sources be larger than 6$\arcsec$ in projection on the sky to be included in our final catalog.  We note that the major and minor axes for many of the \textit{astrodendro}-identified sources in this paper are much smaller than the full extent of their associated pixels on the PP plane.  All of the structures with minor and major axes larger than 6$\arcsec$, however, have total pixel areas larger than the area of the 32$\arcsec$ GBT beam.  Thus, our 6$\arcsec$ major and minor axes cut successfully removes noise spikes while preserving real structures that are larger than the 32$\arcsec$ beam-size of our GBT observations.   


Imposing these parameters and selection criteria results in a dendrogram with 22 top-level ``leaf'' structures, nine mid-level ``branches,'' and four low-level ``trunks.''  Figure \ref{leaves} displays the outlines of the leaves (green) and trunks (cyan) that were identified in the dendrogram overlaid onto the NH$_3$ (1,1) integrated intensity map.  The integrated intensity map was created using the channels surrounding all five hyperfine groups of the NH$_3$ (1,1) emission.  The dendrogram leaves highlight both the peaks in the integrated intensity map, as well as the highest H$_2$ column density portions of the cloud.  Eight of the branches fall within the eastern L1251E/B trunk shown in the upper left panel of Figure \ref{leaves}, indicating the region is highly sub-structured and contains many pockets of dense gas potentially forming within a larger filamentary structure.  The ninth branch is located in the larger trunk in the eastern half of L1251A shown in the middle left panel of Figure \ref{leaves}.  The full tree diagram for the identified sources is shown in the right panel of Figure 5.  The highly sub-structured portion of L1251E/B begins at structure 30 in the diagram.  Visual checks for all leaves were also completed, but no spurious sources were identified that also pass our selection criteria.  Table \ref{Table_NH3} contains the full catalog of leaves identified in the dendrogram which also pass our selection criteria.   

For each 3D structure, we obtain the mean R.A., Dec., major axis ($\sigma_{major}$), minor axis ($\sigma_{minor}$), and position angle ($\theta_{PA}$) projected in the PP plane.  Additionally, we define the projected effective radius of each structure as the geometric mean of the major and minor axes: $R_{eff}$=($\sigma_{major}$ $\sigma_{minor}$)$^{1/2}$.  We note that this effective radius is an upper limit because we have not applied any deconvolution to the measured sizes, which can be unstable when applied to dendrogram-identified objects (see \citealt{Rosolowsky_2008} for a discussion of this issue).  As mentioned above, however, all structures have total areas much larger than the area of the 32$\arcsec$ GBT beam.  Specifically, the median $\pm$ standard deviation for the ratio of the GBT beam area to the areas of the leaves is $0.22 \pm 0.19$.  The low relative size of the beam compared to the size of the identified structures suggests that deconvolution would have a small effect on the structure sizes.  Nevertheless, in Section 3.4 we show that lowering the effective radius for the dendrogram-identified structures does not alter their virial state.  Moreover, in Appendix A, we also demonstrate that using an alternative formulation for the effective radius, based on a structure's total surface area, increases their virial parameters by factors of a few but does not change the main conclusions of the virial analysis presented in this paper.  

Figure \ref{R_vs_aspect} plots effective radius versus aspect ratio ($\sigma_{major}$/$\sigma_{minor}$) for the leaves and parent structures identified in the dendrogram.  The parent structures tend to have larger aspect ratios (median $\sigma_{major}/\sigma_{minor} = 2.7 \pm 0.6$) than the leaves (median $\sigma_{major}/\sigma_{minor} = 1.7 \pm 0.5$), likely a result of the tendency of dense gas to lie along filamentary structures within the cloud \citep{Friesen_2016}.  





\subsection{Source Masses}
We estimate the masses of the ammonia-identified structures based on their corresponding H$_2$ column density as measured by \textit{Herschel} dust continuum observations of Cepheus \citep{DiFrancesco_prep}.  Each source's 2D mask, based on its projection onto the PP plane, is used to define the region over which pixels in the H$_2$ column density map (convolved from a resolution of 18.2$\arcsec$ to 32$\arcsec$ to match the beam-size of the NH$_3$ observations) are integrated and converted to mass.  We assume a distance of 300 pc to Cepheus-L1251 and a mean molecular weight per hydrogen molecule ($\mu_H$) of 2.8 \citep[see, e.g., Appendix A in][]{Kauffmann_2008}.  Figure \ref{R_vs_mass} shows source effective radius versus mass for the leaves and parents identified in our dendrogram analysis.  The structures have a range in mass from 0.9 M$_\odot$ for the smallest top-level structure to 80 M$_\odot$ for the largest parent structure.  A power-law fit to the data produces a best-fit slope of 1.94 $\pm$ 0.18, where the uncertainty on the slope has been estimated from the square-root of the diagonal term of the covariance matrix from the total least squares curve-fitting method.  This slope is consistent with the \cite{Larson_1981} $M \propto R^2$ relation.  In Appendix B, we show that determining the masses of the ammonia-identified leaves based on continuum maps that have been spatially filtered to remove emission from large-scale structures has little impact on their calculated virial parameters.


\subsection{Virial Analysis}
\subsubsection{Ammonia-identified Structures}
The stability of our ammonia-identified structures can be estimated from a virial analysis.  Neglecting external pressure and magnetic fields, the virial parameter ($\alpha_{vir}$) can be used to determine whether a cloud structure is gravitationally bound or unbound.  We follow the virial analysis method outlined in \cite{Friesen_2016} and define the virial mass as:

\begin{equation}
M_{vir} = \frac{5\sigma^{2}R}{aG}
\end{equation} where $\sigma$ is the velocity dispersion of the core, \textit{R} is the core radius (which we set to be $R_{eff}$), \textit{G} is the gravitational constant, and 

\begin{equation}
a = \frac{1 - k/3}{1 - 2k/5}
\end{equation} is a term which accounts for the radial power-law density profile of a core, where $\rho(r) \propto r^{-k}$ \citep{Bertoldi_1992}.  We note that recent virial analyses by \cite{Pattle_2017} and \cite{Kirk_submitted} assume a spherically symmetric Gaussian density distribution (derived in \cite{Pattle_2016}, for reference).  For such a distribution, Equation 2 instead becomes $a = 5 / 6\sqrt{\pi} \sim 0.47$.  In the virial analysis we present here, we adopt Equation 2 with $k = 1.5$ (thus, $a = 1.25$) as the assumed density profile of the observed structures to be consistent with \cite{Friesen_2016}.  In the analysis presented in Section 4.2, in which we add pressure to the virial equation as was done by \cite{Pattle_2017} and \cite{Kirk_submitted}, we assume a Gaussian density distribution.  For both virial analyses presented in this paper, however, we also show the effect that using the alternative density profile would have on the calculated virial parameters.

Both the thermal and non-thermal components of the velocity dispersion for a particle of mean mass are included in $\sigma$.  Using the parameter maps obtained from the NH$_3$ line fitting mentioned in Section 3.1, we calculate $\sigma$ as

\begin{equation}
\sigma^2 = \sigma_{v}^{2} - \frac{k_{B}T}{m_{NH_3}} + \frac{k_{B}T}{\mu_p m_H}
\end{equation} where $k_B$ is Boltzmann's constant, $m_{NH_3}$ is the molecular mass of NH$_3$, $m_H$ is the atomic mass of hydrogen, and $\mu_p$ is the mean molecular mass of interstellar gas \citep[2.33; see, e.g., Appendix A in][]{Kauffmann_2008}.  $\sigma_v$ is found by first creating a 2D mask from each 3D structure's projection onto the PP plane.  This mask is then overlaid onto the velocity dispersion map created from the line fitting procedure of Section 3.1.  All pixels in the velocity dispersion map falling within the mask are used to calculate an average velocity dispersion for the structure, weighted by the NH$_3$ (1,1) integrated intensity maps, which is used as $\sigma_{v}$ in Equation 3.  The same procedure is followed to obtain an average $T_K$ value for each structure, which is used for $T$ in Equation 3.  


The virial parameter, $\alpha_{vir}$, is defined as $\alpha_{vir} = M_{vir} / M$, where $M_{vir}$ is given in Equation 1 and $M$ is the observed mass for the structure found using the method described in Section 3.3.  For $\alpha_{vir} \geq 2$, gas motions alone are presumed to be strong enough to prevent gravitational collapse.  For $\alpha_{vir} < 2$, without the presence of magnetic pressure, structures are unable to provide support against gravitational collapse.  Since the virial parameter assumes spherical symmetry, we only calculate $\alpha_{vir}$ for structures with aspect ratios less than 2.  The top panel of Figure \ref{Mass_vs_virial} displays the observed mass versus measured virial parameter for all the leaves identified in our ammonia dendrogram analysis that pass this aspect ratio criterion.  We find that all the ammonia structures sit below $\alpha_{vir}$ = 2, suggesting they are gravitationally bound structures when magnetic pressure is not considered and a power-law density profile is assumed for the structures. 

The dotted horizontal line in Figure \ref{Mass_vs_virial} shows where $\alpha_{vir}$ = 1 would be located if we assume a Gaussian density profile for the structures rather than a power-law density profile.  Using this alternative density profile assumption and $\alpha_{vir}$ = 1 as the virial stability threshold, only two of the leaves are bound by gravity.  While this change in the virial state of the ammonia-identified structures between the two sets of assumptions highlights the large uncertainties in virial parameter estimations, it may also indicate the leaves are currently at or near a state of virial equilibrium (see Sections 4.1 and 4.2 for further evidence of this scenario).

\subsubsection{Herschel-identified Dense Cores}
In addition to our virial analysis of the ammonia structures identified in this paper, we also derive virial parameters for the \textit{Herschel}-identified dense cores in the Cepheus-L1251 region that were originally identified by \cite{DiFrancesco_prep}.  We treat these \textit{Herschel}-identified dense cores as a separate set of structures because they were identified using higher-resolution observations (18.2$\arcsec$ FWHM at 250 $\mu m$) and a different source extraction algorithm than the ammonia-identified structures.  The \citeauthor{DiFrancesco_prep} catalog features 187 dense cores identified by the \textit{getsources} \citep{Menshchikov_2012} extraction algorithm within the Cepheus-L1251 region as part of the \textit{Herschel} Gould Belt Survey.  Each core in the \citeauthor{DiFrancesco_prep} catalog is classified as either ``protostellar,'' ``starless,'' ``prestellar candidate,'' or ``robust prestellar'' (see Section 2.2 for a discussion of this classification scheme).  The ``prestellar candidate'' and ``robust prestellar'' cores are thought to be gravitationally bound and likely to form stars in the future due to their higher critical Bonnor-Ebert mass ratio ($\alpha_{BE} = M_{BE,crit} / M_{obs}$), which is an appropriate substitute for the virial mass ratio in the absence of spectroscopic information.  The critical Bonnor-Ebert mass \citep{Bonnor_1956} is given by \begin{equation}
M_{BE,crit} = \frac{2.4 c_{s}^2 R}{G}
\end{equation} where $c_s$ is the isothermal sound speed, $R$ is the radius, and $G$ is the gravitational constant.  Thus, dense cores with $\alpha_{BE} < 2 $ should likely have $\alpha_{vir} \lesssim 2$  in the absence of magnetic pressure.  

To test whether or not the \textit{Herschel}-identified cores are truly ``prestellar,'' we use the R.A., Dec., major FWHM, minor FWHM, and position angle provided for each source in the \citeauthor{DiFrancesco_prep} catalog to construct elliptical, 2D masks.  We follow the method adopted by \cite{Kirk_submitted} and set the outline of the mask to be twice the size of the core's FWHM ellipse (i.e., the full extent of the structure), as measured in the \textit{Herschel} H$_2$ column density map.  If the core's ellipse falls within the area mapped in NH$_3$ (1,1) and (2,2) by GAS, and at least one-fourth of the pixels in the 2D mask contain reliable fits to the NH$_3$ (1,1) and (2,2) spectra, we use its 2D mask to obtain mean values for $\sigma_v$ and $T_K$, weighted by the NH$_3$ (1,1) integrated intensity map.  This criterion eliminates 87 of 94 ``starless'' cores, 28 of 40 ``prestellar candidate'' cores,  9 of 42 ``robust prestellar'' cores, and 6 of 11 ``protostellar'' cores from the analysis.  The ``starless'' dense cores are preferentially eliminated due to their lower H$_2$ column densities, where NH$_3$ (2,2) emission was rarely detected, and thus did not produce reliable fits during our line-fitting procedure.  The right column of Figure \ref{leaves} shows the outlines of the 2D elliptical masks used for the \textit{Herschel}-identified cores that match our selection criteria overlaid onto the NH$_3$ (1,1) integrated intensity map.  In many cases, the NH$_3$-identified structures are consistent with the positions of \textit{Herschel}-identified cores.  The higher spatial resolution of the \textit{Herschel} observations, as well as the different source extraction algorithm used for the \textit{Herschel}-identified sources, leads to two or more \textit{Herschel}-identified cores being identified within several of the NH$_3$ leaves.  

The intensity-weighted mean values of $\sigma_v$ and $T_K$, along with the SED-derived mass and deconvolved $R_{eff}$ quoted in the \citeauthor{DiFrancesco_prep} catalog for each core, are used to derive the virial parameter for each source using the same method described above for the ammonia-identified structures.  We note that the mass and effective radius used for the \textit{Herschel}-identified cores in the virial analysis presented in this paper are the same values used by \citeauthor{DiFrancesco_prep} to determine the critical Bonnor-Ebert classification of the cores.  The $R_{eff}$ of the \textit{Herschel}-identified cores have a mean and standard deviation of 0.029 $\pm$ 0.014 pc, which is consistent with the $R_{eff}$ of the ammonia-identified leaves ($R_{eff, mean}$ = 0.023 $\pm$ 0.008 pc).  In Appendix A, we show that using the surface area of the \textit{Herschel}-identified cores to determine their effective radius increases their virial parameters by factors of a few.  Similarly, in Appendix B, we show that the virial parameters for the \textit{Herschel}-identified dense cores are significantly lower when their observed mass is calculated by summing all the pixels in the \textit{Herschel}-derived H$_2$ column density map that fall within their elliptical mask.  These lower virial parameters are driven by the larger core masses measured directly from the H$_2$ column density map, which includes large-scale structure that was filtered out by the \textit{getsources} extraction algorithm prior to the original mass estimate for the \textit{Herschel}-identified cores. 


The top panel of Figure \ref{Mass_vs_virial} displays the observed mass versus virial parameter for the \textit{Herschel}-identified cores overlaid atop the results for the ammonia-identified structures.  All of the ``starless'' and ``prestellar candidate'' cores have $\alpha_{vir} \gtrsim 4$, while the ``robust prestellar'' cores are split almost evenly above and below $\alpha_{vir} = 2$.  When assuming a Gaussian density profile for the structures and setting $\alpha_{vir} = 1$ as the virial stability threshold, only one \textit{Herschel}-identified core is gravitationally bound.  This result suggests that many of the \textit{Herschel}-identified cores that have been deemed ``prestellar'' based on a critical Bonnor-Ebert sphere analysis, may actually be gravitationally unbound structures.  Additionally, the cores with $\alpha_{vir} < 2$ tend to be located over the brightest NH$_3$ (1,1) emission.  The right column of Figure \ref{leaves} shows the \textit{Herschel}-identified cores that have $\alpha_{vir} < 2$ as solid ellipses and those with $\alpha_{vir} \geq 2$ as dotted ellipses.  None of the cores with $\alpha_{vir} < 2$ correspond to regions that contain the lowest detectable levels of NH$_3$ (1,1) emission.  

\subsection{Elongated Structure Virial Parameter}
The aforementioned virial analysis assumes spherical symmetry for structures.  This assumption means elongated and filamentary structures must be analyzed independently to reduce the risk of misrepresenting their virial parameters.  Thus, we use instead the mass per unit length of the ammonia-identified leaves and parents with aspect ratios $\geq$ 2 to calculate a filament virial parameter, $\alpha_{vir, fil}$.  We define the filament virial mass per unit length as \citep{Fiege_2000} 

\begin{equation}
m_{vir, fil} = \frac{2\sigma^{2}}{G} ~,
\end{equation} where the velocity dispersion, $\sigma$, is calculated from Equation 3 using the same method described in Section 3.4.  We obtain the observed mass, $M$, for the structures in the same way described in Section 3.3.  The length of the structures, $L$, is set to twice the FWHM of its major axis:

\begin{equation}
L = 2\sqrt{2\ln2}~\sigma_{major} ~.
\end{equation}

The filament virial parameter then becomes $\alpha_{vir, fil} = m_{vir, fil} / (M/L)$.  In the bottom panel of Figure \ref{Mass_vs_virial}, we plot $M/L$ versus $\alpha_{vir, fil}$ for all leaves and parent structures with aspect ratios $\geq$ 2.  All such structures fall below $\alpha_{vir, fil}$ = 2.  The high spatial correlation of dense cores with the positions of filamentary structures in dust continuum observations \citep[e.g.,][]{Konyves_2015}, combined with the larger prevalence of elongated structures being gravitationally bound, provides further evidence in support of star formation being driven by the gravitational collapse and fragmentation of filaments within molecular clouds.   

It should be noted that while dendrograms are viable options for uncovering hierarchies of related structures, they are not optimized for detecting elongated, filamentary structures.  While the structures we include in our elongated structure analysis do indeed have large aspect ratios, they are not necessarily the parsec-length filamentary networks found in dust continuum observations of star-forming regions by dedicated filament-finding algorithms \citep{Konyves_2015}.  As a result, we are likely excluding a large number of filaments in our analysis, particularly those at low surface brightness that the dendrogram has excluded from the hierarchy tree.  


\subsection{Source Temperatures and Column Densities}
Along with the kinematics of the ammonia-identified structures and \textit{Herschel}-identified dense cores, we also compare their temperatures and column densities.  Our NH$_3$ (1,1) and (2,2) line fitting procedure produces $T_K$ and NH$_3$ column density ($N_{para-NH_3}$) maps over the portion of Cepheus-L1251 mapped by our GBT observations.  We also utilize the $T_{dust}$ and H$_2$ column density maps derived by \cite{DiFrancesco_prep} for the entire Cepheus-L1251 cloud.  These maps were convolved and re-gridded to match the resolution and pixel scale of the NH$_3$ observations, allowing a pixel-by-pixel comparison between the NH$_3$ and dust continuum measurements.  

Figure \ref{Thist} shows a scatter plot of $T_{dust}$ versus $T_K$ for all pixels that fall within either an ammonia-identified leaf or \textit{Herschel}-identified core.  A Gaussian kernel density estimate of the data is also shown as a background colorscale in Figure \ref{Thist}.  Despite many of the pixels falling within a small range of temperatures and a large amount of scatter, a slight positive correlation does exist between the two temperature measurements.  The distributions of $T_{dust}$ and $T_{K}$ have median $\pm$ standard deviation values of 11.7 $\pm$ 1.1 K and 10.3 $\pm$ 2.0 K, respectively.  The side histograms in Figure \ref{Thist} show temperature distributions for all pixels falling within the individual structure categories (i.e., the ammonia-identified leaves and the \textit{Herschel}-identified dense core sub-categories).  The overall median values appear to be representative of each individual structure type, considering all the sub-categories show similar distributions of $T_{dust}$ and $T_{K}$.  These results suggest the dust and gas in the structures is likely coupled, which is predicted to occur at H$_2$ volume densities $n(H_2) > 10^{4.5}$ \citep{Goldsmith_2001}.  The slightly higher (but not significant) median $T_{dust}$ value may be caused by heated material on the outskirts of the cloud which falls along the line of sight to the observed structures.  The kinetic temperatures, derived from the dense gas emission originating in the center of the cloud, are less affected by the warmer outer layers of the cloud, which may explain the lower median $T_{K}$ value.

The abundance of para-NH$_3$ is defined as $\chi_{NH_3} = (N_{para-NH_3}/N_{H_2})$.  We divide the $N_{para-NH_3}$ map presented in this paper by the \cite{DiFrancesco_prep} \textit{Herschel}-derived H$_2$ column density map to obtain a map of $\chi_{NH_3}$ (shown in the right column of Figure \ref{Params2}).  Figure \ref{Mass_new3} shows a histogram of $\chi_{NH_3}$ for all pixels in the $N_{para-NH_3}$ map (shown as a black line) along with a similar histogram for only the pixels which fall within either an ammonia-identified leaf or \textit{Herschel}-identified dense core (grey bars).  Both distributions peak at $\chi_{NH_3} \sim 10^{-8}$ and have a range between $\chi_{NH_3} \sim 10^{-9} - 10^{-7}$.  Assuming a typical ortho-to-para ratio of 1:1 for NH$_3$, this abundance range is consistent with measurements of the \textit{total} NH$_3$ abundance observed in other nearby star-forming environments, which is generally found to be between $\sim 10^{-9}$ and $\sim 10^{-8}$ \citep{Hotzel_2001, Tafalla_2006, Crapsi_2007, Friesen_2009, Battersby_2014} but has been found to be as low as $7 \times 10^{-10}$ in B68 \citep{DiFrancesco_2002} and as high as a few times 10$^{-7}$ in Serpens South \citep{Levshakov_2013} and infrared dark clouds \citep{Ragan_2011}.  The bottom panel of Figure \ref{Mass_new3} plots $\chi_{NH_3}$ histograms for each sub-category of structures included in our analysis.  No significant shifts in abundance are observed for any of the structure categories.

\section{Discussion}

\subsection{Comparison with Serpens South Virial Analysis}
For comparative purposes, we modeled our virial analysis on the methodology adopted by \cite{Friesen_2016}, who analyzed dendrogram-identified NH$_3$ structures in the Serpens South region.  Cepheus-L1251 and Serpens South provide an interesting comparison because they have similar total masses ($\sim 700-800$ M$_\odot$), yet the latter is a much more clustered star-forming environment with higher H$_2$ column densities than the former.  Nevertheless, similar to our results in Cepheus-L1251, \cite{Friesen_2016} found that nearly all of the ammonia-identified structures (both leaves and parents) in Serpens South had $\alpha_{vir}$ $<$ 2.  \cite{Friesen_2016} also found that the elongated, filamentary structures in their sample had the lowest virial parameters and the region's magnetic field is unable to provide sufficient support to prevent their gravitational collapse.  This result lends credence to the radial infall motions observed both along and onto the main filament in Serpens South \citep{Friesen_2013, Kirk_2013}.  Despite the lack of large-scale magnetic field measurements in Cepheus, our virial analysis of L1251, which also showed that nearly all the ammonia-identified structures in the region are gravitationally bound when magnetic pressure is neglected and a power-law density profile is assumed for the structures, matches the results of \cite{Friesen_2016}.  

The similarity in virial parameters for structures in Serpens South and Cepheus-L1251 is an intriguing result, especially when considering the two clouds constitute drastically different star-forming environments.  For instance, Serpens South resides in the Aquila Rift cloud complex, which \cite{Konyves_2015} showed had an H$_2$ column density PDF peak (i.e., the most common column density in the region) of $4-5 \times 10^{21}$ cm$^{-2}$.  In contrast, \cite{DiFrancesco_prep} found that Cepheus has a comparatively much lower amount of material at high column density, with an H$_2$ column density PDF peak of only $\sim 1 \times 10^{21}$ cm$^{-2}$.  Furthermore, Serpens South has a higher average gas mass surface density compared to Cepheus-L1251.  We use the \textit{Herschel} H$_2$ column density maps derived by \cite{Konyves_2015} for Serpens South and by \cite{DiFrancesco_prep} for Cepheus-L1251 to derive total masses for the two clouds.  In a 0.1 square degree area centered on Serpens South, we find its mass to be $\sim 800$ M$_\odot$ assuming a distance of 260 pc \citep{S_2003}.  Alternatively, if we adopt the more recently measured distance of 430 pc to Serpens South \citep{Dzib_2011, Ortiz_2017}, its total mass becomes $\sim 2200$ M$_\odot$.  Assuming a distance of 300 pc for Cepheus-L1251 and using the 0.36 square degree area observed for our NH$_3$ observations, we find its total mass to be $\sim 680$ M$_\odot$.  We convert these values to mass surface densities and find $\Sigma \sim 370$ M$_\odot$ pc$^{-2}$ for Serpens South and  $\sim 70$ M$_\odot$ pc$^{-2}$ for Cepheus-L1251.  Even if we calculate the mass within a 0.1 square degree box centered on L1251E/B, the most active and clustered portion of L1251, we find $\Sigma \sim 100$ M$_\odot$ pc$^{-2}$.  Thus, independent of the cloud-scale environment, it appears that dense ammonia structures are gravitationally bound when magnetic pressure is not considered and a power-law density distribution is assumed for the structures.


An additional, noteworthy difference between Serpens South and Cepheus-L1251 is the \textit{number} of gravitationally bound NH$_3$ structures in each cloud.  \cite{Friesen_2016} found 85 NH$_3$ leaves in Serpens South compared to the 22 leaves we identify in Cepheus-L1251 over a larger area.  Similarly, Serpens South has a higher amount of YSOs compared to Cepheus-L1251.  We use the \cite{Dunham_2015} YSO catalog, which includes Class 0$-$III sources identified by the \textit{Spitzer} ``cores-to-disks" and ``Gould Belt" Legacy surveys, to determine the number of protostars within each region.  We find that 133 Class 0$-$III sources fall within the 0.1 square degree area in Serpens South used above to calculate the cloud's mass, while Cepheus-L1251 contains only 36 Class 0$-$III sources within the 0.36 square degree area observed for our NH$_3$ observations.  These discrepancies show that Serpens South is producing both gravitationally bound structures and protostars at a higher rate than Cepheus-L1251.  This production difference may be linked to the clustered star-forming environment that is found in Serpens South, which is likely more efficient at funneling gas onto forming dense cores than the more isolated environs of Cepheus-L1251.  While stars and dense cores are still being formed in Cepheus-L1251, its lower overall column densities might be preventing a higher rate of dense core and protostar production.  

Yet another dissimilarity between Serpens South and Cepheus-L1251 is the alignment between their NH$_3$ leaves and \textit{Herschel} cores.  For instance, 75 of the 85 NH$_3$ leaves ($\sim88 \%$) in Serpens South were found to have a \textit{Herschel} core counterpart with a center within one $R_{eff}$ of the leaf center.  In Cepheus-L1251, the same matching criterion reveals that only 4 of its 22 NH$_3$ leaves ($\sim18 \%$) have a \textit{Herschel} core counterpart.  Even when using 32$\arcsec$ (the FWHM beam-size of the NH$_3$ observations) as the maximum angular offset for the cross-match, which is larger than the $R_{eff}$ of all sources, only 13 of 22 NH$_3$ leaves ($\sim60 \%$) in Cepheus-L1251 have a matched \textit{Herschel} core.  In either case, the higher ratio of ammonia leaves that contain continuum counterparts observed in Serpens South may be due to its highly clustered environment.  Alternatively, chemical effects related to higher density could be causing the continuum in Serpens South to be a better probe of NH$_3$ gas when compared with Cepheus-L1251.

Despite their different levels of correspondence between dense gas and continuum structures, the NH$_3$ in both Serpens South and Cepheus-L1251 is tracing a higher fraction of prestellar and protostellar cores than the continuum.  Namely, the \textit{Herschel} core counterpart for 66 of the 75 matched NH$_3$ leaves ($88 \%$) in Serpens South were classified as either prestellar or protostellar.  Likewise, all 13 of the matched NH$_3$ leaves in Cepheus-L1251 had either a prestellar or protostellar counterpart.  This result provides further evidence that ammonia is predominately tracing dense, gravitationally bound structures regardless of cloud-scale environment.  





\subsection{Role of External Pressure}
Although the virial analysis presented in this paper has shown the gravitational boundedness of structures in L1251, we have not yet considered the role of external pressure by the ambient cloud.  \cite{Seo_2015} combined CO and NH$_3$ observations to show that the majority of NH$_3$ dendrogram-identified leaves in Taurus B211/213 that were unbound gravitationally are confined by the external pressure exerted by the filaments and clouds within which the leaves reside.  Similarly, \cite{Pattle_2017} combined $^{13}$CO observations with a JCMT GBS-identified dense core catalog for several Cepheus Flare regions (including L1251) and determined that the majority of Cepheus dense cores are pressure-confined rather than gravitationally bound.  Our virial analysis for L1251, which was able to incorporate the internal line widths of structures that the \cite{Pattle_2017} analysis lacked, supports their result since we have shown that the \textit{Herschel}-identified structures in the region are mostly gravitationally unbound when magnetic pressure is neglected.  Furthermore, \cite{Kirk_submitted} use GAS observations of Orion A to determine the virial states of dense cores identified by the JCMT Gould Belt Survey and found that while few cores in the region are bound by self-gravity, nearly all are pressure-confined when taking into consideration the weight of the encompassing molecular cloud.  Therefore, many of the structures we find within L1251 that are gravitationally unbound could also be predominantly pressure-confined.  We calculate below the effect of external pressure from the L1251 molecular cloud  upon the virial stability of its structures.  

We follow the methodology outlined in \cite{Pattle_2015} and \cite{Kirk_submitted} to express the virial contributions of external pressure, self-gravity, and thermal plus non-thermal motions using the following energy density equations:

\begin{equation}
\Omega_P = -4 \pi P R^3
\end{equation}

\begin{equation}
\Omega_G = \frac{-1}{2\sqrt{\pi}}\frac{GM^2}{R}
\end{equation}

\begin{equation}
\Omega_K = \frac{3}{2}M\sigma^2
\end{equation} where $M$ is the observed structure mass, $R$ is the effective radius, $G$ is the gravitational constant, $\sigma^2$ is the same as Equation 3, and $P$ is pressure.  As noted in Section 3.4, Equation 8 assumes that the structures have a Gaussian density profile.  This contrasts with the power-law density distribution assumed by \cite{Friesen_2016} that we adopted in Section 3.4.  Due to this discrepancy, we calculate $\Omega_G$ using each density profile and compare the resulting virial parameters in the forthcoming discussion. 

We estimate the pressure on the structures consists of two main components: 1) the turbulent pressure ($P_T$) exerted by turbulent motions within the parental cloud and 2) the pressure supplied by the weight of the parental cloud ($P_W$).\footnote{We do not consider the pressure exerted by filaments, as was included by \cite{Kirk_submitted} in their Orion A virial analysis, due to the following reasons: 1) There is less obvious filamentary structure in Cepheus-L1251 compared to Orion A. 2) \citeauthor{Kirk_submitted} estimated the filament pressure did not contribute significantly to the total external pressure on cores in Orion A.}  Thus, $P = P_T + P_W$, where the turbulent component is given by:
\begin{equation}
P_T=\mu_H m_H\times \rho_{^{13}CO}\times \sigma ^2_{^{13}CO} ~,
\end{equation}$\sigma_{^{13}CO}^2$ is the line width of emission from $^{13}$CO $(1-0)$, a transition that traces the outer envelopes of dense cores, and $\rho_{^{13}CO}$ is the volume density at which the $^{13}$CO $(1-0)$ emission originates.  Here, we use a previously measured value of $\sigma_{^{13}CO}$ = 0.8 km s$^{-1}$ obtained by \cite{Yonekura_1997} for a single pointing at 2.4 arcminute resolution in the eastern portion of Cepheus-L1251.  In the absence of other data, we assume this single line width for all structures in our analysis, which is consistent with the virial analysis presented in \cite{Pattle_2017}.  We also set $\rho_{^{13}CO}$ = $5.5 \times 10^{3}$ cm$^{-3}$, which is the average of the typical density range traced by $^{13}$CO $(1-0)$ \citep[$n \sim 10^{3} - 10^{4}$ cm$^{-3}$; ][]{DiFrancesco_2007}.  The cloud weight pressure term is given by:

\begin{equation}
P_W = \pi G \bar{N} N (\mu_H m_H)^2
\end{equation} where $\bar{N}$ is the mean cloud column density and $N$ is the column density at the structure \citep[e.g., ][]{Mckee_1989, Kirk_2006, Kirk_submitted}.  We note that Equation 11 applies only to spherical clouds with a density profile given by $\rho \propto r^{-k}$, with $k = 1$ \citep[see Appendix C in][]{Kirk_submitted}.  When $1 < k < 3$, an additional positive term is added to the expression for $P_W$, making the estimate presented here a lower limit.

To isolate the large-scale features of Cepheus-L1251 that are contributing to the cloud weight pressure, we calculate $\bar{N}$ and $N$ using a spatially filtered version of the \cite{DiFrancesco_prep} H$_2$ column density map.  Following a method outlined in \cite{Kirk_submitted}, we use the \textit{a Trous} wavelet transform to determine the column density contribution at each spatial scale\footnote{We use the \textit{atrous.pro} IDL script developed by Erik Rosolowsky that is available at \url{https://github.com/low-sky/idl-low-sky/blob/master/wavelet/atrous.pro}}.  The wavelet transform smooths the image on pixel scales with increasing powers of 2$^N$ and subtracts the contribution of that scale from the image until the pixel scale exceeds the image size.  Since the clumps within which dense cores reside have sizes larger than a few tenths of a parsec, we add the column density contribution of the scales $\geq$ 128 pixels, which corresponds to $\sim$ 0.6 pc.  We calculate $\bar{N}$ using the portion of the spatially filtered map that was observed for our NH$_3$ observations, which produces a mean of 2.1$\times$10$^{21}$ cm$^{-2}$.  

Under this formalism, a structure is in virial equilibrium when $2\Omega_K$ = $-1(\Omega_G + \Omega_P$).  We define all structures with an observed mass larger than their virial mass to be ``sub-virial.''  Conversely, any structure with a mass smaller than its corresponding virial mass is designated ``super-virial.''  The left panel of Figure \ref{virial_plane}, which we refer to as the ``virial plane,'' displays the balance between the three aforementioned energy densities for the same structures included in the virial analysis presented in Section 3.4.  Structures to the right of the vertical line in Figure \ref{virial_plane} are sub-virial, while structures on the left are super-virial. Structures that lie below the horizontal dotted line in Figure \ref{virial_plane} are dominated by pressure over gravity, while structures above the line are dominated by gravity over pressure.  When adding pressure to the virial equation and assuming a Gaussian density profile for the structures (shown in the left panel of Figure \ref{virial_plane}), many of the \textit{Herschel}-identified cores are revealed to be dominated by pressure rather than gravity.  Specifically, fourteen of the 57 \textit{Herschel}-identified cores are dominated by gravity over pressure, with only two of those being sub-virial.  For the other 43 pressure-dominated \textit{Herschel}-identified cores, 16 are super-virial and 27 are sub-virial.  On the other hand, all the ammonia-identified leaves are dominated by their gravitational potential energy.  

In the right panel of Figure \ref{virial_plane}, we show how the virial plane would shift if the density profile for the structures is assumed to be a power-law consistent with the virial analysis presented in Section 3.4.  Since the density profile is incorporated into the $\Omega_G$ term, the virial parameters of the gravitationally-dominated structures are more sensitive to these assumptions.  When a Gaussian density profile is assumed, nearly all of the gravitationally-dominated structures are super-virial.  Conversely, when a power-law density profile is assumed, nearly all the gravitationally-dominated structures become sub-virial.  While this result demonstrates the uncertainty surrounding virial parameter measurements, it may also indicate that these gravitationally-dominated structures, many of which are ammonia-identified leaves, are currently in or near a state of virial equilibrium.  Such a scenario would explain why the NH$_3$ emission is less well matched to the dense cores in L1251, as was discussed in Section 4.1, since structures in virial equilibrium are unable to undergo the gravitational collapse required to form cores and protostars.


Our results after adding pressure to the virial equation contrast with the aforementioned work of \cite{Pattle_2017}.  \cite{Pattle_2017} found that all 34 of the sources they identified in L1251 from the region's 850 $\mu$m emission were sub-virial and all but four of those 34 were pressure-dominated.  The discrepancy between our results likely arises because the virial parameters presented in \cite{Pattle_2017} were upper limits due to the lack of non-thermal line width measurements for their cores.  Our spectroscopic information shows that the structures in Cepheus-L1251 are dominated by non-thermal motions, which results in higher $\Omega_K$ terms (i.e., structures become more super-virial) than those reported by \cite{Pattle_2017}. 


Similarly, our results differ from those of \cite{Kirk_submitted}, which did incorporate GAS NH$_3$ spectroscopic information into their virial parameter calculations.  They find that most of the cores in Orion A are pressure-dominated and sub-virial, with a much lower fraction of gravitationally-dominated structures than we observe in Cepheus-L1251.  The giant molecular cloud environment of Orion A is likely the culprit causing the higher rate of pressure-dominated and sub-virial cores in the \cite{Kirk_submitted} analysis.  For instance, \cite{Kirk_submitted} measure the average column density of the large-scale structures in Orion A to be over an order of magnitude larger ($\bar{N}$ = 3.9$\times$10$^{22}$ cm$^{-2}$) than our estimate for Cepheus-L1251 ($\bar{N}$ = 2.1$\times$10$^{21}$ cm$^{-2}$).  Assuming the column density measured at each core (i.e., $N$ in Equation 11) in Orion A is also roughly an order of magnitude larger than those measured for the structures in Cepheus-L1251, that would cause the $P_W$ estimates in Orion A to be over two orders of magnitude larger than those of Cepheus-L1251.  Such an increase in pressure would move many of the structures in our analysis into the sub-virial and pressure-dominated regime of the virial plane.

The low column densities and high non-thermal motions in Cepheus-L1251 cause the turbulent pressure to be the dominant term in $\Omega_{P}$ for our analysis, with $P_T$ being a factor of $2-9$ larger than $P_W$ for all structures.  This result displays the need for similar resolution $^{13}$CO measurements across Cepheus-L1251, which are required to understand whether or not the 0.8 km s$^{-1}$ estimate of $\sigma_{^{13}CO}$ adequately represents the motions of the parental cloud nearest to each structure identified in our analysis.  We also note that the internal magnetic fields of the structures, which may provide additional support against gravitational collapse, could cause the virial terms presented on the x-axis of Figure \ref{virial_plane} to move toward being super-virial.  Additional sources of external pressure may also be acting on the observed structures, however, which would move the data points in Figure \ref{virial_plane} towards the sub-virial regime on the right side of the virial plane.  Such sources of pressure include radiation from the interstellar field \citep{Seo_2016} and the weight of the filaments in which the observed structures are embedded \citep{Kirk_submitted}.


\subsection{Presence of CCS and HC$_5$N}
Although our analysis has focused on the identification of dense gas structures in Cepheus-L1251, the GAS observations also reveal significant amounts of CCS $(2_0-1_0)$ and HC$_5$N $(9-8)$ emission towards the highest H$_2$ column density portions of the region.  The presence of carbon-chain molecules, including CCS and HC$_5$N, is thought to indicate the early stages of chemical evolution in dense cloud cores \citep{Benson_1983, Suzuki_1992, Friesen_2013}.  As cores evolve, large carbon-chain molecules deplete onto the surface of dust grains and their production rate decreases as higher fractions of the available carbon atoms react to form CO \citep{Caselli_1999}.  When densities are large enough for CO to be heavily frozen onto dust grains, nitrogen-bearing molecules, such as NH$_3$ and N$_2$H$^+$, increase their abundance.  Thus, the presence or absence of CCS and HC$_5$N can provide insight into the chemical evolution of the structures identified within Cepheus-L1251.  


Here, we perform Gaussian line-fitting on the CCS $(2_0-1_0)$ and HC$_5$N $(9-8)$ maps observed toward Cepheus-L1251 to understand which parts of the region contain significant emission from these carbon-chain molecules.  The CCS $(2_0-1_0)$ and HC$_5$N $(9-8)$ maps observed by GAS toward Cepheus-L1251 contain median rms noise in the off-line channels of 0.23 K and 0.11 K, respectively.  CCS $(2_0-1_0)$ was only observed in the central beam of the seven beam GBT K-band Focal Plane Array, hence its higher rms values.  Due to the faint nature of the CCS and HC$_5$N emission, we first convolve each map to an angular resolution of 64$\arcsec$ (FWHM).  A Gaussian profile is then fit to all pixels with SNR $>$ 3, where the SNR is defined as the ratio of the peak brightness temperature of the emission line to the standard deviation of the off-line channels.  The Levenberg-Marquardt implementation of non-linear least squares is adopted for the fitting.  Synthetic Gaussian cubes, along with individual parameter maps for peak brightness temperature, $V_{LSR}$, and $\sigma$, are obtained from the best-fit results.  1-$\sigma$ statistical uncertainties on the best-fit parameters in each pixel are estimated from the diagonal terms in the covariance matrix.

Figure \ref{CCS_HC5N} displays integrated intensity maps constructed from the synthetic Gaussian cubes created from our line-fitting of the CCS $(2_0-1_0)$ and HC$_5$N $(9-8)$ emission.  For comparative purposes, the plots shown in Figure \ref{CCS_HC5N} match the NH$_3$ (1,1) integrated intensity fields displayed in Figure \ref{leaves}.  Although the CCS $(2_0-1_0)$ and HC$_5$N $(9-8)$ emission tend to correlate with the positions where NH$_3$ (1,1) and (2,2) are detected, spatial differences are seen between the three molecular species.  The positions of the dendrogram-identified NH$_3$ leaves found in this paper are also overlaid onto Figure \ref{CCS_HC5N}.  While some of the NH$_3$ structures coincide with CCS $(2_0-1_0)$ and HC$_5$N $(9-8)$ emission, others contain no detectable emission in either carbon-chain molecule or contain emission from only one of the carbon-chain molecules.  Specifically, 17 of 22 ammonia-identified leaves contain detectable CCS $(2_0-1_0)$ emission in at least one-third of the pixels falling within their 2D mask.  Using the same criteria, 16 of 22 ammonia-identified leaves contain detectable HC$_5$N $(9-8)$ emission.  

Maps of the ratio of the integrated intensity of CCS $(2_0-1_0)$ and HC$_5$N $(9-8)$ to NH$_3$ (1,1) are shown in Figure \ref{CCS_HC5N_ratios}.  In the optically thin limit, these intensity ratios serve as proxies for the abundance ratios of CCS and HC$_5$N relative to NH$_3$.  L1251E/B, located in the eastern-most portion of Cepheus-L1251, contains the lowest relative ratios of both CCS $(2_0-1_0)$ and HC$_5$N $(9-8)$ emission relative to NH$_3$ (1,1) compared with L1251C and L1251A in the central and western portions of the cloud, respectively.  L1251E/B also corresponds with the most star and core formation throughout Cepheus-L1251.  The lower relative levels of carbon-chain emission from L1251E/B may be indicating it is the most chemically evolved sub-region within Cepheus-L1251.  Similarly, the sub-region's higher levels of star-formation activity suggest it is dynamically more evolved as well.  

In Figure \ref{sigma_carbons}, we compare the measured line widths of the CCS $(2_0-1_0)$ and HC$_5$N $(9-8)$ emission to those measured in the NH$_3$ (1,1) map, which was also convolved to a resolution of 64$\arcsec$ and re-run through the GAS line-fitting pipeline.  Comparisons are made for all pixels meeting the following criteria: 1) the pixel falls within the 2D mask of an ammonia-identified leaf and 2) the pixel contains reliable fits for all three transitions.  The top panel of Figure \ref{sigma_carbons} shows that the line width distributions for each transition peak at different values, with the NH$_3$ peak occurring at $\sigma \sim 0.15$ km s$^{-1}$ and the HC$_5$N peak occurring at $\sigma \sim 0.2$ km s$^{-1}$.  The CCS line width distribution is much broader than those of the other two molecules, but its median value of $\sigma \sim 0.18$ km s$^{-1}$ lies between the medians of the other molecules.  The small peaks near 0.3 km s$^{-1}$ in the NH$_3$ and HC$_5$N distributions are driven by several leaves in L1251E/B, the most active region throughout L1251.

In the bottom panel of Figure \ref{sigma_carbons}, we plot a histogram of the difference between the NH$_3$ line width and the HC$_5$N or CCS line width ($\Delta\sigma$) for each individual pixel.  The $\Delta\sigma$ distribution for ${\rm HC_5N} - {\rm NH_3}$ peaks at 0.05 km s$^{-1}$ and shows that the majority of the pixels in the map have $\Delta\sigma \gtrsim 0.05$  km s$^{-1}$.  This difference is equivalent to about a single resolution element in the spectrum (i.e., $\sim$ 0.07 km s$^{-1}$).  The $\Delta\sigma$ distribution for ${\rm CCS} - {\rm NH_3}$ is also skewed to values greater than zero, but the distribution is much broader than that of ${\rm HC_5N} - {\rm NH_3}$.  We estimate uncertainties on $\Delta\sigma$ by adding in quadrature the 1-$\sigma$ uncertainties on the best-fit values for $\sigma$ at each pixel.  The median uncertainty on $\Delta\sigma$ is $\sim$ 0.02 km s$^{-1}$ for both ${\rm HC_5N} - {\rm NH_3}$ and ${\rm CCS} - {\rm NH_3}$, which suggests the observed values of $\Delta\sigma$ are significantly different from zero for most pixels.  

This result is interesting when considering CCS and HC$_5$N are $\sim$ 3 and $\sim$ 4 times more massive than NH$_3$, respectively.  If all three molecules are tracing the same volume of gas imbued with common turbulent motions, one would naively expect NH$_3$ to have the largest line widths of the three tracers due to its lower molecular mass.  Namely, the lighter NH$_3$ molecules would, on average, have greater thermal velocities than the heavier molecules. The larger line widths for the carbon-based species, however, suggest they are not tracing the same gas as NH$_3$.  Instead, the carbon-based molecules may be tracing the outer layers of the cloud, which are likely more turbulent than its central regions.  This claim supports previous observations by \cite{Pineda_2010} that suggest the dissipation of turbulence within dense cores provides the conditions necessary for their gravitational collapse to form protostars.  The de-coupling of the carbon-based molecular species observed in this paper from NH$_3$ could be caused by the depletion of the carbon-based species in the central, colder regions of the cloud.  This scenario would also explain the offsets observed between the NH$_3$ (1,1) emission peaks and those of the carbon-based tracers.

In addition, we also observe significant centroid velocity offsets (up to $\sim$ 0.5 km s$^{-1}$) between the emission from NH$_3$ and the carbon-based molecules for several sources.  These offsets can clearly be seen in Figure \ref{vlsr_carbons}, which plots the $\Delta V_{LSR}$ distribution for all pixels falling within an ammonia-identified leaf.  While most of the $V_{LSR}$ measurements between the three transitions are in agreement, there is a considerable number of pixels with $\Delta V_{LSR} > 0.5$ km s$^{-1}$ for HC$_5$N and $\Delta V_{LSR} \sim 0.3$ km s$^{-1}$ for CCS.  Such large offsets are significant considering the median uncertainty on $\Delta V_{LSR}$, estimated using the same method discussed above for the uncertainty on $\Delta\sigma$, is $\sim$ 0.02 km s$^{-1}$ for both ${\rm HC_5N} - {\rm NH_3}$ and ${\rm CCS} - {\rm NH_3}$.  Furthermore, the pixels with high $\Delta V_{LSR}$ values are not dispersed evenly throughout the map, but rather isolated within specific leaves.  This isolation can clearly be seen in the top right panel of Figure \ref{vlsr_carbons}, which plots the weighted average $\Delta V_{LSR}$ for all pixels falling within the 2D mask of each individual ammonia-identified leaf, weighted by the peak brightness temperature of each respective emission line.  While most of the leaves hover around $\Delta V_{LSR}$ = 0 km s$^{-1}$, sources 12 and 16 show $\Delta V_{LSR} > 0.5$ km s$^{-1}$ for HC$_5$N and $\Delta V_{LSR} > 0.3$ km s$^{-1}$ for CCS.  The weighted average spectra for each source, plotted in the bottom panels of Figure \ref{vlsr_carbons} and weighted by the peak brightness temperature of each respective emission line, show that the calculated offsets do indeed correspond to real features in the spectra.

For sources 12 and 16, the largest centroid velocity offsets are observed between NH$_3$ and HC$_5$N, which is the heavier of the two observed carbon-based species.  \cite{Ohashi_2016} observed a similar offset between NH$_3$ (1,1),  $^{13}$CO $(1-0)$, and CS $(2-1)$ emission toward a starless dense core in the Orion A molecular cloud named TUKH122.  There, they found the NH$_3$ (1,1) emission had a centroid velocity 0.7 km s$^{-1}$ lower than the $^{13}$CO $(1-0)$ and CS $(2-1)$ emission.  Similarly, \cite{Swift_2005} observed a 0.16 km s$^{-1}$ offset between the centroid velocities of NH$_3$ (1,1) and C$_2$S $(3_2-2_1)$ emission towards a pre-protostellar core in L1551.  They interpret their result as arising from either an infalling or outflowing, lower-density envelope surrounding the high-density gas forming the protostar.  The former is a more likely explanation for sources 12 and 16 in our analysis, since both are likely prestellar cores because they lack a \textit{Herschel}-identified or \textit{Spitzer}-identified protostellar counterpart.

\begin{figure}[ht]
\epsscale{1.0}
\plotone{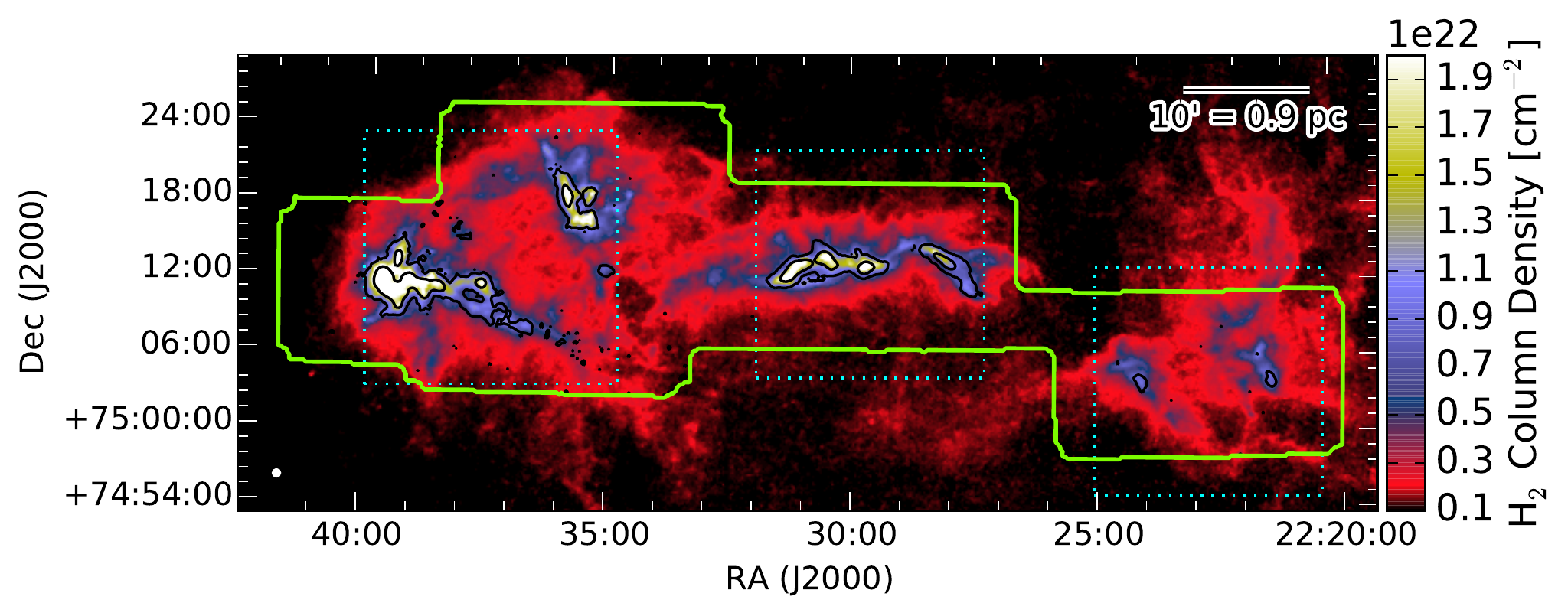}
\caption{Background colorscale shows the H$_2$ column density map of Cepheus-L1251 derived by \cite{DiFrancesco_prep}.  The green contour outlines the region observed in NH$_3$ (1,1) and (2,2) by GAS \citep{Friesen_submitted} using the Green Bank Telescope.  The black contours show the NH$_3$ (1,1) integrated intensity at 0.5 and 3.5 K km s$^{-1}$.  The cyan dotted lines denote the extents of the three zoomed panels shown in Figure 2.  The 32$\arcsec$ FWHM beam-size is shown by the white circle in the lower left corner.}
\label{obs}
\end{figure}

\begin{figure}[ht]
\epsscale{1.0}
\plottwo{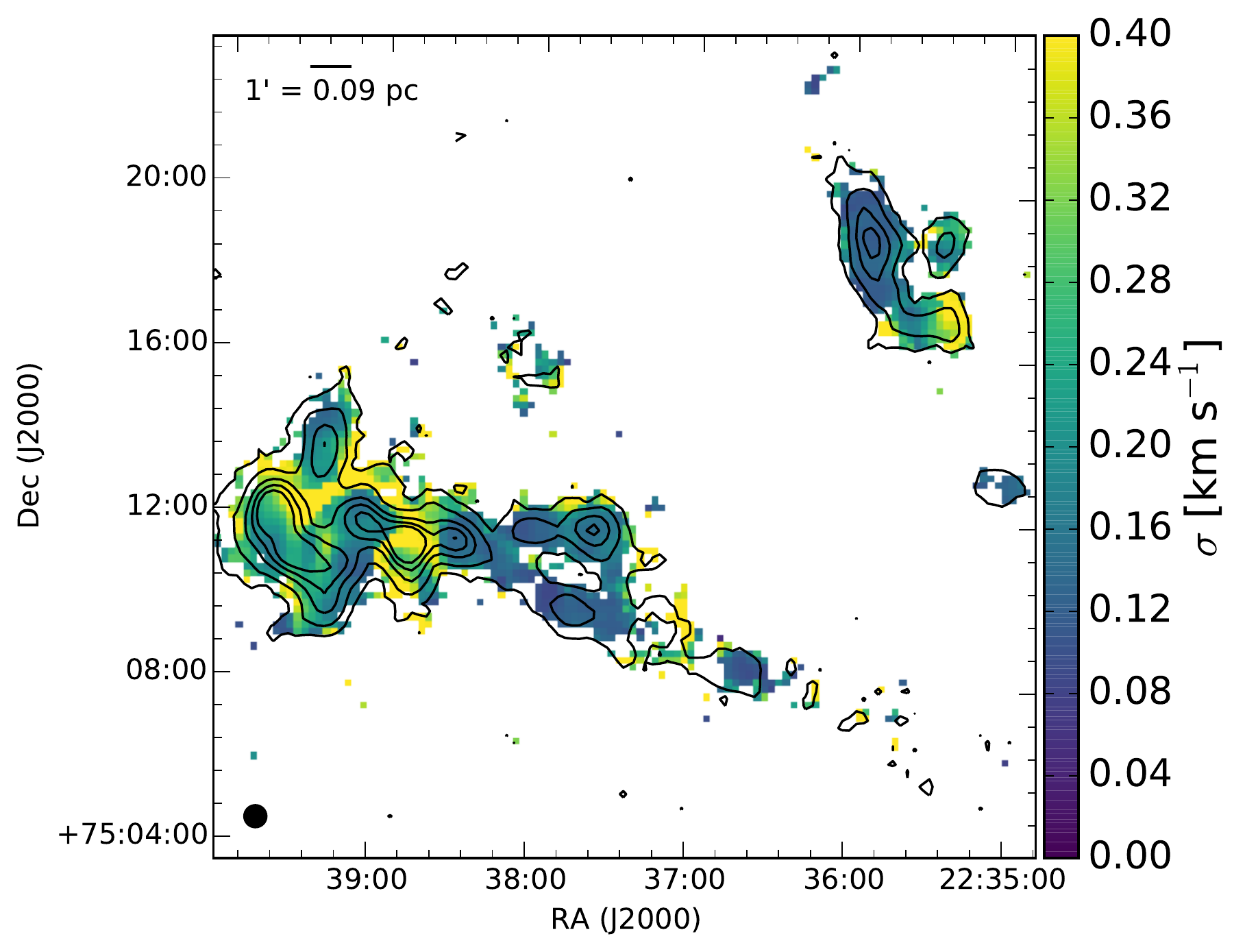}{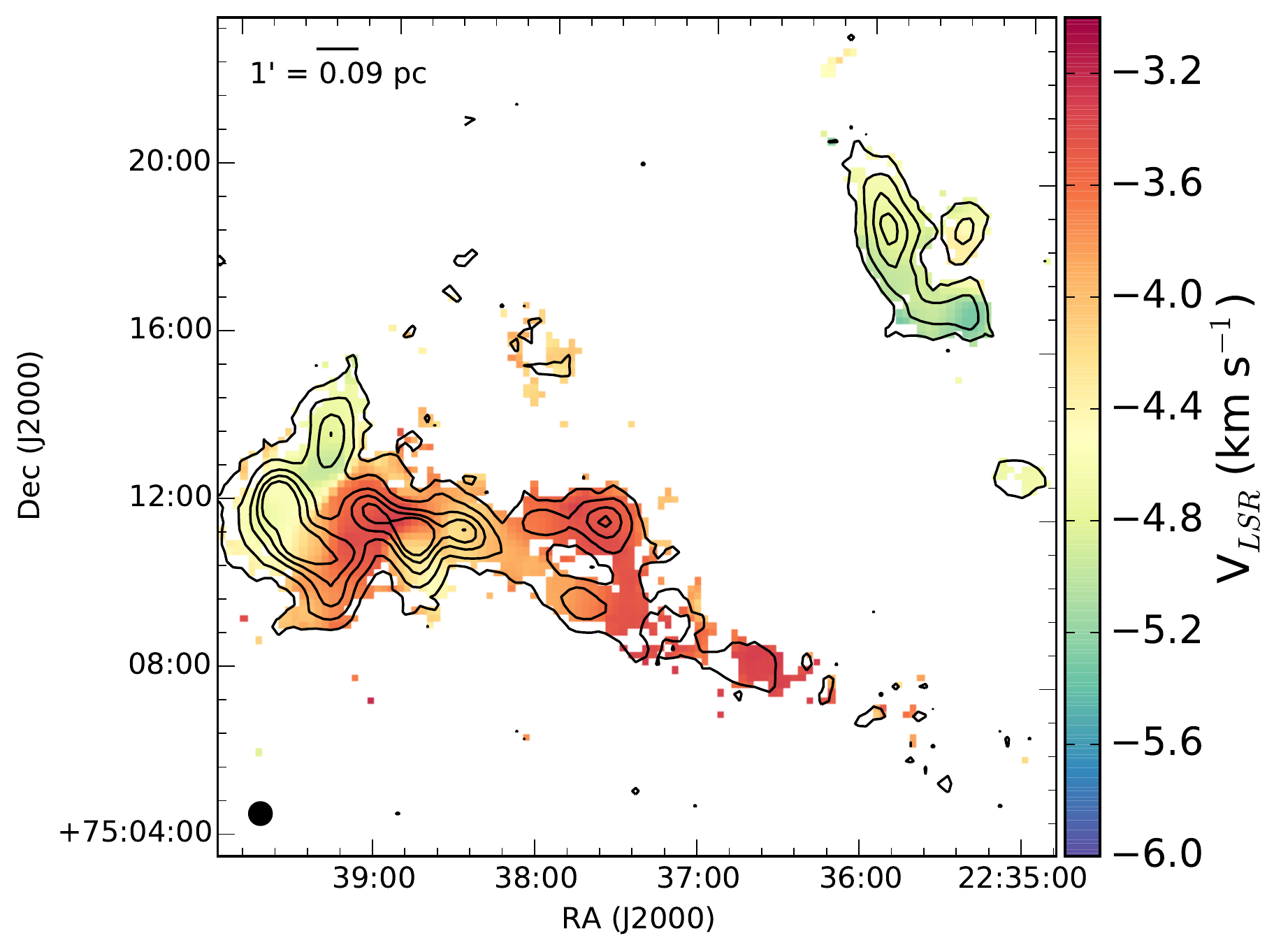} \\
\plottwo{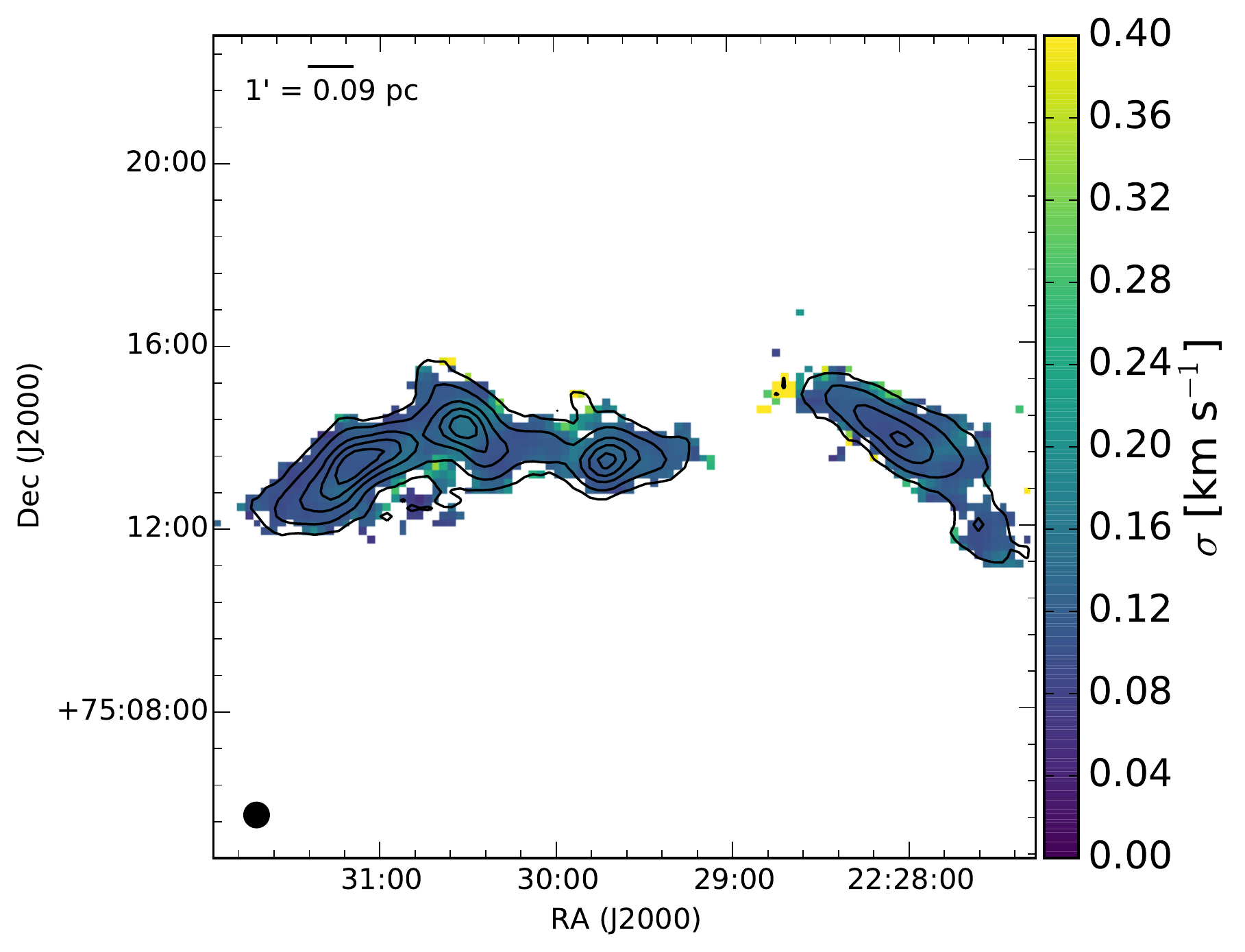}{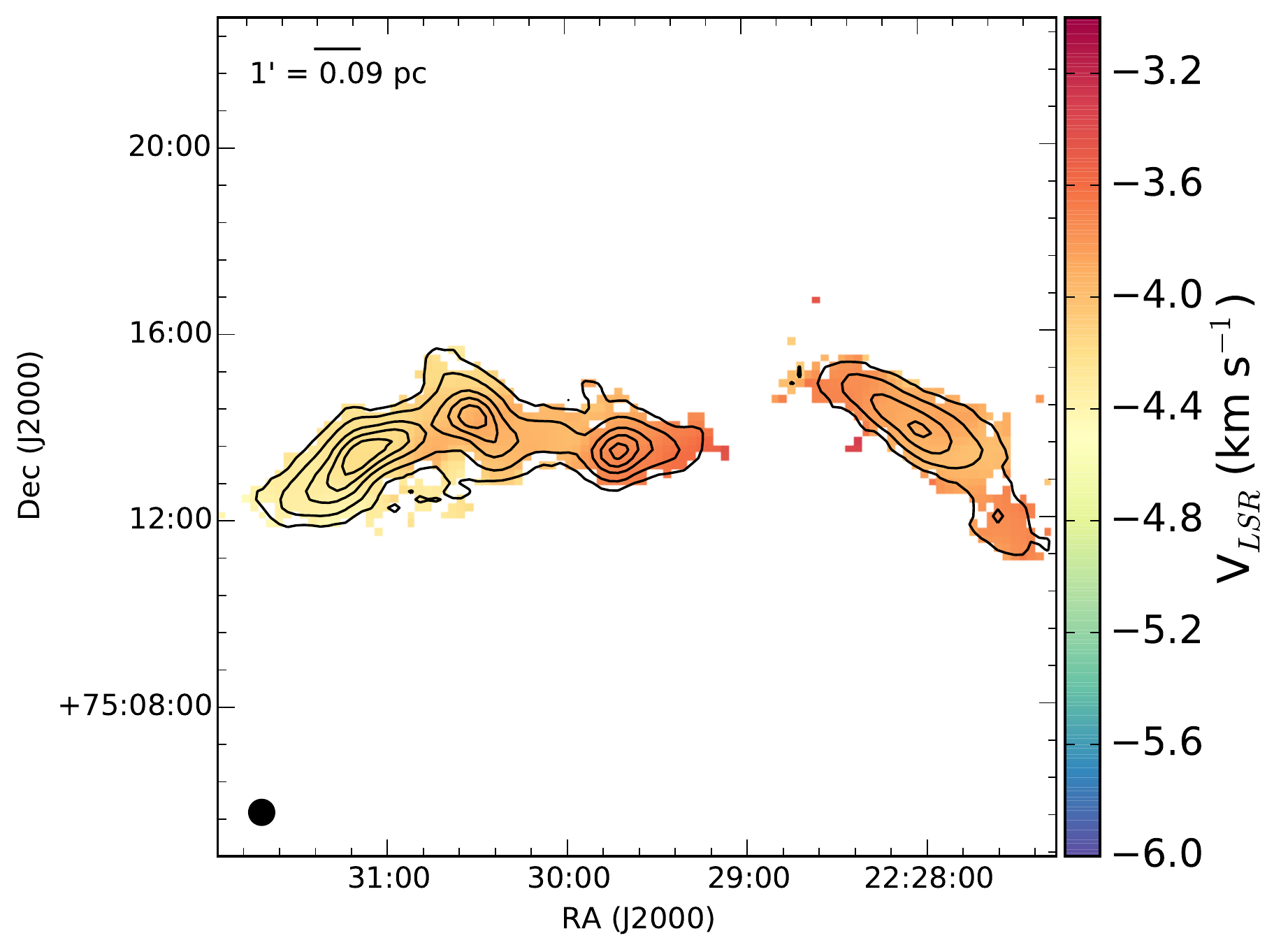} \\
\plottwo{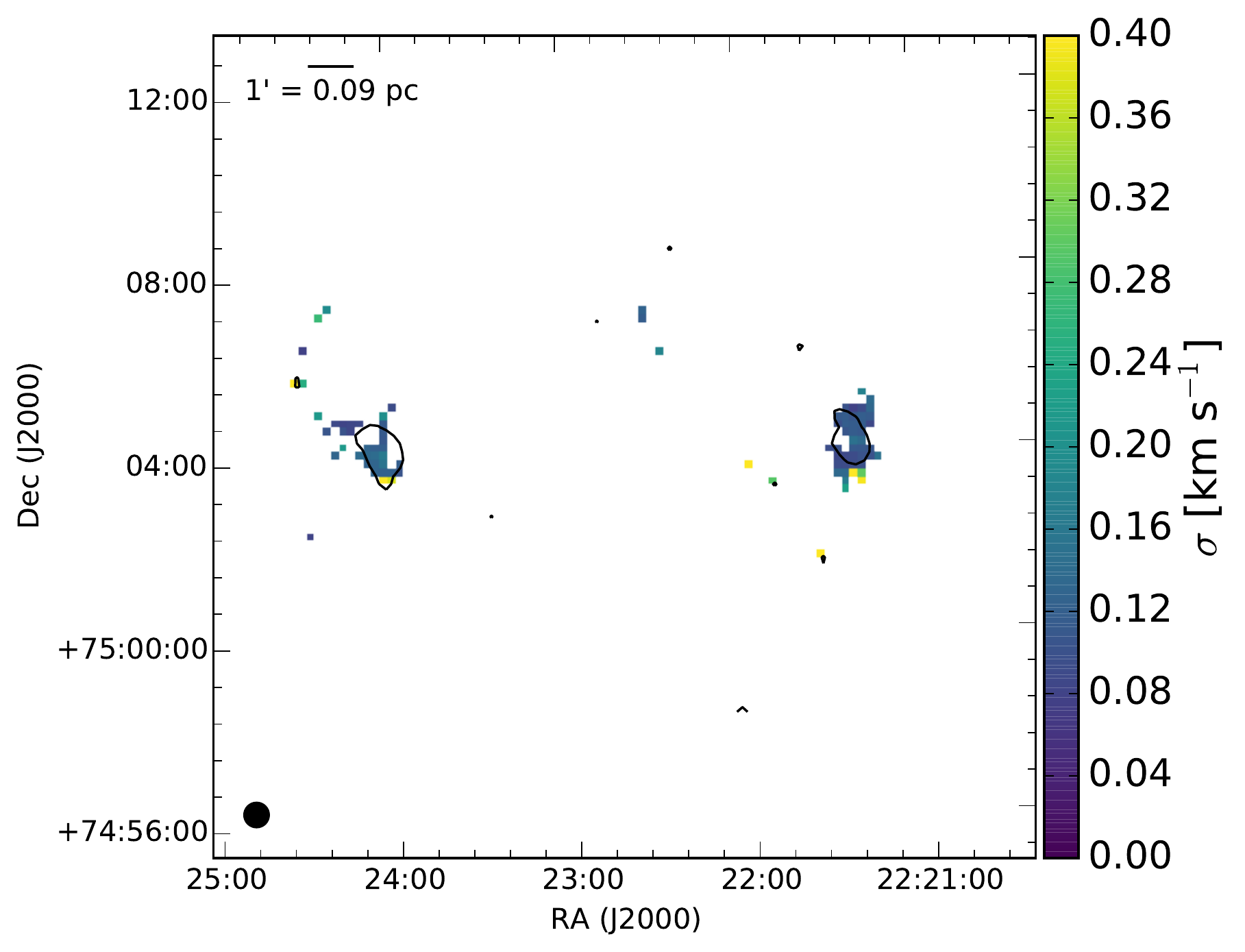}{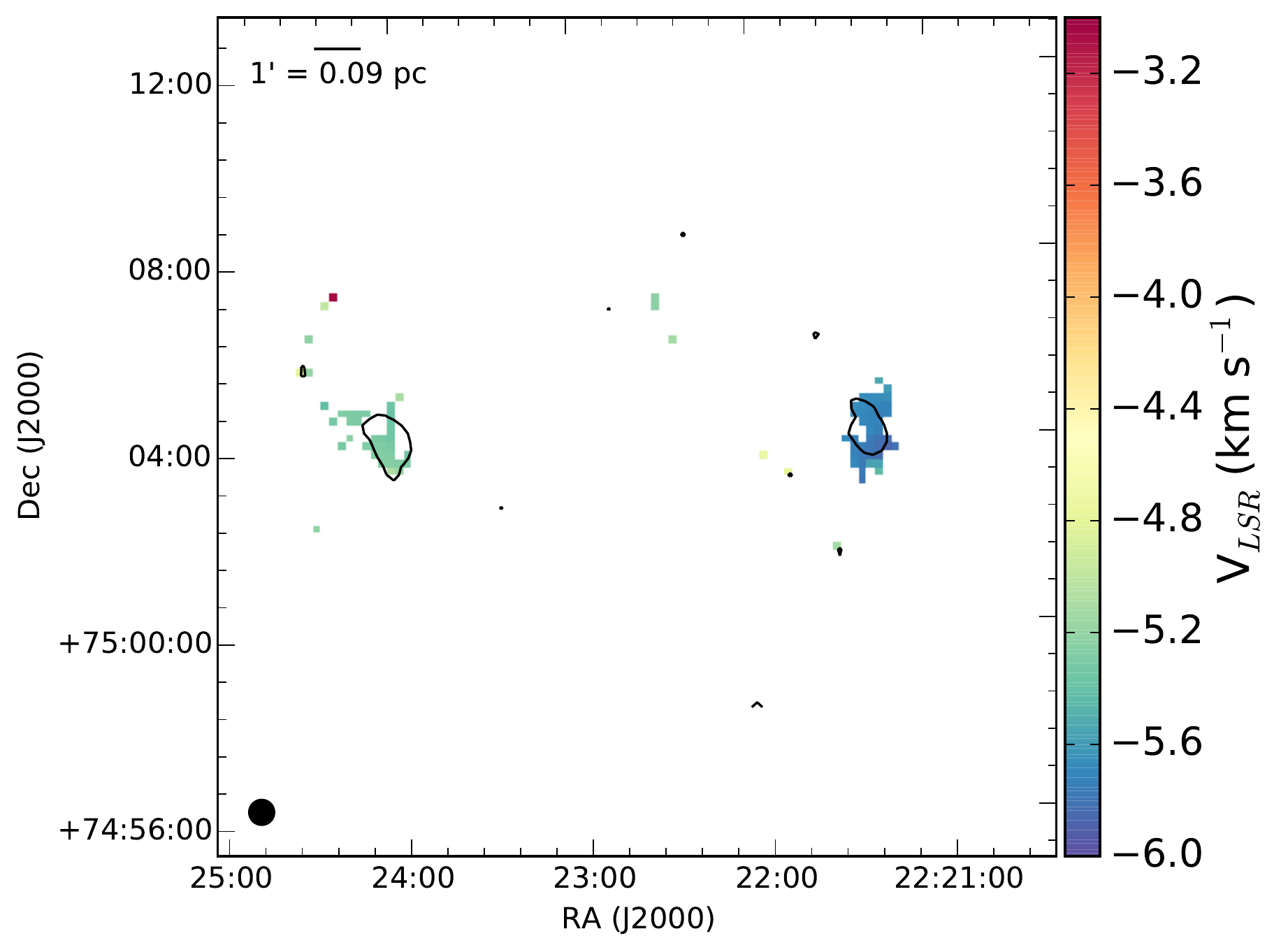}
\caption{Velocity dispersion (left column) and centroid velocity (right column) of the NH$_3$ (1,1) emission from Cepheus-L1251.  Black contours represent the NH$_3$ (1,1) integrated intensity at 0.5, 1.5, 3.5, 5.5, and 7.5 K km s$^{-1}$.  The panels display the eastern (L1251E/B, top panel), central (L1251A, middle panel), and western (L1251C, bottom panel) portions of L1251.  The 32$\arcsec$ FWHM beam-size is shown as a black circle in the lower left corner of each panel.}
\label{Params1}
\end{figure}

\begin{figure}[ht]
\epsscale{1.0}
\plottwo{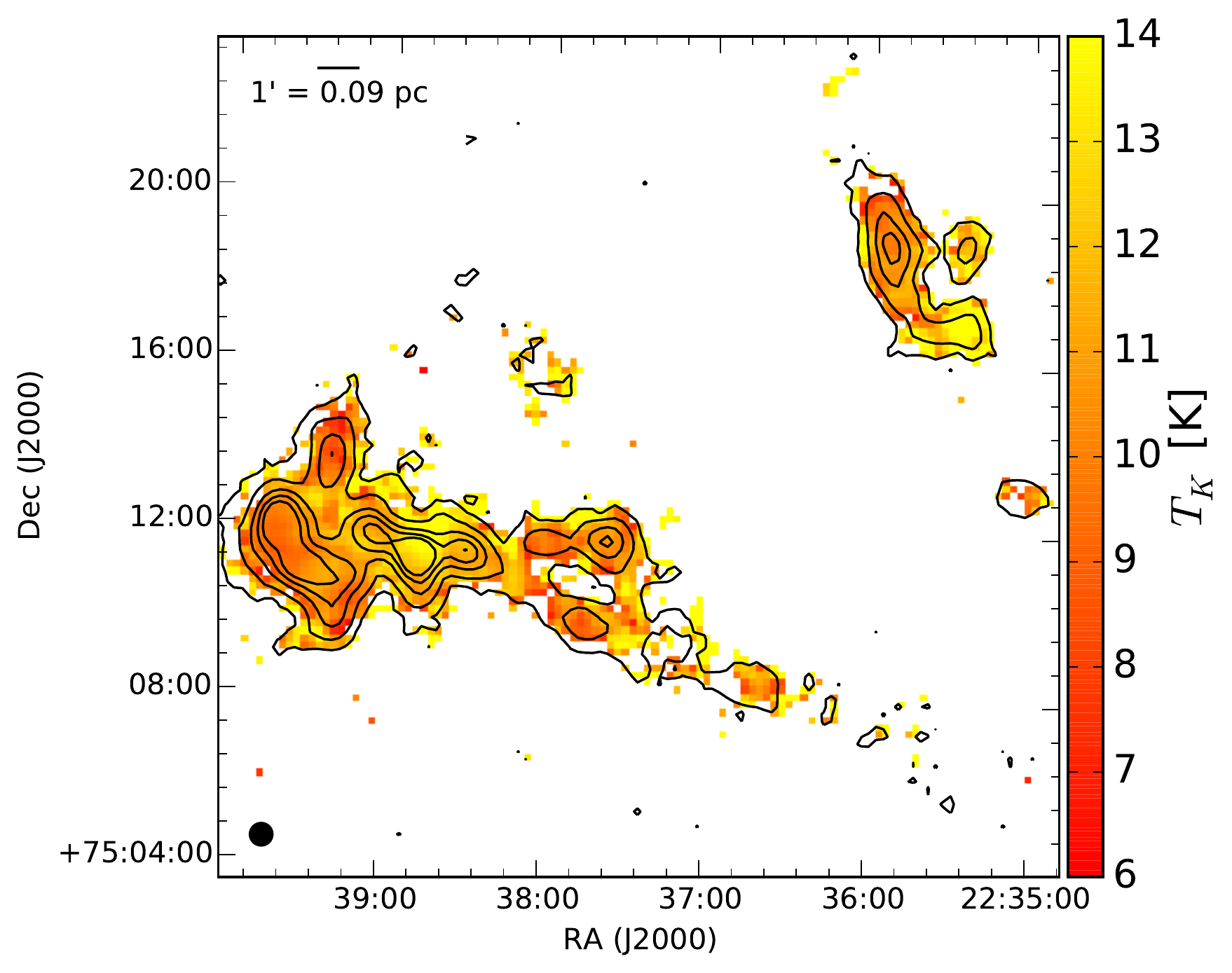}{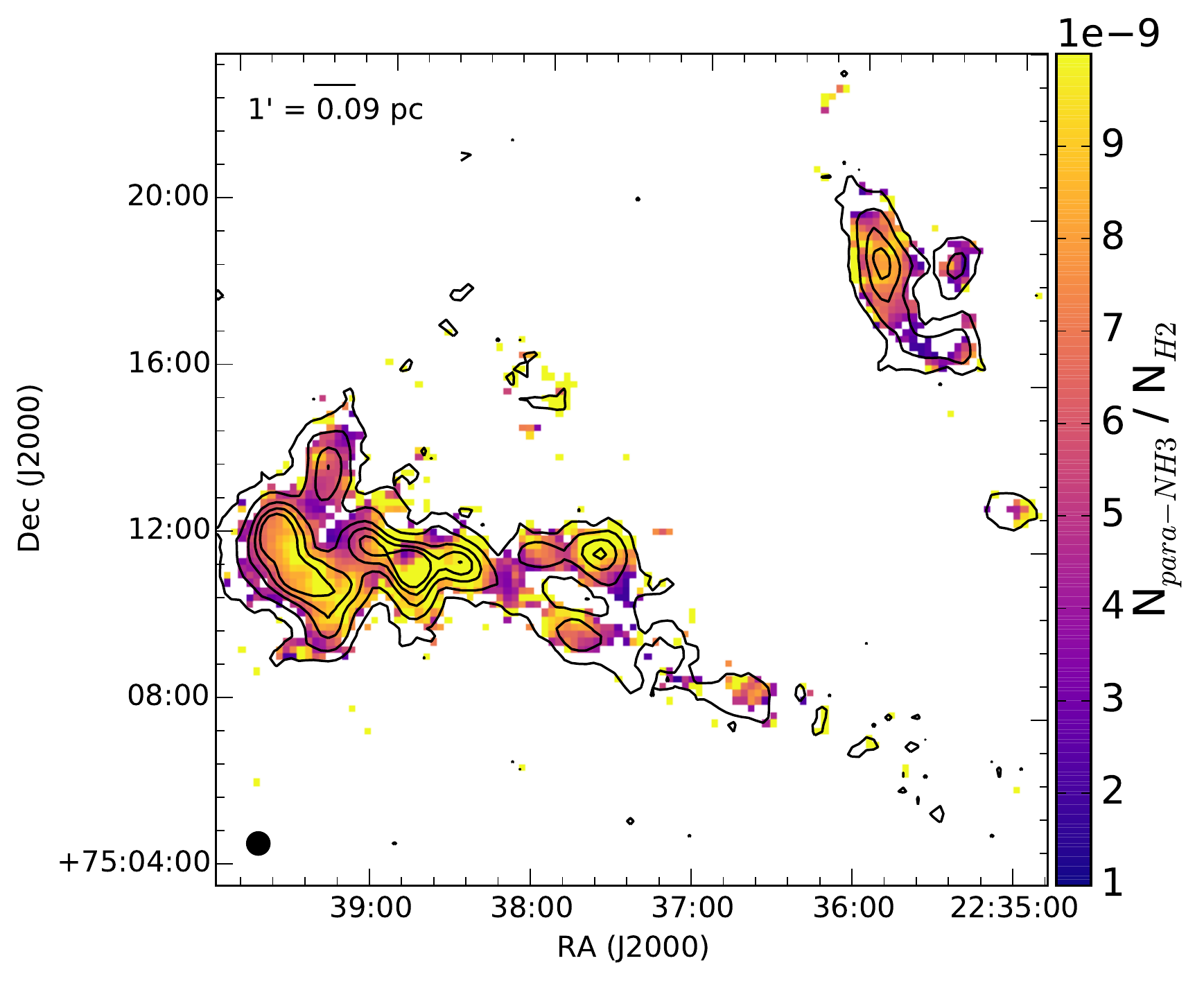} \\
\plottwo{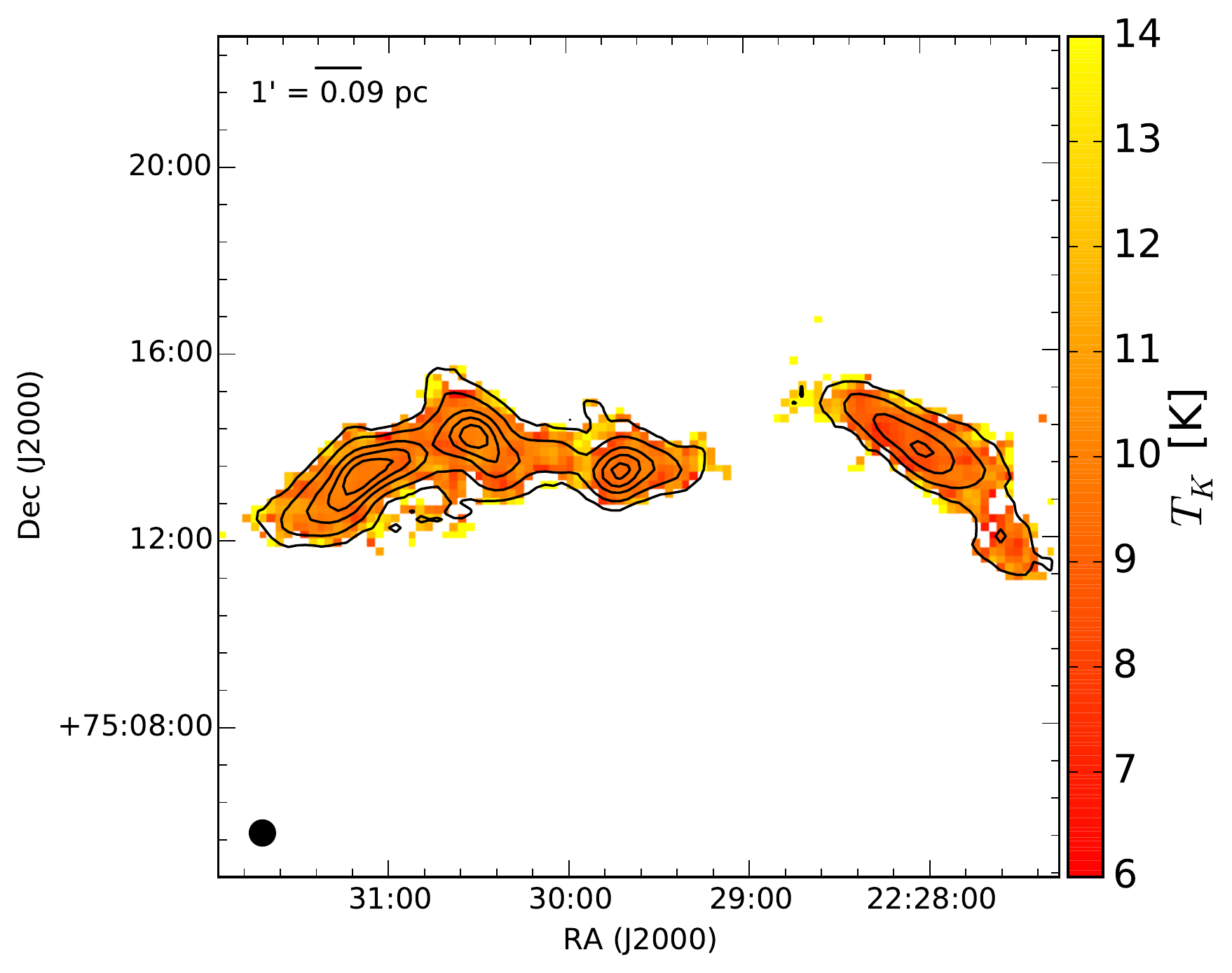}{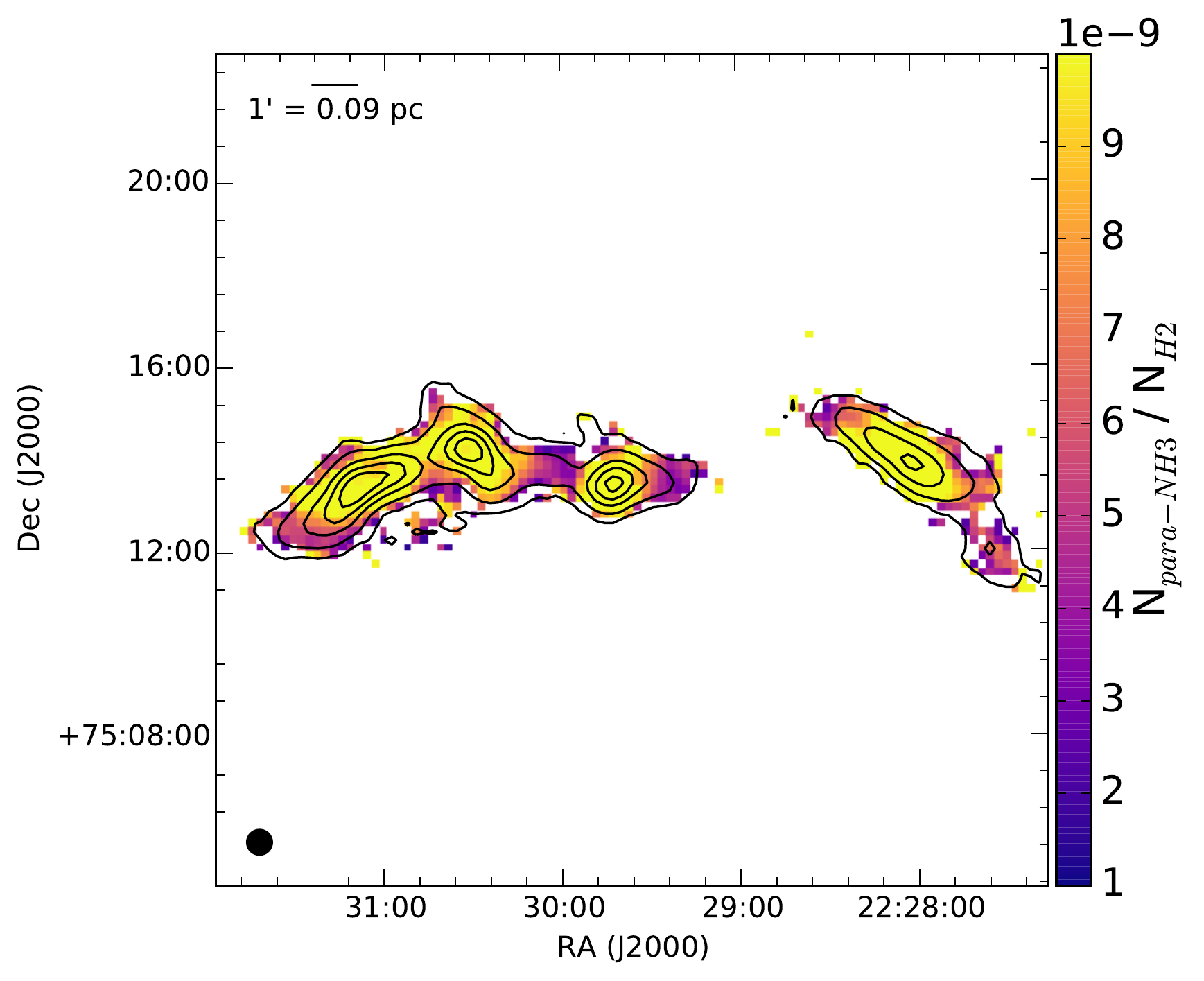} \\
\plottwo{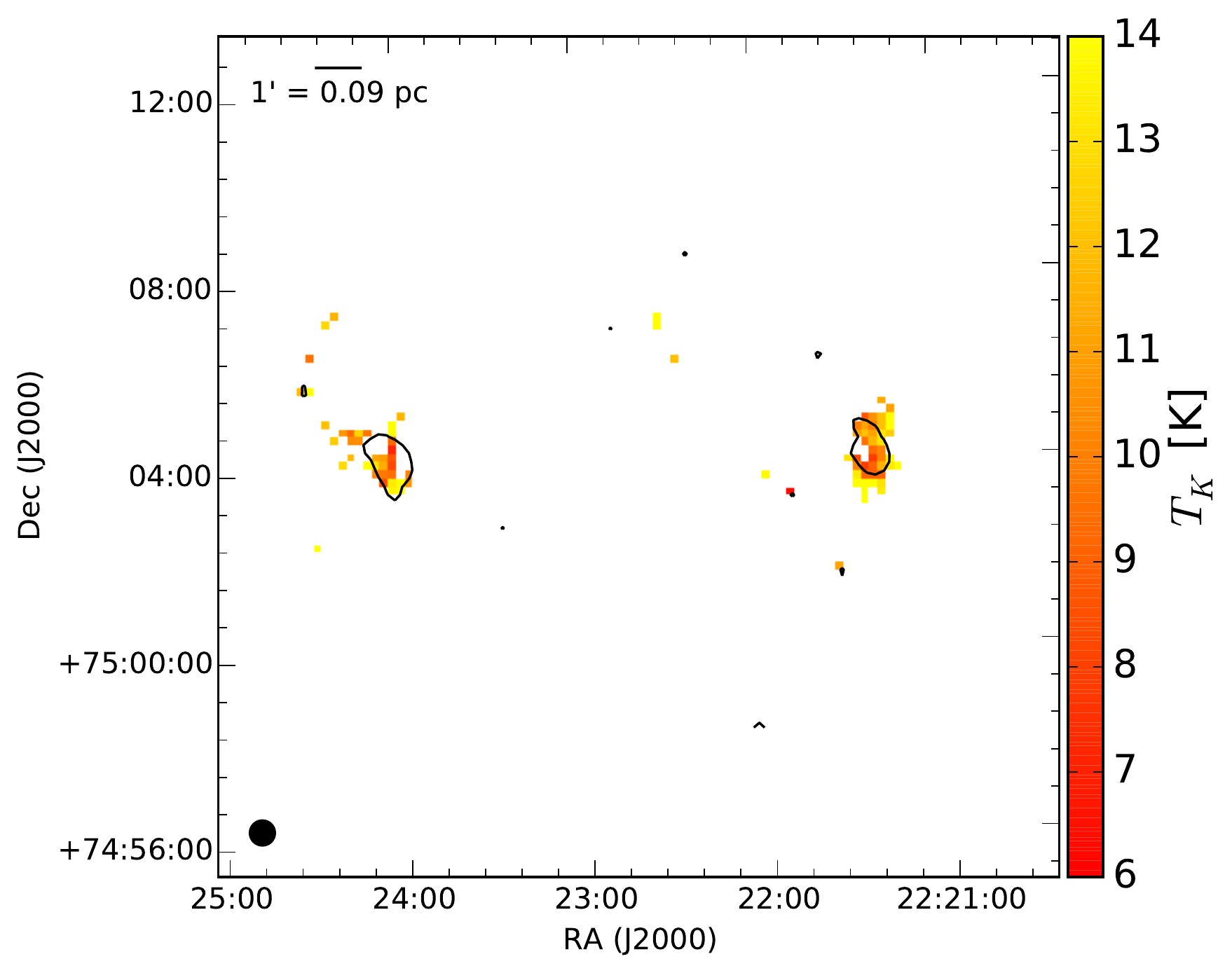}{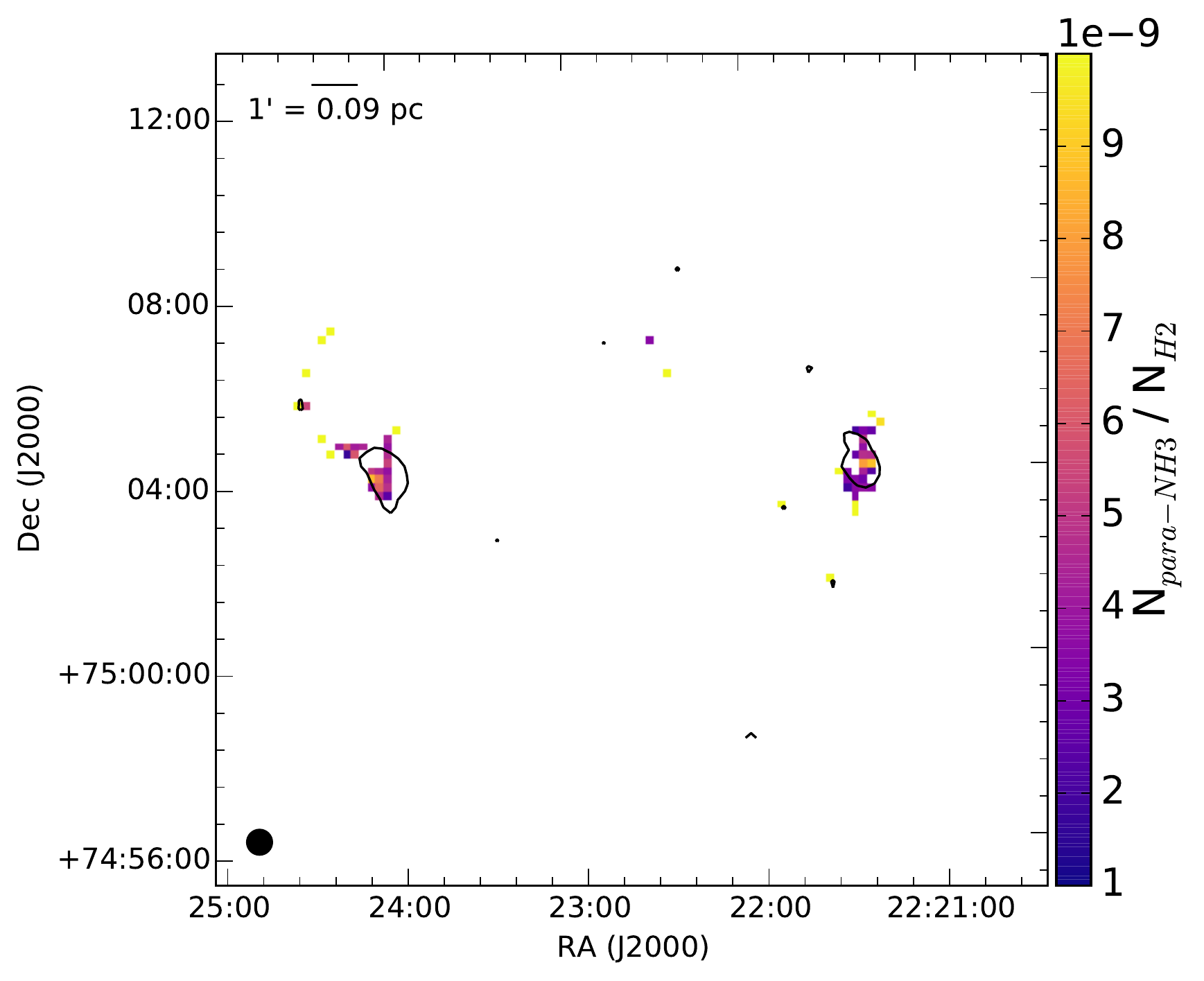}
\caption{Same as Figure \ref{Params1} but for kinetic temperature (left column) and para-NH$_3$ abundance (right column).}
\label{Params2}
\end{figure}

\begin{figure}[ht]
\epsscale{1.0}
\plottwo{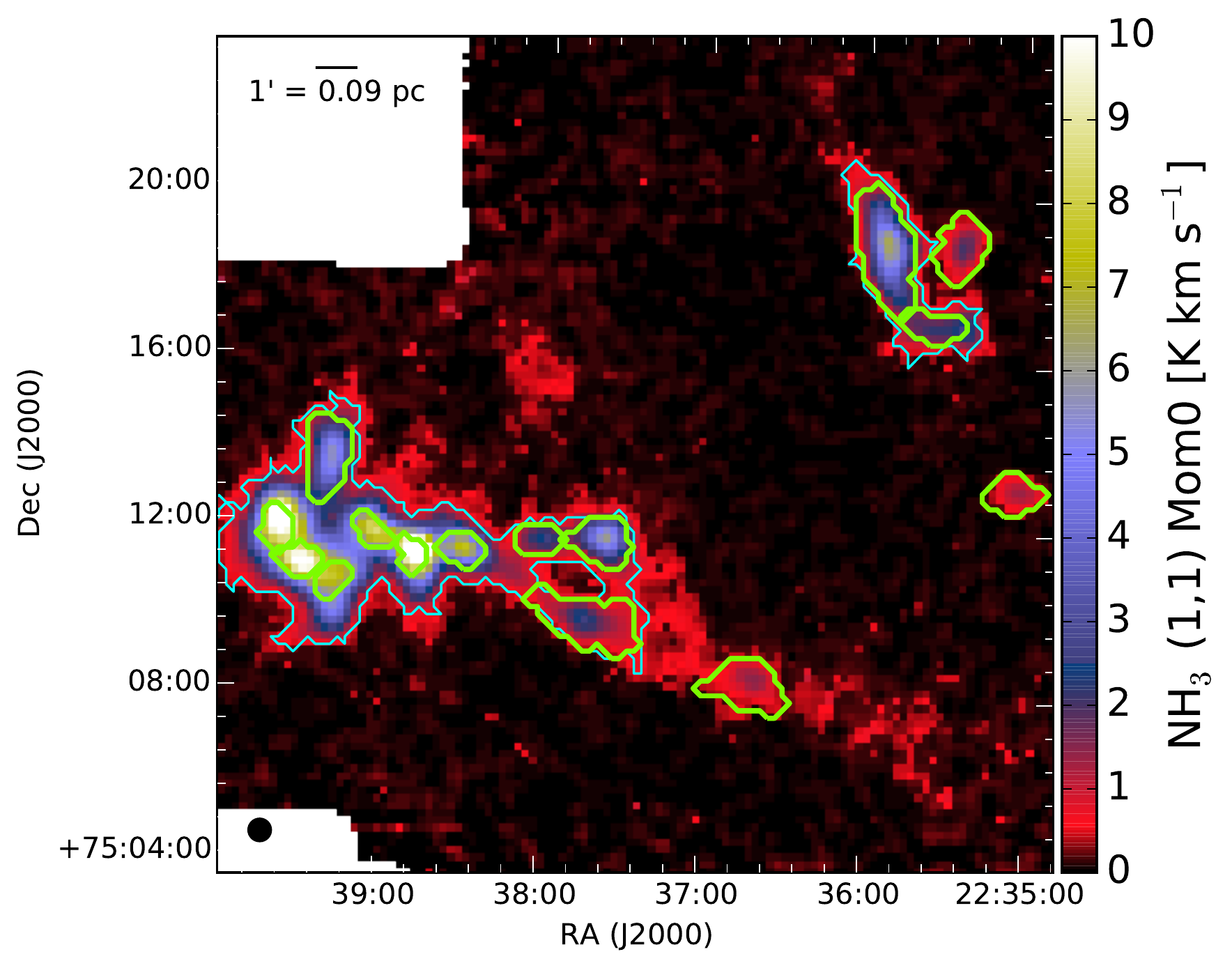}{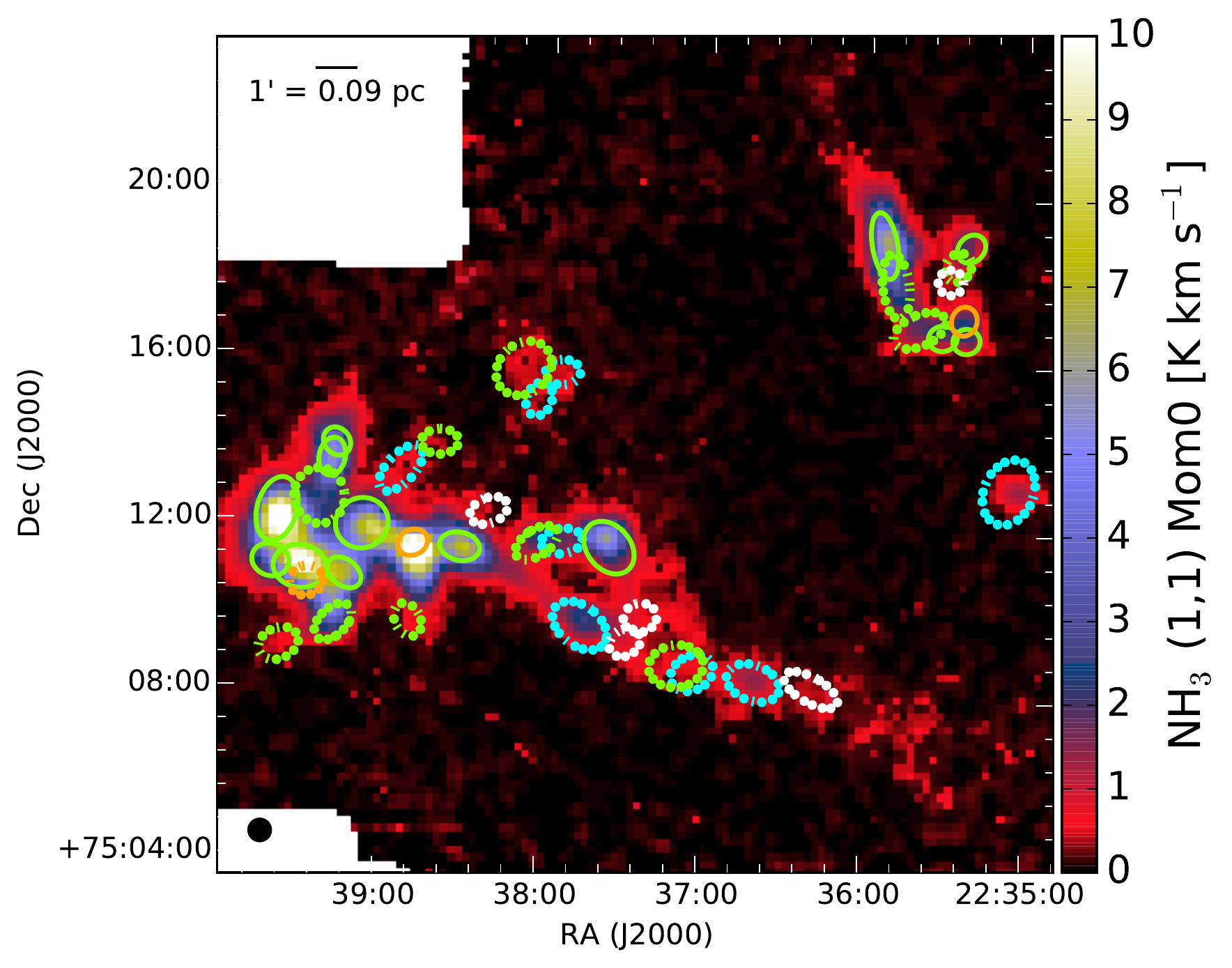} \\
\plottwo{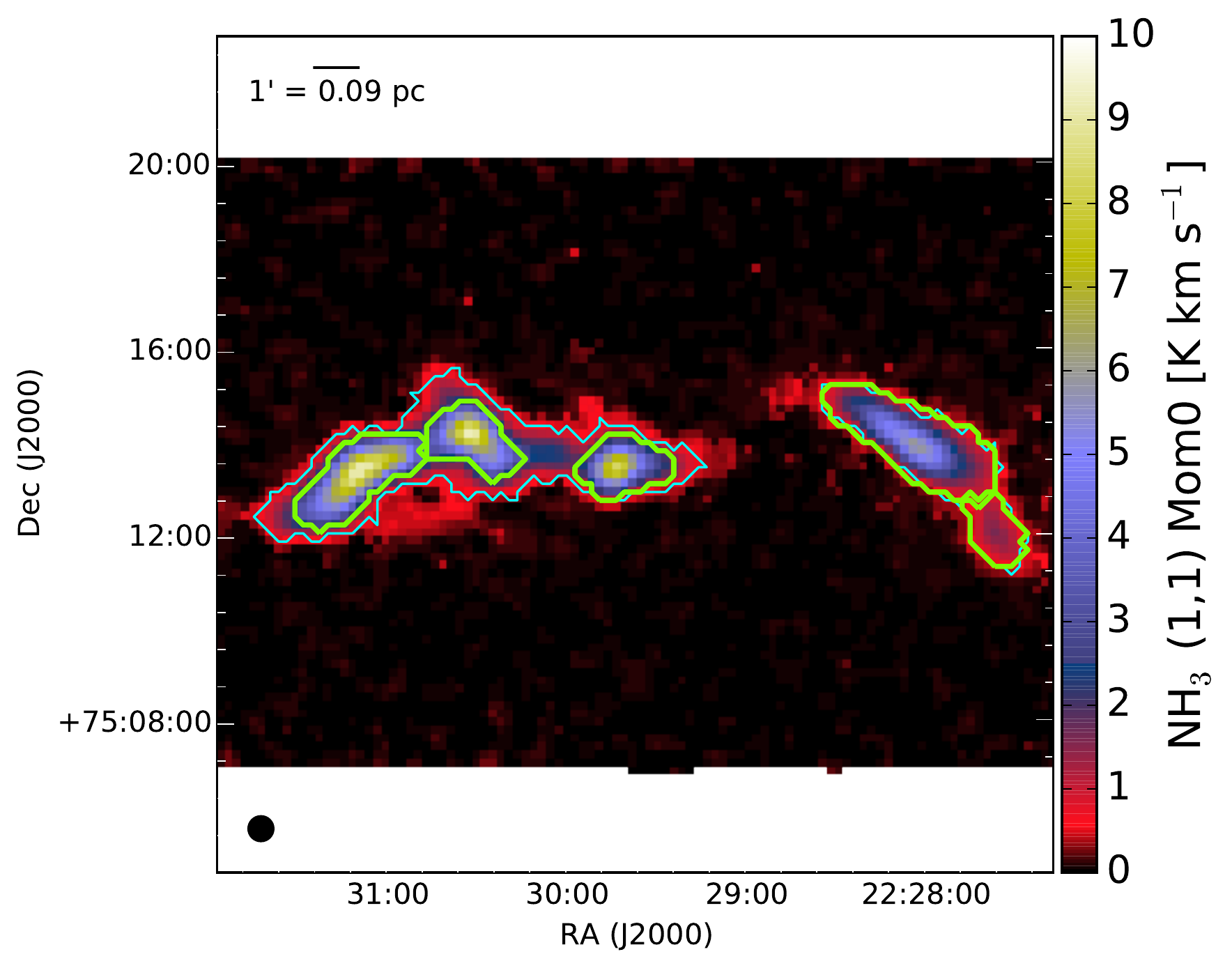}{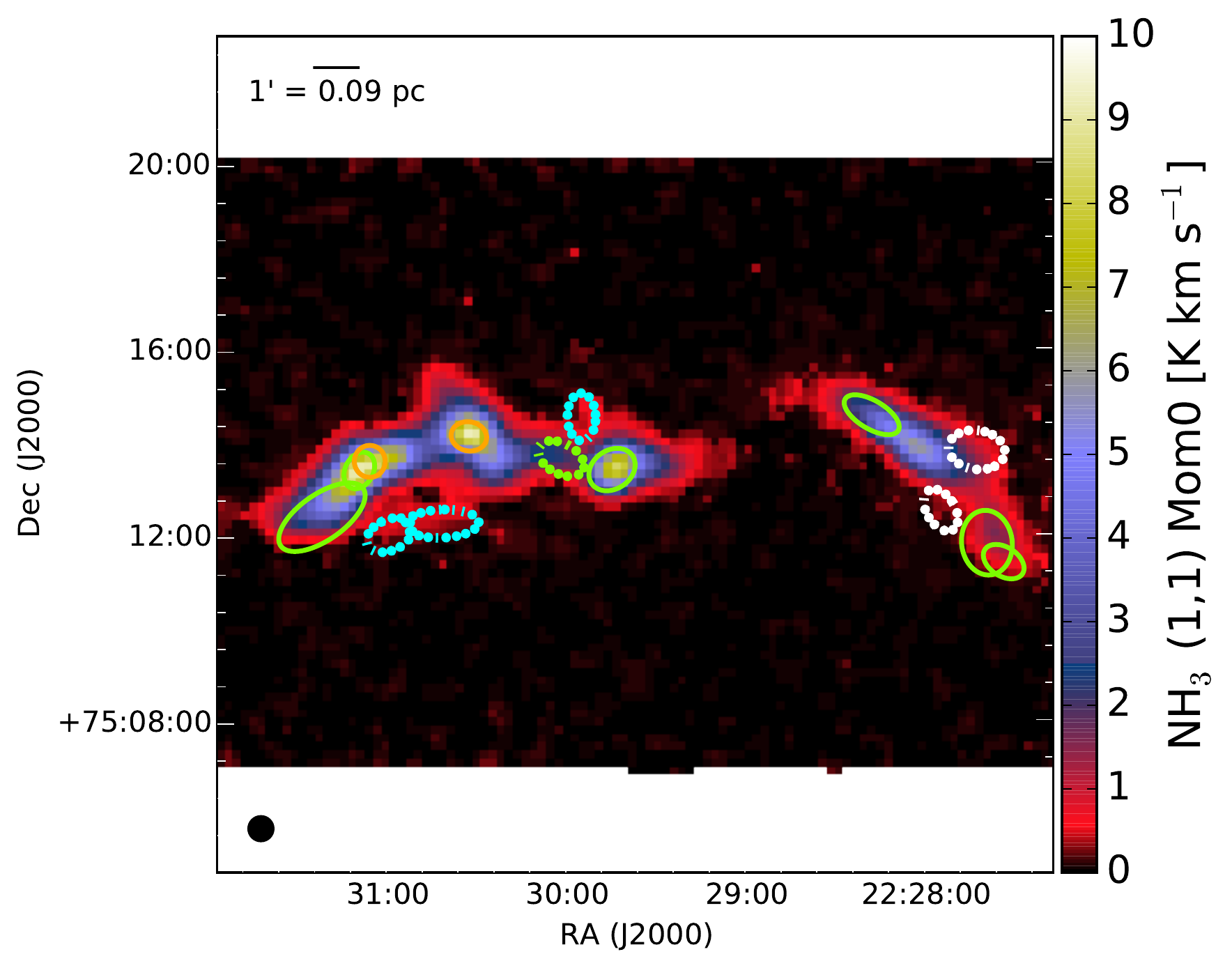} \\
\plottwo{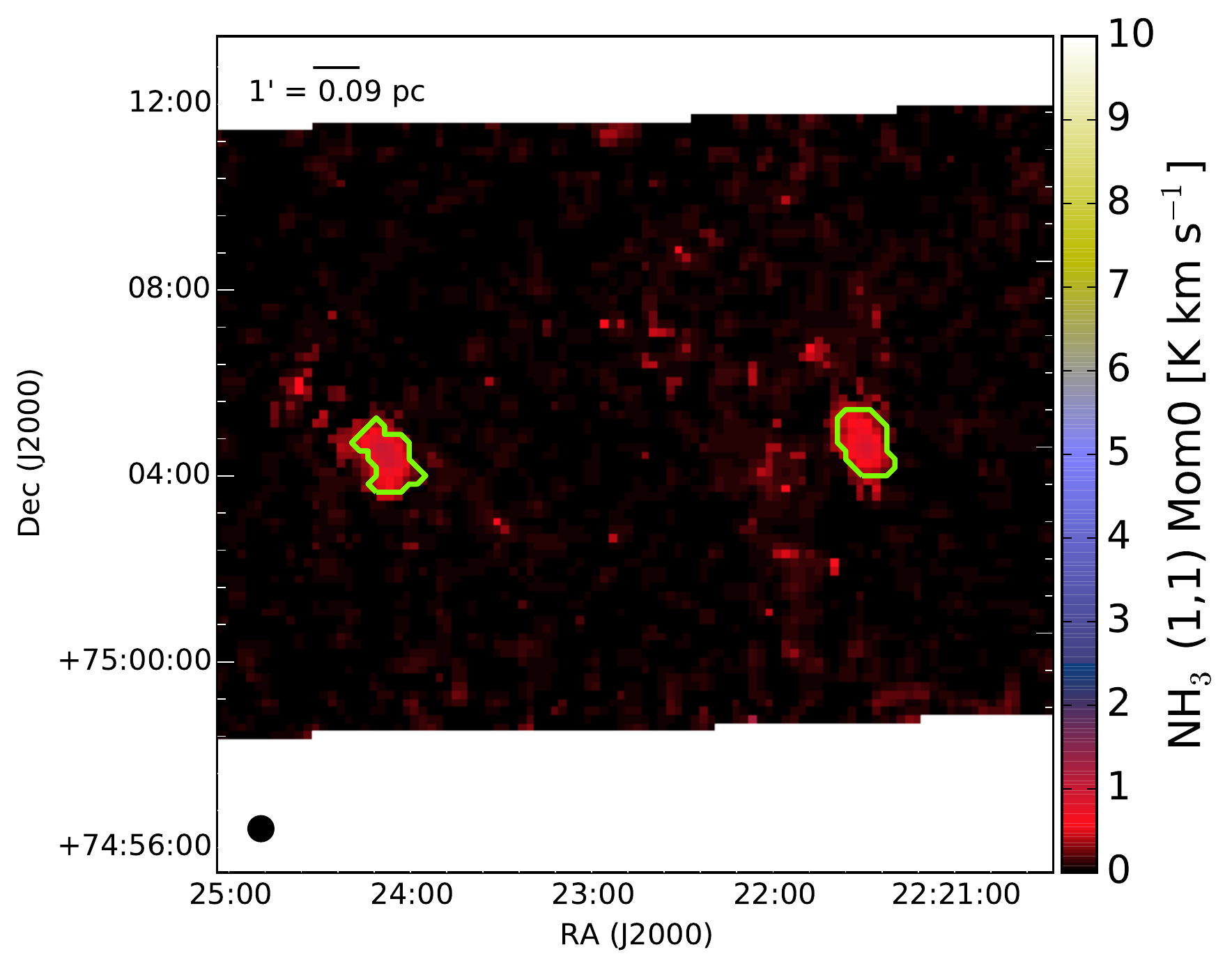}{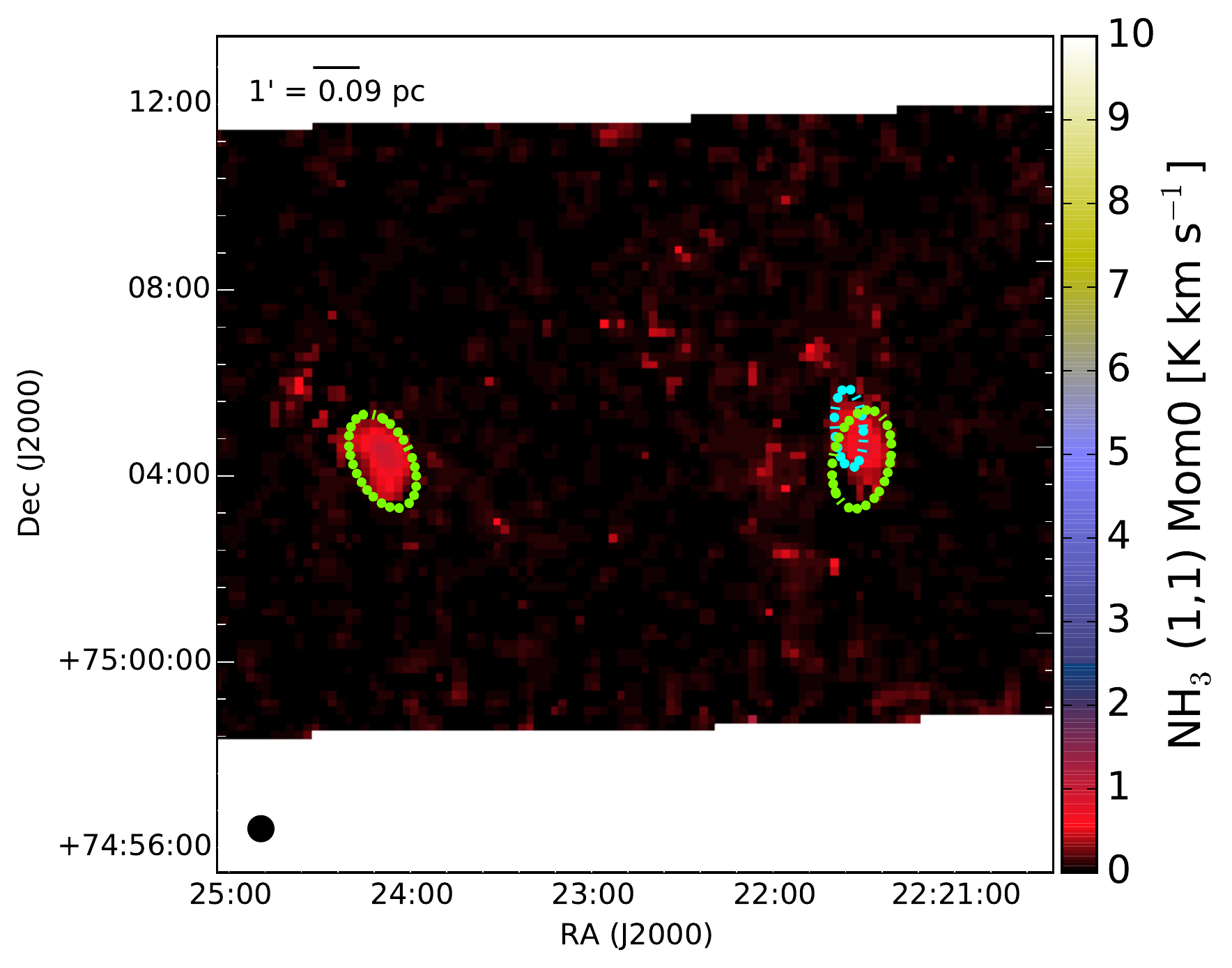}
\caption{Left column: Top-level, ``leaf'' structures (outlined in green) identified from our dendrogram analysis of the NH$_3$ (1,1) emission from Cepheus-L1251.  The background image is the NH$_3$ (1,1) integrated intensity map and the panels outline the same fields shown in Figure \ref{Params1}.  The four lowest-level, ``trunks'' identified by the dendrogram are outlined in cyan.  Right column: Positions of \textit{Herschel}-identified dense cores that have reliable kinematic measurements (see text).  Each ellipse represents twice the FWHM for the source.  Robust prestellar cores are shown in green, prestellar candidate cores in cyan, starless cores in white, and protostellar cores in orange.  Structures with $\alpha_{vir} < 2$ are shown in solid lines, while those with $\alpha_{vir} \geq 2$ are shown by dotted lines.}
\label{leaves}
\end{figure}

\begin{figure}[ht]
\epsscale{1.2}
\plottwo{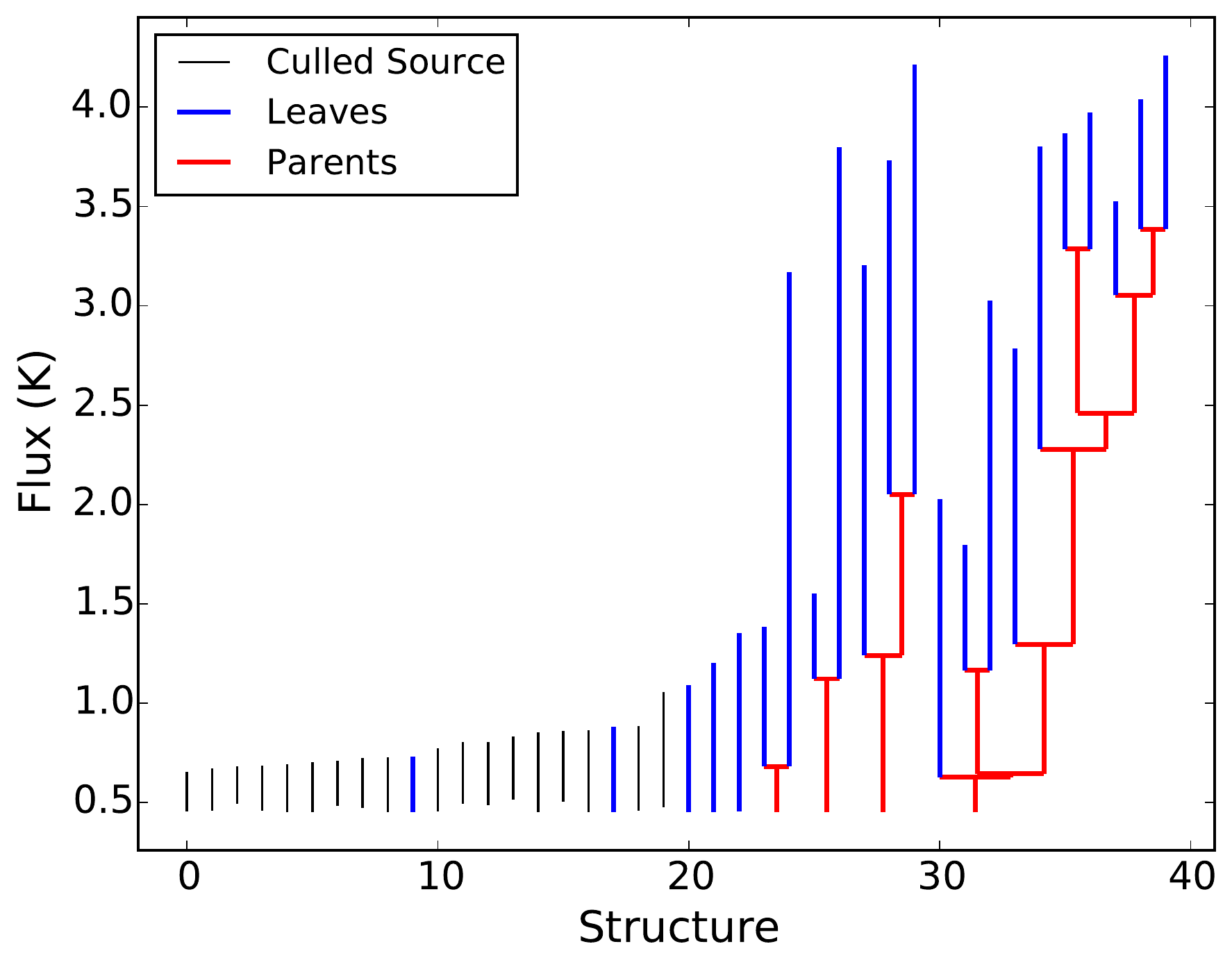}{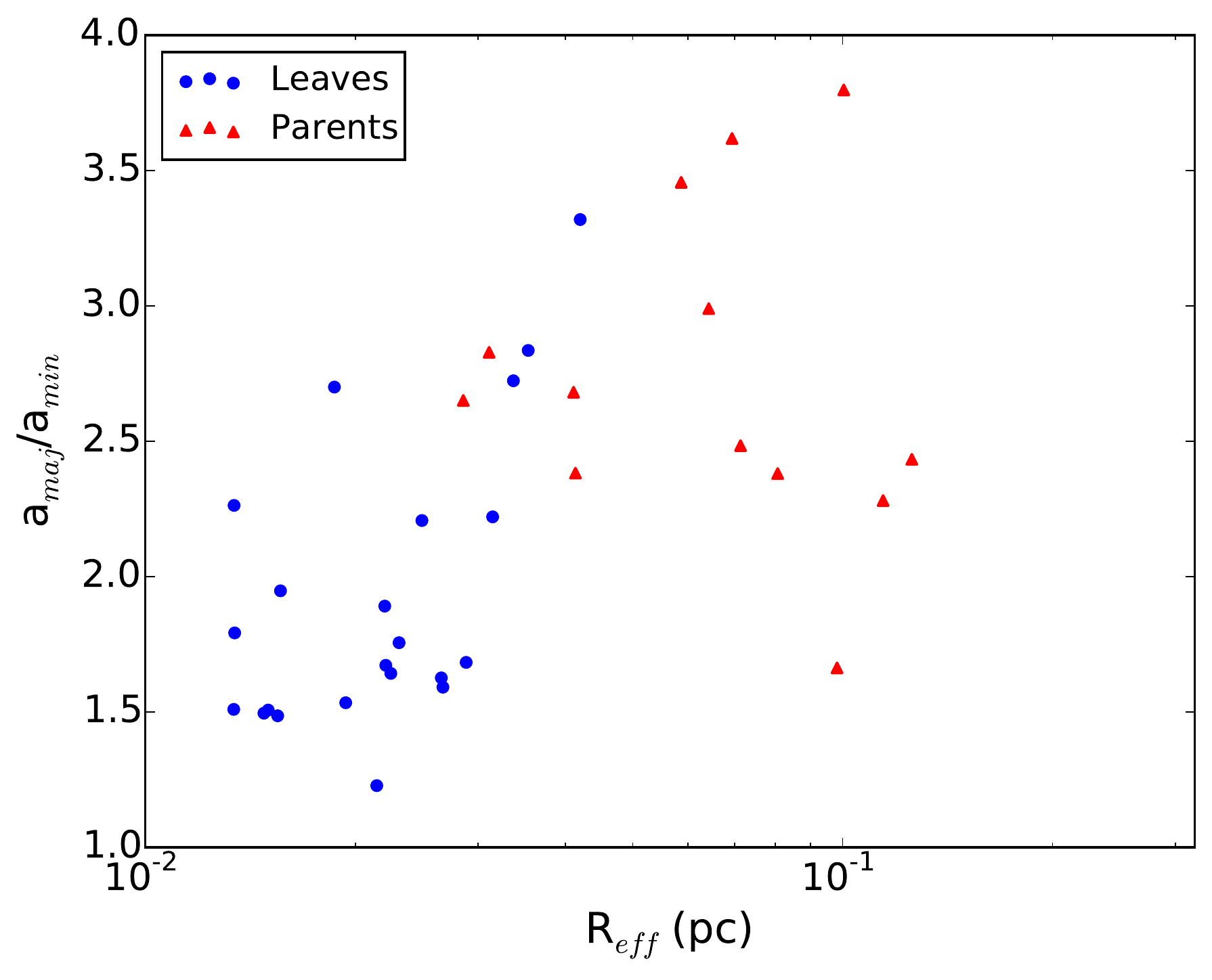}
\caption{Left panel: Dendrogram tree identified for our ammonia observations of Cepheus-L1251.  Black denotes sources that have major and minor axes smaller than 6$\arcsec$ in projection on the sky and were not included in our analysis.  Blue and red show the ``leaf'' and ``parent'' structures, respectively, that are larger than 6$\arcsec$ and were included in our analysis.  Right panel: Effective radius versus aspect ratio for the leaves (blue dots) and parents (red triangles) shown in the left panel. The parent structures include both the ``branches'' and ``trunks'' in the dendrogram.}
\label{R_vs_aspect}
\end{figure}

\begin{figure}[ht]
\epsscale{0.6}
\plotone{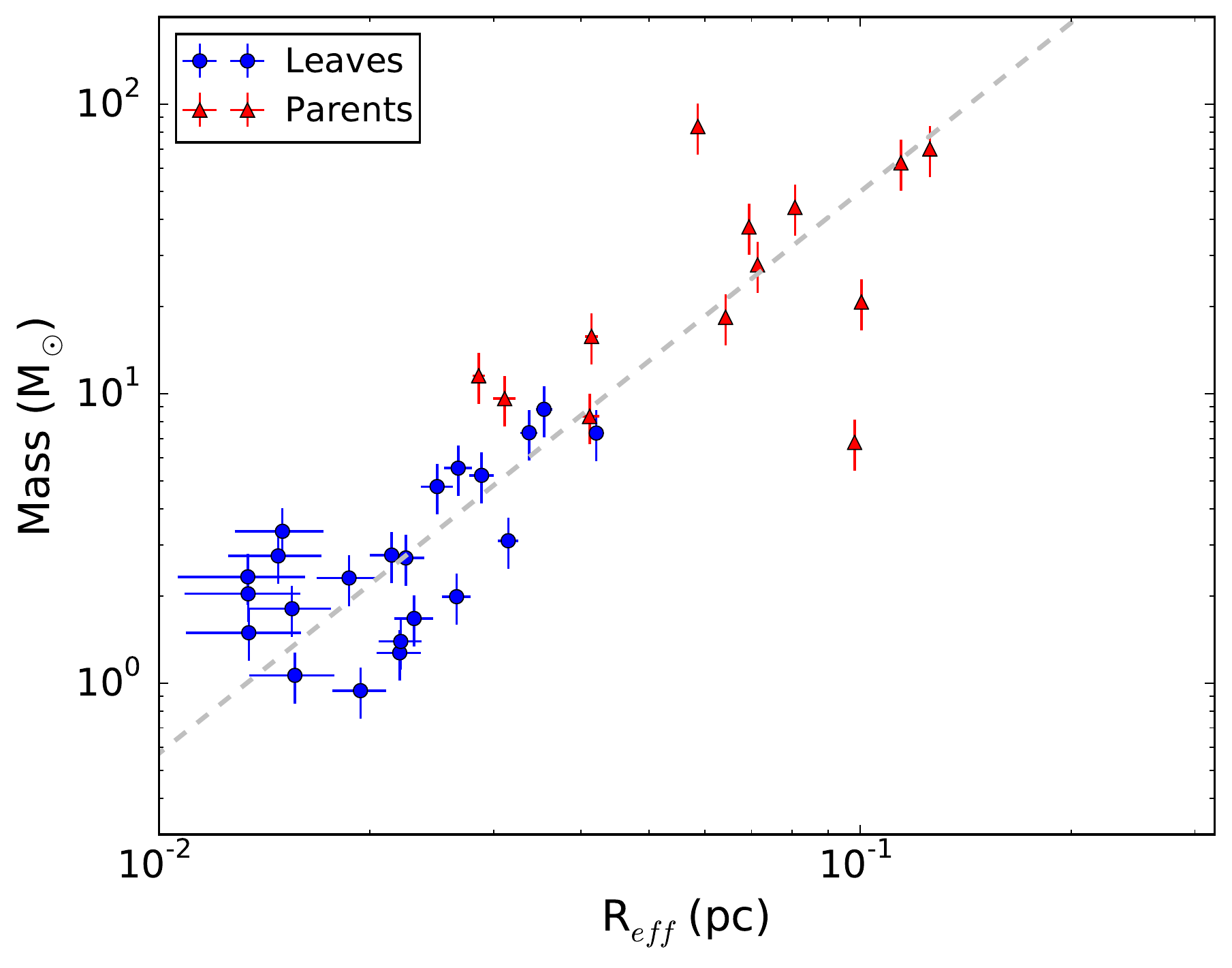}
\caption{Effective radius versus observed mass for the same structures as in Figure \ref{R_vs_aspect}.  The dashed line is a power-law fit to the combined leaf and parent data points.  The best-fit slope is found to be 1.94 $\pm$ 0.18.  Errorbars in the y-axis direction indicate 20$\%$ uncertainty on the H$_2$ column densities used to calculate mass.  Errorbars in the x-axis direction represent $\sqrt{A_{pix}/\pi N_{pix}}$, where $A_{pix}$ is the area of a pixel in the NH$_3$ (1,1) emission map and N$_{pix}$ is the number of pixels falling within the structure.}
\label{R_vs_mass}
\end{figure}

Alternatively, the relative motions we observe between the low and high density gas could be related to the positions of these particular sources within the molecular cloud.  Sources 12 and 16 are the two eastern-most leaves in our sample (shown in the far-left side of the top left panel of Figure \ref{leaves}).  As can be seen in Figure \ref{obs}, the positions of these sources correspond to a sharp change in H$_2$ column density.  Thus, sources 12 and 16, which lie on the outskirts of the high-density central regions of Cepheus-L1251, may be more exposed to the turbulent outer layers of the parental molecular cloud.  Furthermore, these sources lie at the edge of a ``comet-like'' feature in the east of Cepheus-L1251 (also shown in Figure \ref{obs}), which may indicate some external forces (e.g., winds or radiation) moving in the east to west direction are producing a larger influence upon the outer gas layers in that portion of the cloud.  Chemical differentiation of carbon-based species has been observed toward sharp edges of H$_2$ column density \citep{Spezzano_2016}, which may also explain why the centroid velocity offsets we observe are more extreme for the emission from HC$_5$N than that of CCS.  In total, these results suggest such centroid velocity offsets between certain molecular transitions may be a common occurrence at sharp transitions between high and low density gas. 


\section{Summary}
We have performed a dendrogram analysis on Green Bank Ammonia Survey observations of the Cepheus-L1251 molecular cloud to identify hierarchical dense gas structures that may form stars in the future.  Our final catalog consists of 22 top-level structures, which reside within 13 lower-level parent structures.  We also use the ammonia data to characterize the gas properties of a dense core population identified by \cite{DiFrancesco_prep} using \textit{Herschel} photometric observations of thermal dust continuum emission across the region.  The results of our analysis for the ammonia-identified and \textit{Herschel}-identified structures are summarized below:

1.  The observed masses for the top-level ammonia-identified sources, estimated from a dust continuum-derived H$_2$ column density map for the region, range from 0.9 M$_\odot$ for the smallest top-level structure to 80 M$_\odot$ for the largest parent structure.  The top-level structures are predominantly compact, with $R_{eff}$ $<$ 0.05 pc, and show a strong spatial correspondence with the highest H$_2$ column density portions of the observed field.

2.  The virial parameters for the top-level ammonia-identified structures suggest they are gravitationally dominated, yet may be in or near a state of virial equilibrium.  Conversely, the majority of the \textit{Herschel}-identified dense cores are pressure-confined, sub-virial objects rather than gravitationally bound structures.  These results appear to hold for multiple sets of assumptions on the density profile, radius, and mass of the structures.  Our NH$_3$ results are also consistent with a similar virial analysis conducted by \cite{Friesen_2016} for the more active star-forming Serpens South region, despite Cepheus-L1251 and Serpens South being drastically different environments (as indicated by their significantly different cloud mass surface densities, column densities, numbers of YSOs, and numbers of dense cores).  As such, it appears that, independent of the cloud-scale environment, dense ammonia structures are gravitationally dominated, while most \textit{Herschel}-identified cores are pressure-confined.



3. All of the elongated, filamentary, ammonia-identified structures with aspect ratios greater than 2 appear to be gravitationally bound.  This finding lends credence to the idea that filamentary collapse and fragmentation gives rise to dense core and protostar formation.  

4.  The median value of $T_{dust}$ amongst the ammonia- and \textit{Herschel}-identified structures is 11.7 $\pm$ 1.1 K, while the median value of $T_{K}$ is 10.3 $\pm$ 2.0 K.  These results suggest the dust and dense gas within the structures are coupled.  The slightly higher median $T_{dust}$ value is potentially caused by heated material on the outskirts of the cloud, which is not traced by NH$_3$, that falls along the line of sight to the observed structures.   


5.  The abundance of para-NH$_3$, $\chi_{NH_3} = (N_{para-NH_3}/N_{H_2})$, is measured for all structures and found to peak just below $\chi_{NH_3}$ = 10$^{-8}$, which is consistent with $\chi_{NH_3}$ measurements from many other star-forming regions.  The ammonia-identified leaves and \textit{Herschel}-identified cores show similar $\chi_{NH_3}$ distributions, with no significant variations between the different structure types.

6.  17 of 22 ammonia-identified leaves contain detectable CCS $(2_0-1_0)$ emission in at least one-third of the pixels falling within their 2D mask.  Similarly, 16 of 22 ammonia-identified leaves contain detectable HC$_5$N $(9-8)$ emission.  The line widths measured for the carbon-based molecular tracers, particularly HC$_5$N $(9-8)$, are generally higher than the line width measured from NH$_3$ (1,1).  This difference suggests the carbon-based molecules are tracing the more turbulent outer layers of the molecular cloud where they have not yet suffered depletion at higher densities and colder temperatures. 

7.  Two of the ammonia-identified leaves (sources 12 and 16 in Table \ref{Table_NH3}) show centroid velocity offsets of $\Delta V_{LSR} > 0.5$ km s$^{-1}$ between NH$_3$ (1,1) and HC$_5$N $(9-8)$ and $\Delta V_{LSR} > 0.3$ km s$^{-1}$ between NH$_3$ (1,1) and CCS $(2_0-1_0)$.  The position of these structures on the outskirts of the high-density regions of the cloud suggest they may be located at a sharp transition between the turbulent parental cloud and quiescent star-forming region.  

\section*{Acknowledgments}
We thank the entire \textit{Herschel} Gould Belt Survey team for their efforts obtaining and preparing the data used for the dense core catalogs utilized in this paper.  We particularly appreciate the efforts of Bilal Ladjelate, Vera K{\"o}nyves, and Alexander Men'shchikov for their contributions to the getsources extractions used for our analysis.  JK, JDF, ER, MCC, and CDM acknowledge the financial support of a Discovery Grant from NSERC of Canada.  RKF is a Dunlap Fellow at the Dunlap Institute for Astronomy $\&$ Astrophysics. The Dunlap Institute is funded through an endowment established by the David Dunlap family and the University of Toronto.  PC, JP, AP, and ACT acknowledge the financial support of the European Research Council (ERC; project PALs 320620).  The National Radio Astronomy Observatory is a facility of the National Science Foundation operated under cooperative agreement by Associated Universities, Inc.  This research made use of astrodendro, a Python package to compute dendrograms of Astronomical data (\url{http://www.dendrograms.org/}).  This research also made use of Astropy (\url{http://www.astropy.org}), a community-developed core Python package for Astronomy \citep{Astropy_2013}.

{\it Facility:} \facility{GBT, Herschel}

\section*{Appendix}
\begin{appendix}
\section{Effective Radius}

As discussed in Section 3.4, the virial parameter presented in this paper depends upon the radius ($R$) of the structure being analyzed.  In this paper, we chose to use $R = R_{eff} = (\sigma_{major} \sigma_{minor})^{1/2}$ (i.e., the geometric mean of the major and minor axes) for the ammonia-identified leaves to be consistent with the virial analysis presented in \cite{Friesen_2016}.  Similarly, the effective radius used for the \textit{Herschel}-identified dense cores is the deconvolved geometric mean of their FWHM axes, which was chosen to be consistent with the effective radius used for the Bonnor-Ebert core classification presented in \cite{DiFrancesco_prep}.  Previous studies that have involved virial analyses, however, have determined the radii of sources in other ways.  For instance, \cite{Kauffmann_2013} adopt $R = (A/\pi)^{1/2}$ as the effective radii used in their virial analysis of low- and high-mass star-forming regions (where $A$ represents the structure's area on the PP plane).

To investigate the dependence of our virial parameter calculations upon our chosen formulation for the effective radius, we re-calculate the virial parameters of both the ammonia-identified structures and \textit{Herschel}-identified dense cores using $R = (A/\pi)^{1/2}$ as input into Equation 1.  The altered virial parameters are used to create Figure \ref{Reff_new}, which plots updated versions of Figures \ref{R_vs_aspect}, \ref{R_vs_mass}, and \ref{Mass_vs_virial}, showing the impact the changes in effective radius make upon our conclusions.  The new radii formulation increases the $R_{eff}$ of all the leaves, while it increases some of the parent structures and decreases others.  The amorphous shapes that some of the parent structures exhibit are likely not well-characterized by the assigned major and minor axes, causing a more extreme change in their $R_{eff}$ estimates between the two formulations.  The mean effective radii of the ammonia-identified leaves and Herschel-identified cores increase to 0.052 $\pm$ 0.019 pc and 0.040 $\pm$ 0.011 pc, respectively, when using the altered formulation.


The virial parameters of the ammonia-identified leaves increase as a result of their larger effective radii.  The bottom panel of Figure \ref{Reff_new}, however, shows that while their virial parameters increase, nearly all the ammonia-identified leaves remain below $\alpha_{vir} = 2$.  Similarly, the top panel of Figure \ref{virial_plane2} shows that most of the ammonia-identified leaves move into the sub-virial, pressure-dominated portion of the virial plane when using the altered $R_{eff}$.  

The change in effective radii, and hence virial parameter, for the \textit{Herschel}-identified cores is similar to the effect observed for the ammonia-identified leaves.  The radii of the \textit{Herschel}-identified cores increase by factors of a few, likely because deconvolution is not being applied to the core sizes in this altered analysis, which increases their virial parameters by factors of a few.  Regardless, the majority of the \textit{Herschel}-identified cores remain in the sub-virial and pressure-dominated portion of the virial plane.    

 
\section{Mass}
The method used for obtaining the observed mass of structures also has a bearing on the calculated virial parameters.  For instance, our analysis of the \textit{Herschel}-identified dense cores in Cepheus-L1251 uses the masses presented in \cite{DiFrancesco_prep}.  The masses presented in that paper were obtained from SED fits to the integrated dust continuum fluxes for each source after emission from large-scale background structures (i.e., the filaments within which the cores reside) was subtracted.  Such filtering reduces the total amount of flux associated with each core, causing derived masses to be significantly lower than if all the emission along the line of sight to the core were used in the mass derivation.  

\begin{figure}[ht]
\epsscale{0.6}
\plotone{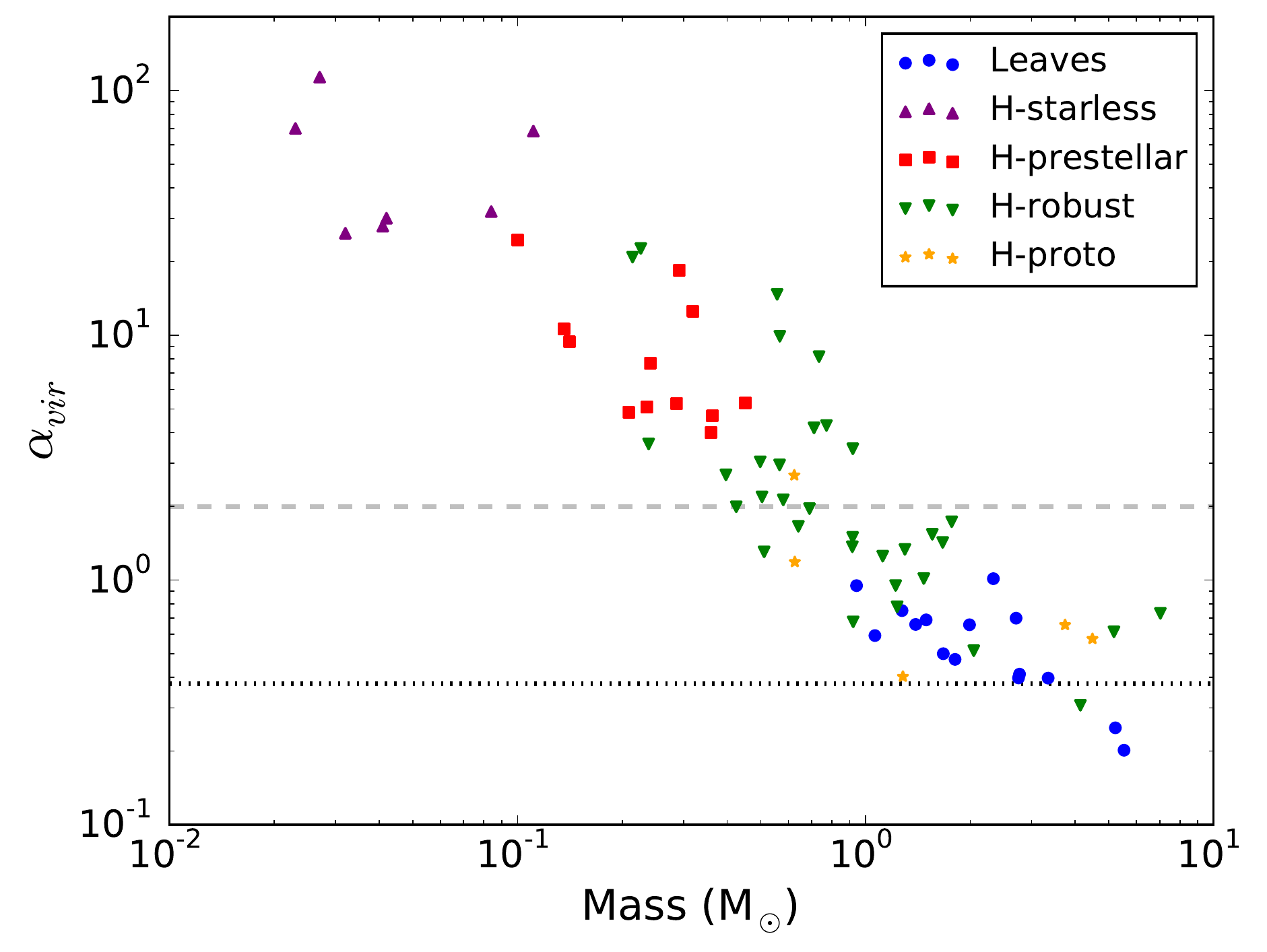}
\plotone{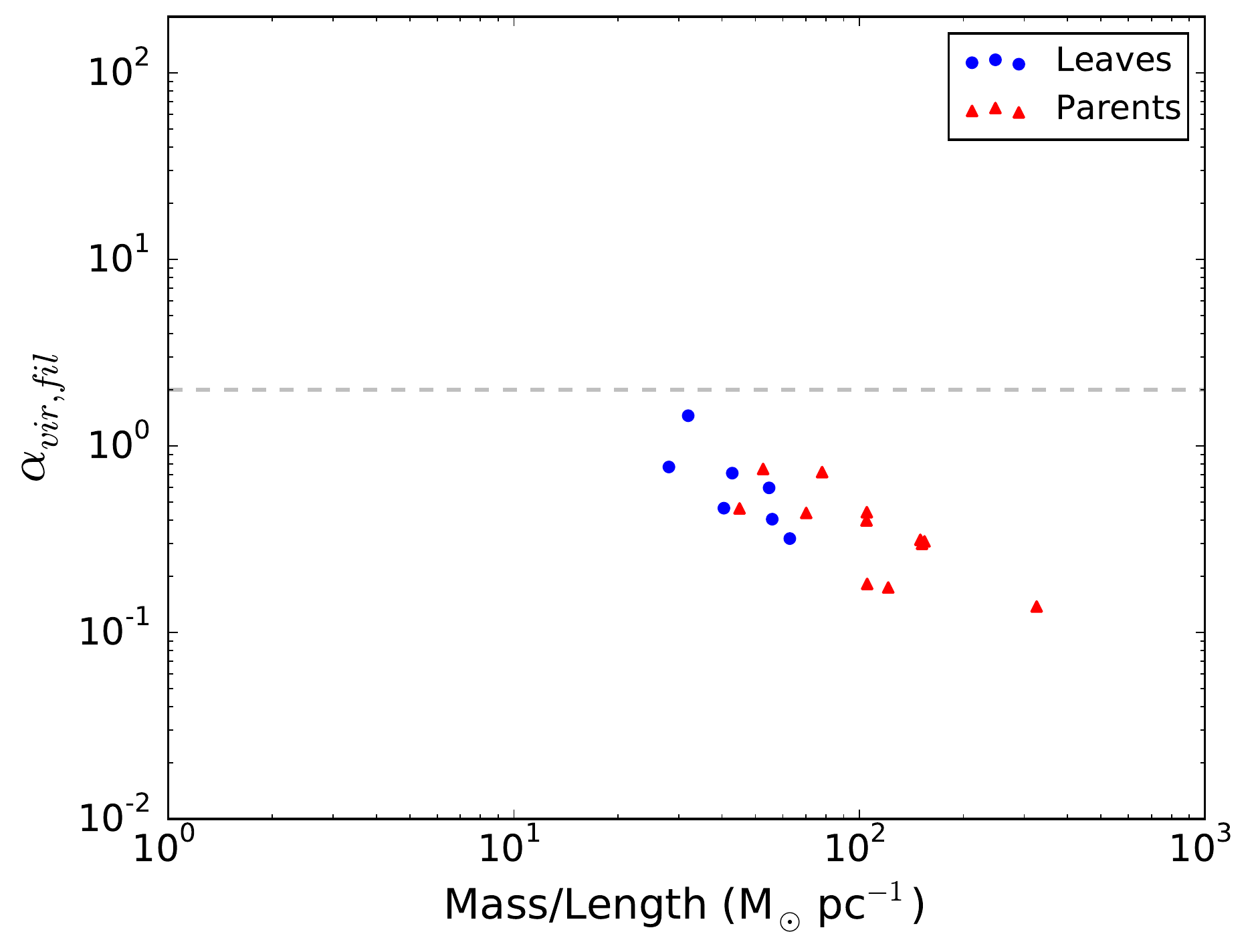}
\caption{Top: Virial parameter, $\alpha_{vir}$, versus observed mass for the NH$_3$ (1,1) top-level structures identified in our dendrogram analysis (blue), as well as the \textit{Herschel}-identified ``starless'' (purple), ``prestellar candidate'' (red), ``robust prestellar'' (green), and ``protostellar'' (orange) dense cores from \cite{DiFrancesco_prep}.  The dashed line denotes $\alpha_{vir}$ = 2, above which structures are gravitationally unbound in the absence of magnetic pressure and assuming a power-law density profile for the structures.  The dotted line represents where $\alpha_{vir}$ = 1 would occur if a Gaussian density profile is assumed for the structures (for reference, this line is consistent with the vertical dotted line shown in the left panel of Figure \ref{virial_plane}).  Bottom: Filamentary virial parameter, $\alpha_{vir, fil}$, versus mass per unit length for the structures identified in our dendrogram analysis with $a_{maj}/a_{min} \geq 2$.  Blue denotes top-level, leaf structures.  Red signifies lower-level, parent structures within which the leaves reside.  }
\label{Mass_vs_virial}
\end{figure}

\begin{figure}[ht]
\epsscale{1.0}
\plotone{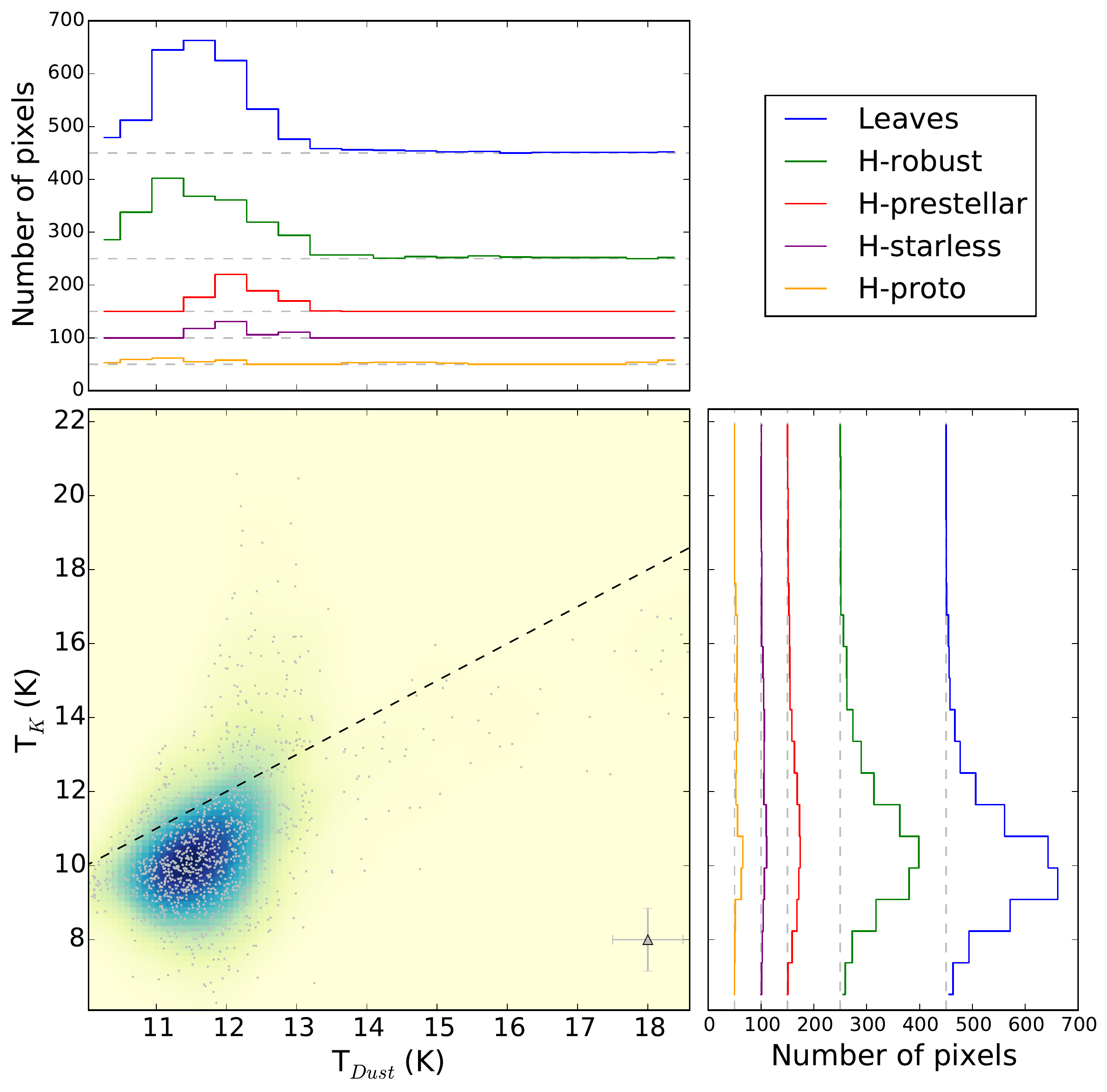}
\caption{Dust versus kinetic temperature for all pixels that fall within the 2D mask of either a \textit{Herschel}-identified dense core or ammonia-identified leaf.  The background colorscale represents a Gaussian kernel density estimate of the actual measurements, which are shown as grey dots.  Typical errorbars for the data are shown in the bottom right corner and the black dashed line shows the 1:1 line between the two parameters.  The histograms show the distributions of $T_K$ (right) and $T_{dust}$ (top) for the pixels that fall within each structure type (ammonia-identified leaves in blue, \textit{Herschel} robust prestellar cores in green, prestellar candidates in red, starless cores in purple, and protostellar cores in orange).  The orange, purple, red, green, and blue histograms are offset from the zero position on the $y$-axis by 50, 100, 150, 250, and 450, respectively.  The dashed grey lines beneath each histogram represent the y=0 position for the distribution.}
\label{Thist}
\end{figure}

\begin{figure}[ht]
\epsscale{0.6}
\plotone{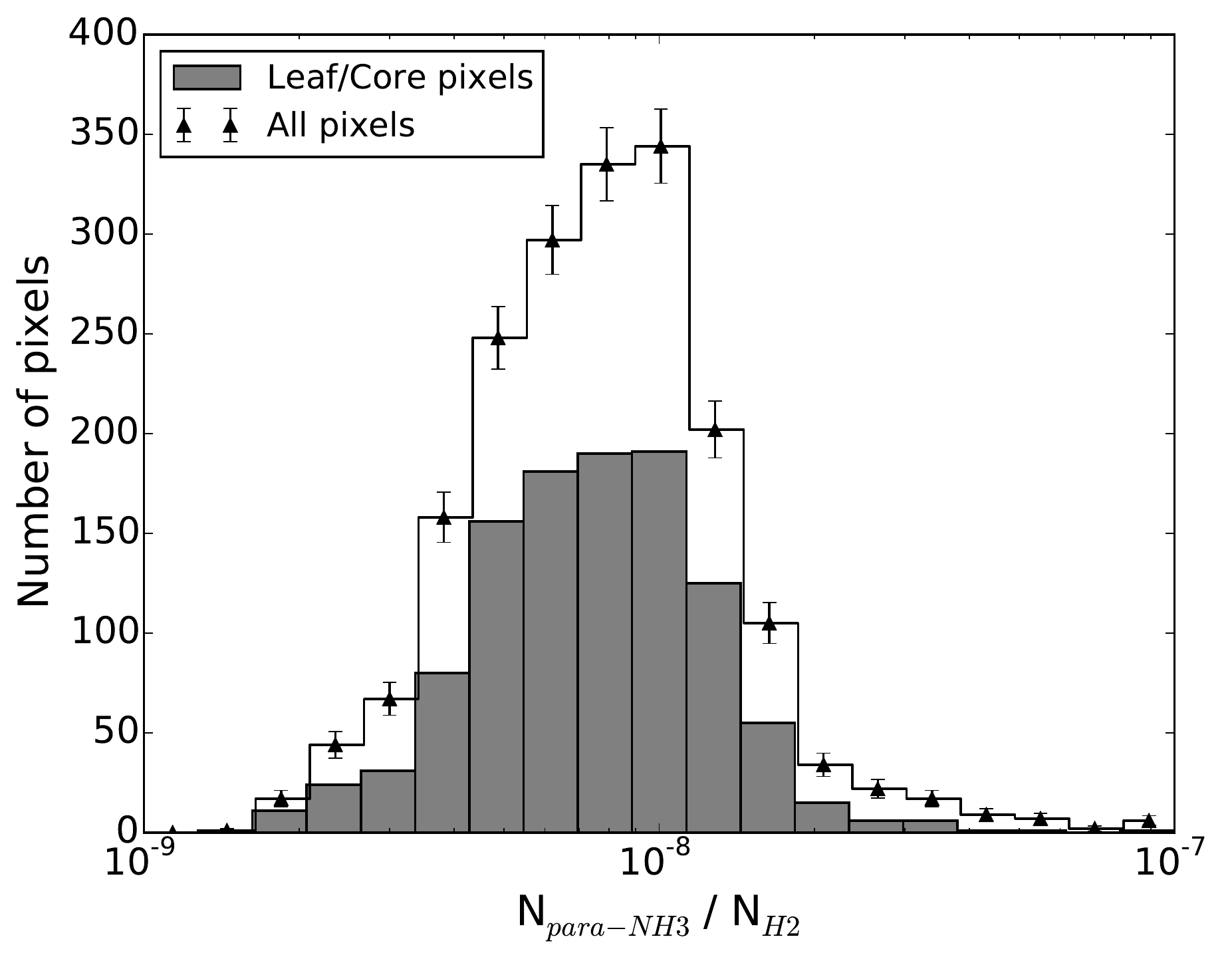}
\plotone{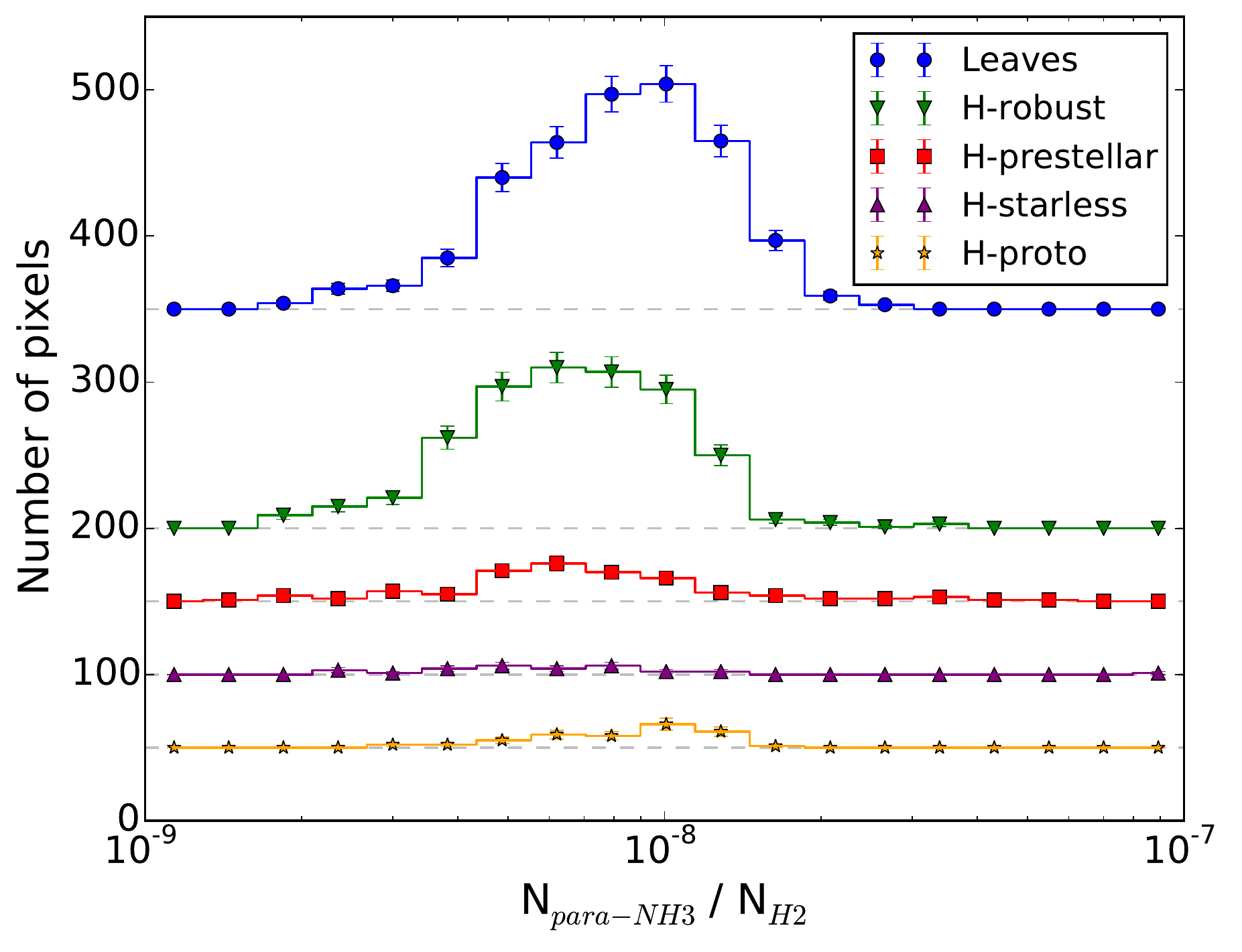}
\caption{Top: Histogram of para-NH$_3$ abundance (N$_{para-NH3}$ / N$_{H2}$) for all pixels with reliable kinematic measurements (black line) and all pixels that fall within either an ammonia-identified leaf or \textit{Herschel}-identified dense core (grey bars).  Bottom: Histograms of para-NH$_3$ abundance for all pixels falling within each structure type (colors represent the same structures as those shown in Figure \ref{Thist}).  The orange, purple, red, green, and blue histograms are offset from the zero position on the $y$-axis by 50, 100, 150, 200, and 350, respectively.  The dashed grey lines represent the y=0 position for each histogram. }
\label{Mass_new3}
\end{figure}

\begin{figure}[ht]
\epsscale{1.1}
\plottwo{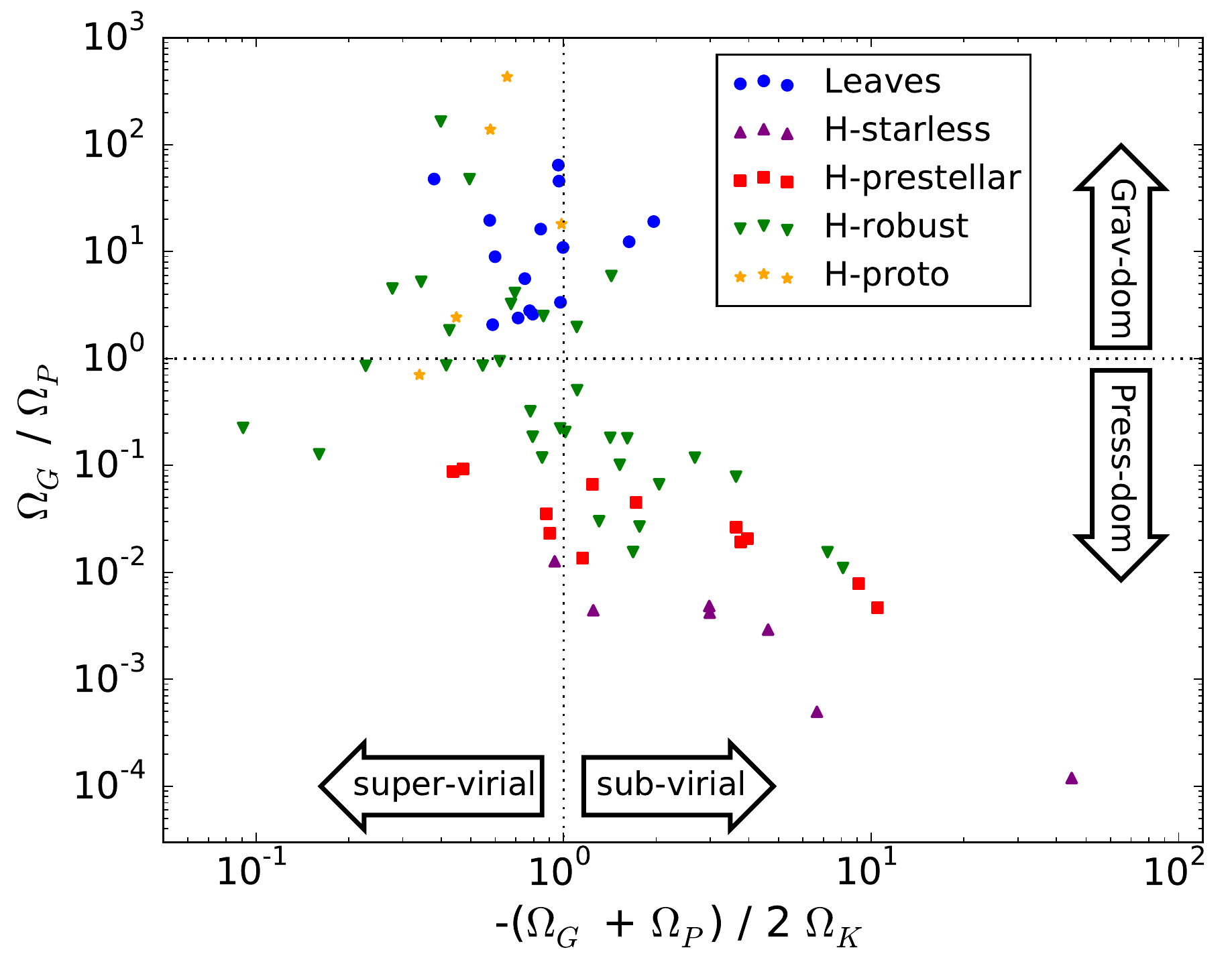}{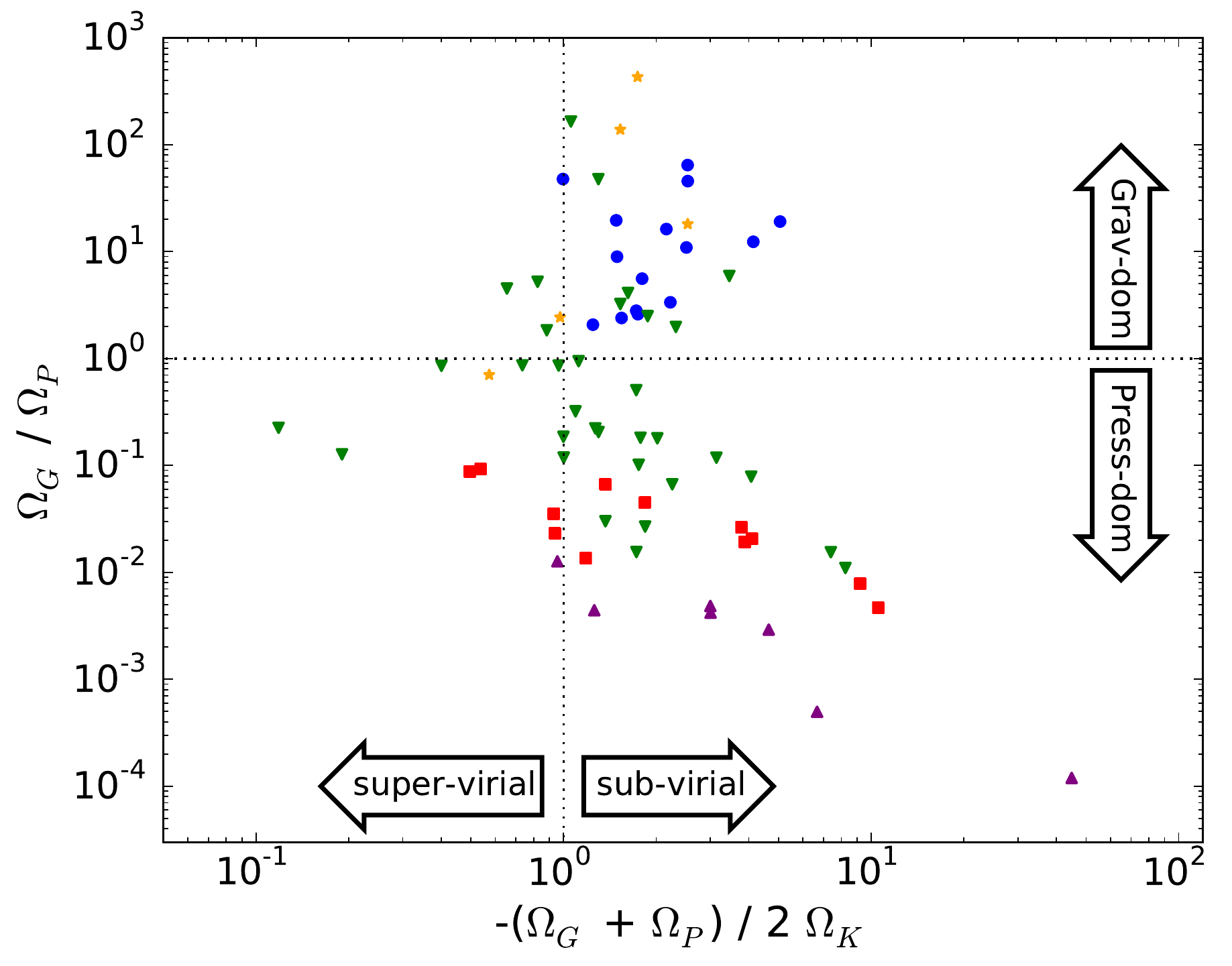}
\caption{Left: Virial plane, including the contributions of gravity, pressure, and kinetic energy, for the same structures shown in the top panel of Figure \ref{Mass_vs_virial}.  Structures to the right of the vertical line are sub-virial, while structures on the left are super-virial.  Structures below the horizontal line are dominated by pressure over gravity, while structures above the line are dominated by gravity over pressure.  The density profile used in $\Omega_G$ is assumed to be Gaussian, consistent with \cite{Kirk_submitted}.  Right: Virial plane for the same structures as in the left panel when a power-law density profile is assumed for $\Omega_G$, consistent with \cite{Friesen_submitted}.}
\label{virial_plane}
\end{figure}

\begin{figure}[ht]
\epsscale{1.0}
\plottwo{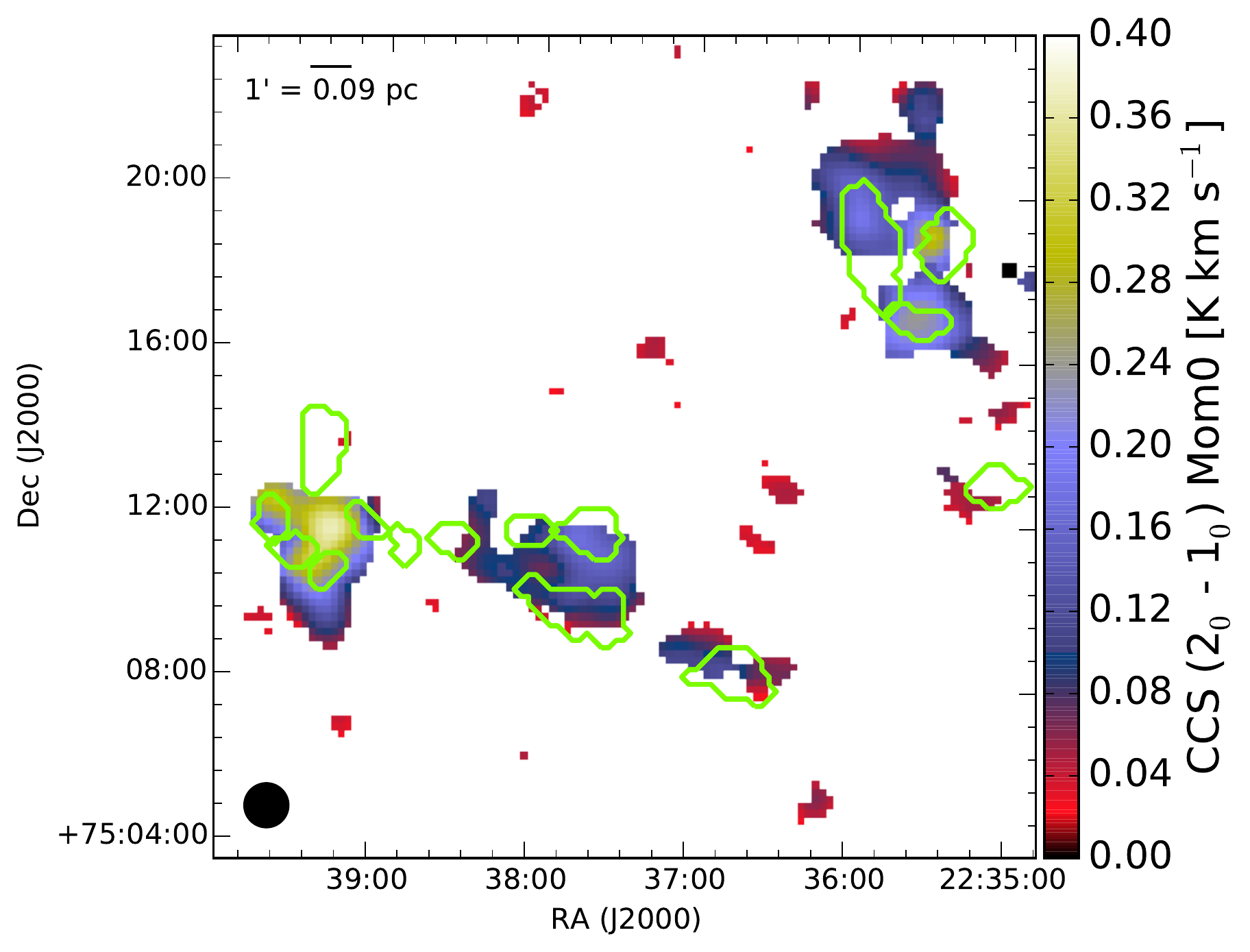}{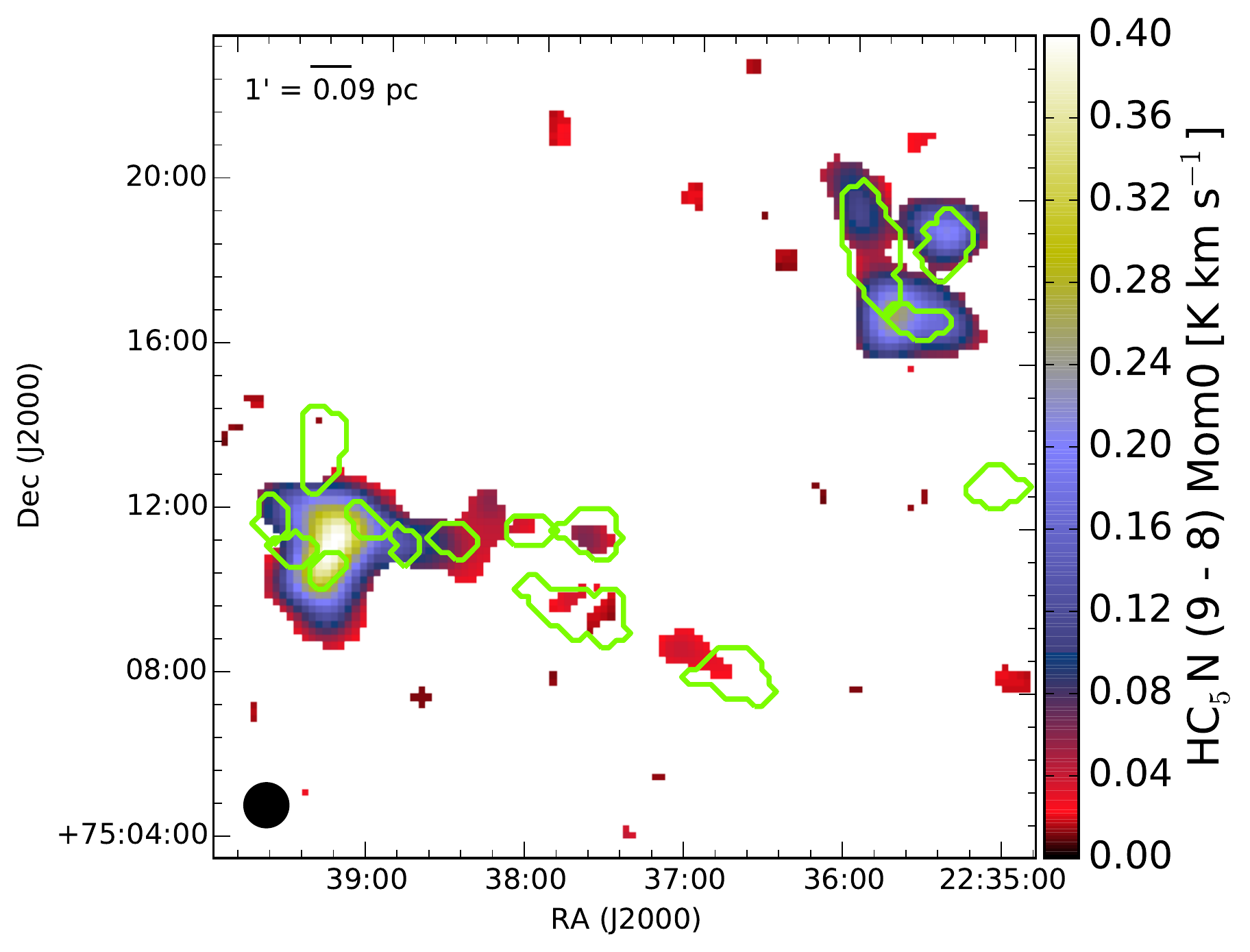} \\
\plottwo{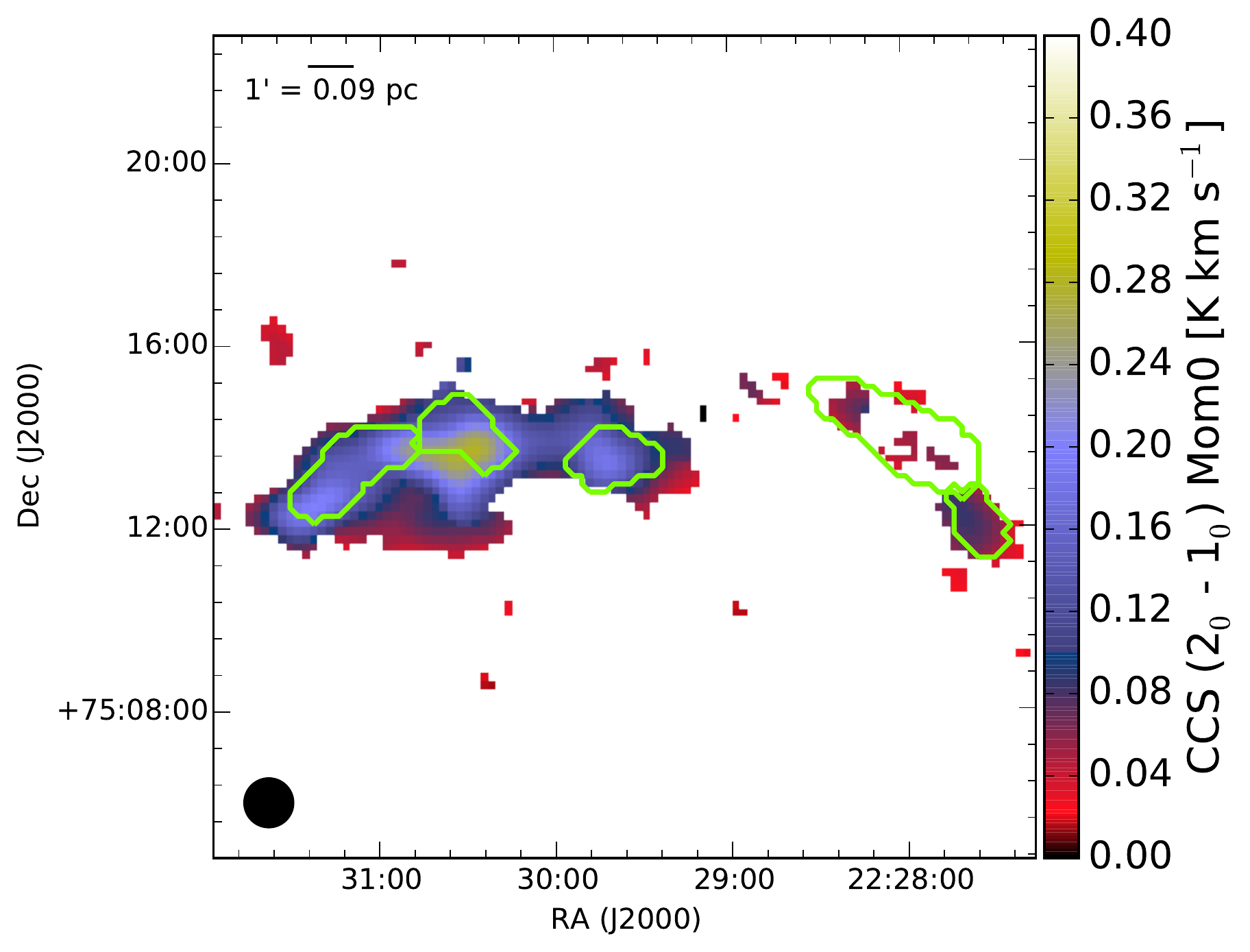}{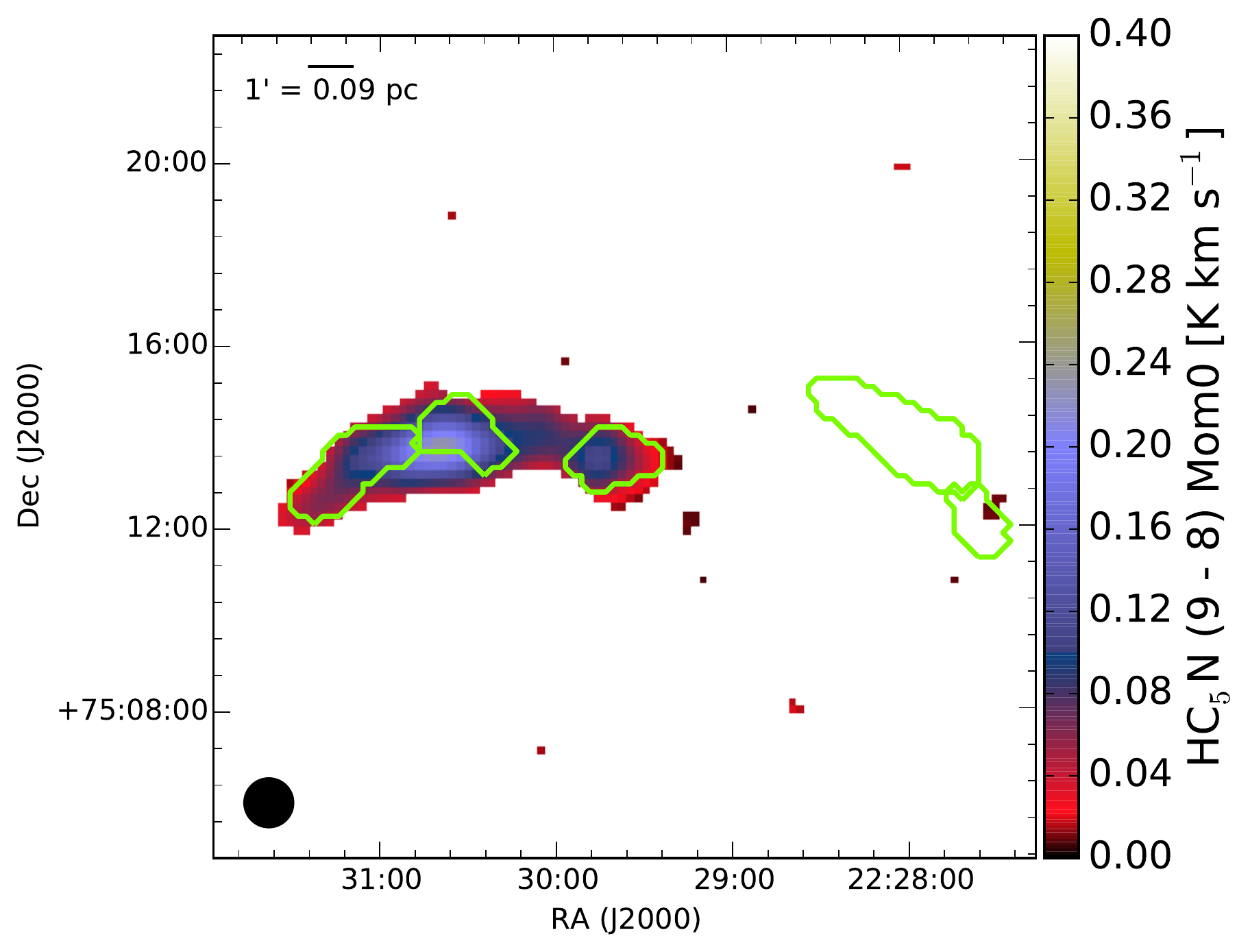} \\
\plottwo{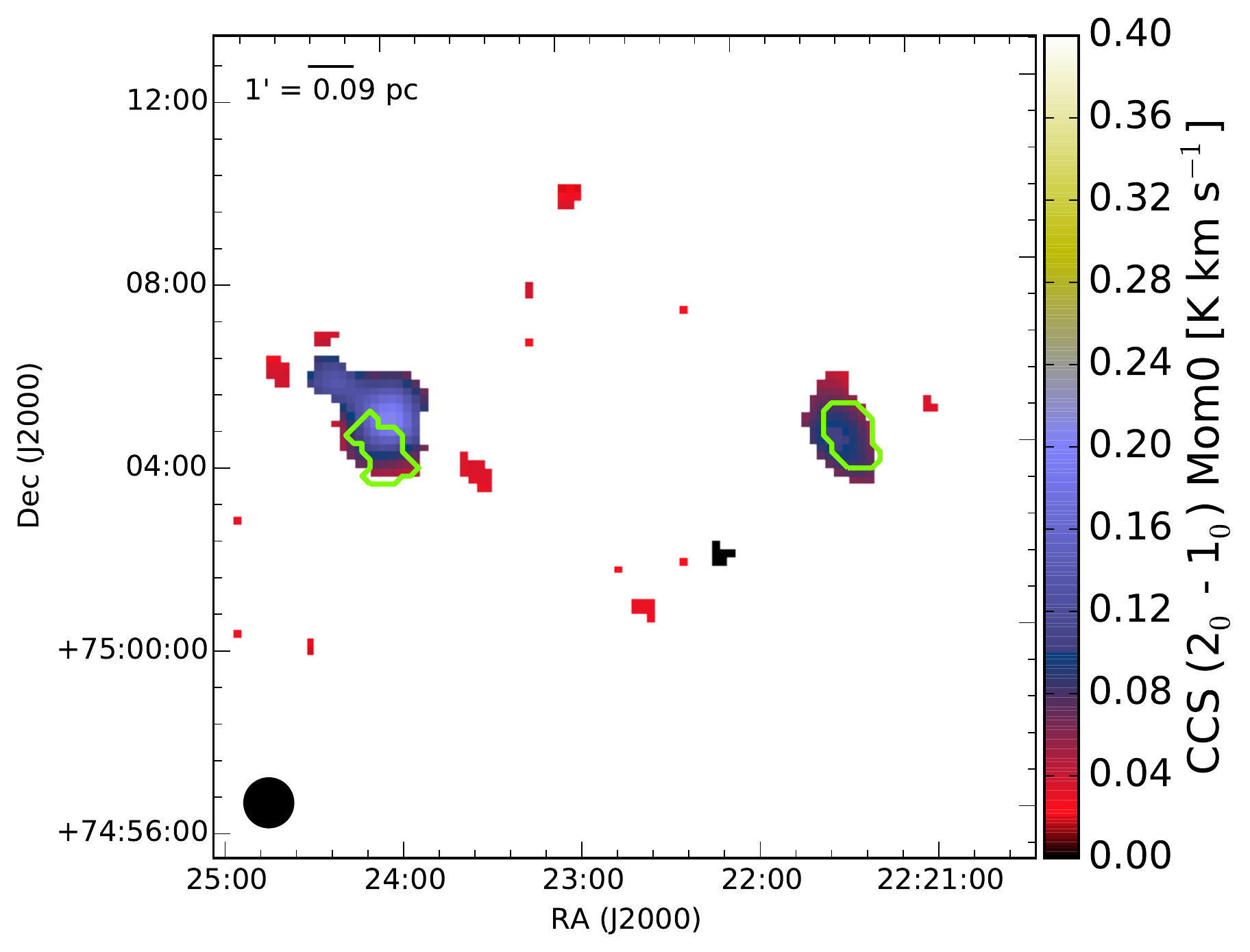}{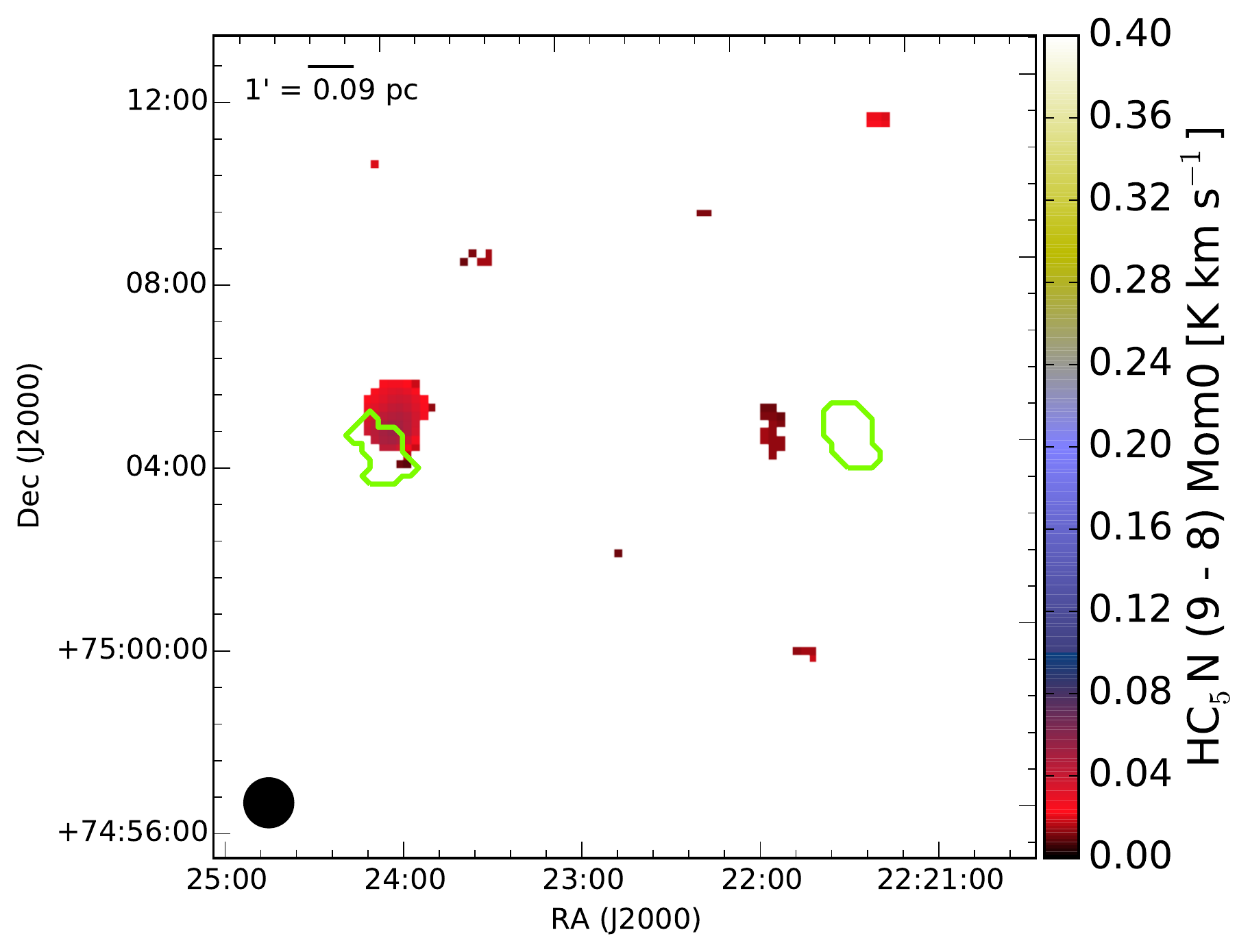}
\caption{Comparison of CCS $(2_0-1_0)$ (left column) versus HC$_5$N $(9-8)$ (right column) emission observed towards Cepheus-L1251.  The colorscale shows the integrated intensity map created from Gaussian fits to the real molecular emission that was first convolved to a spatial resolution of 64$\arcsec$ (see text for details; new beam-size shown in lower left corner of all plots).  The ammonia-identified leaves detected in this paper are outlined in green.  Each row shows the same fields displayed in Figure \ref{leaves}.  }
\label{CCS_HC5N}
\end{figure}

\begin{figure}[ht]
\epsscale{1.0}
\plottwo{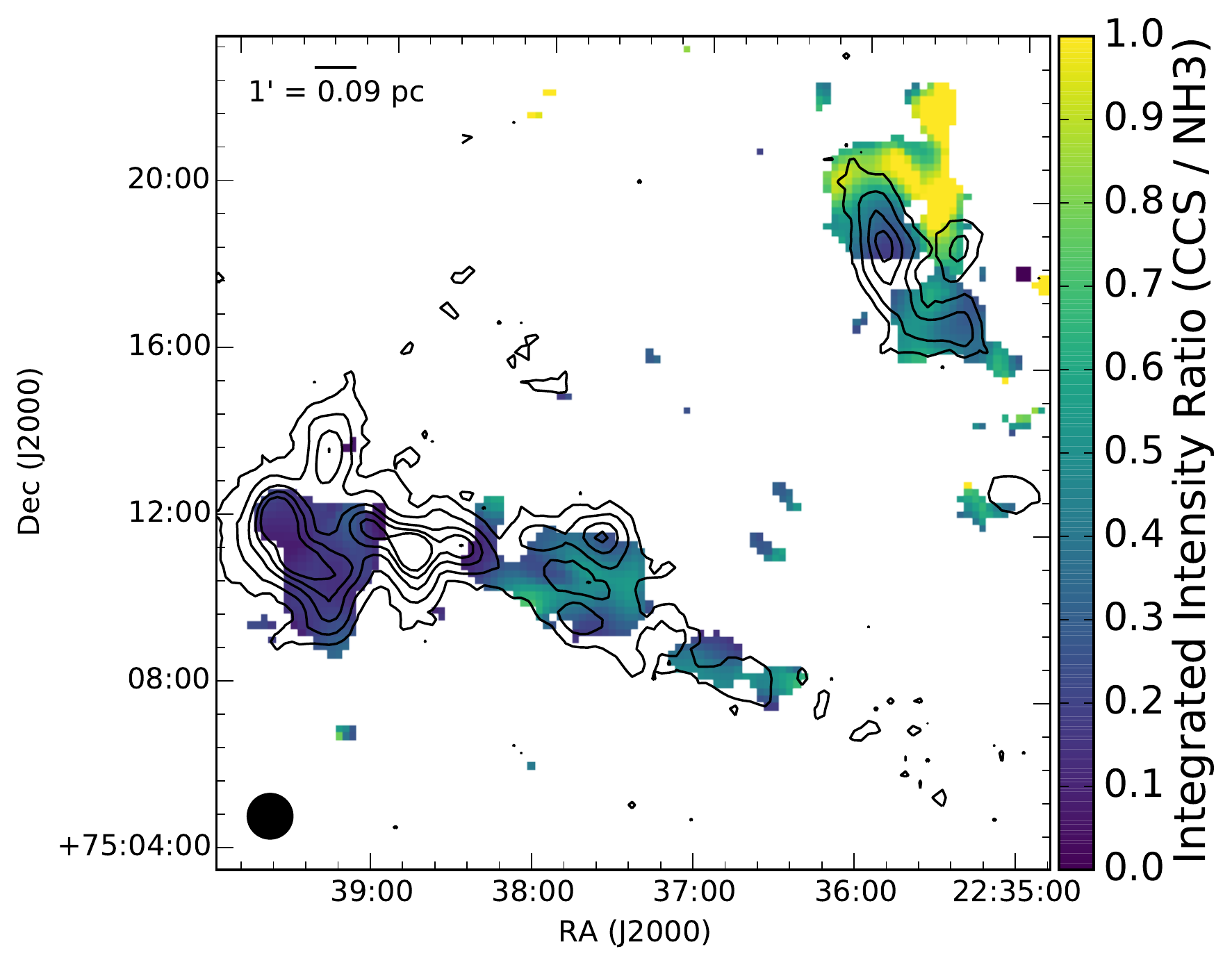}{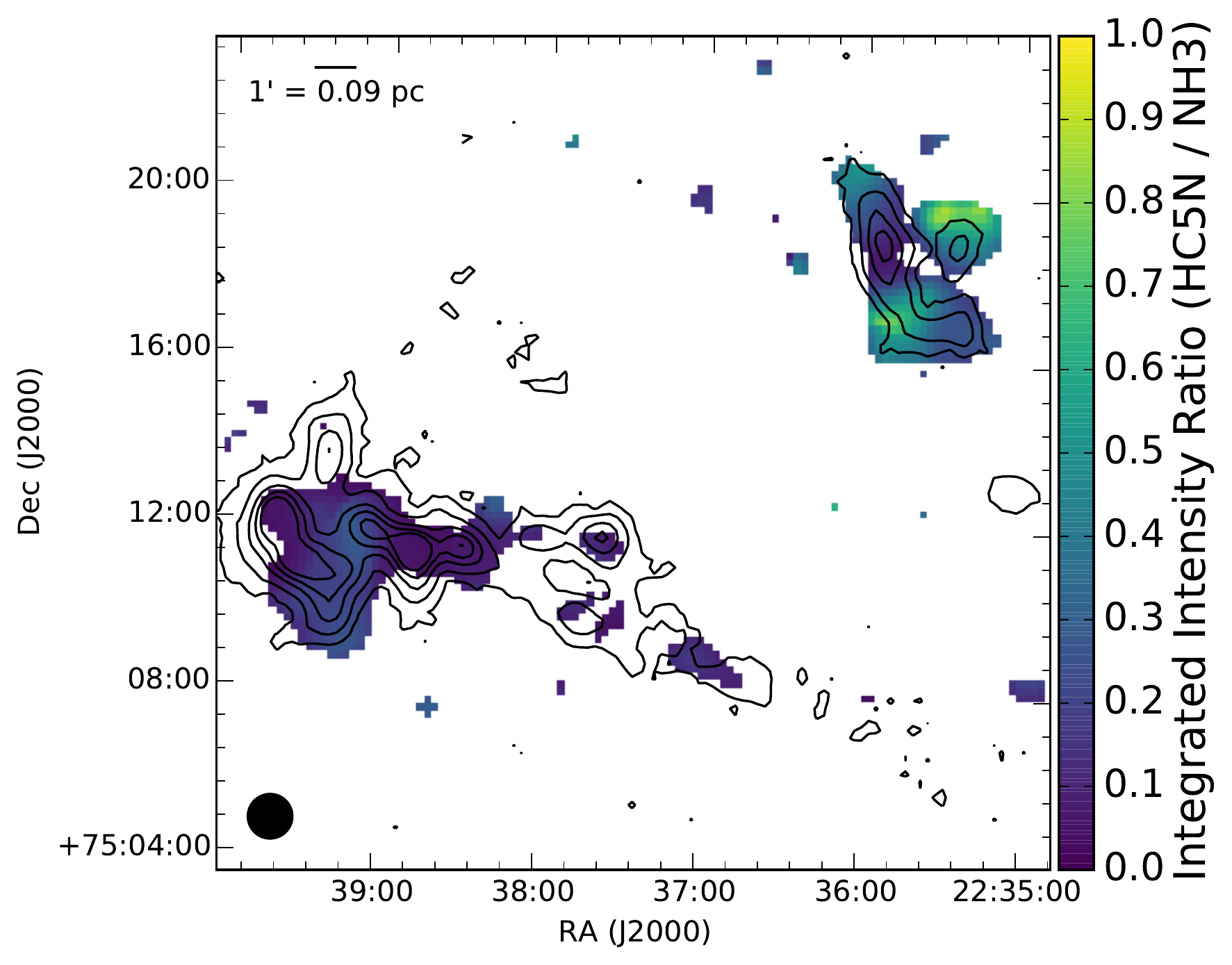} \\
\plottwo{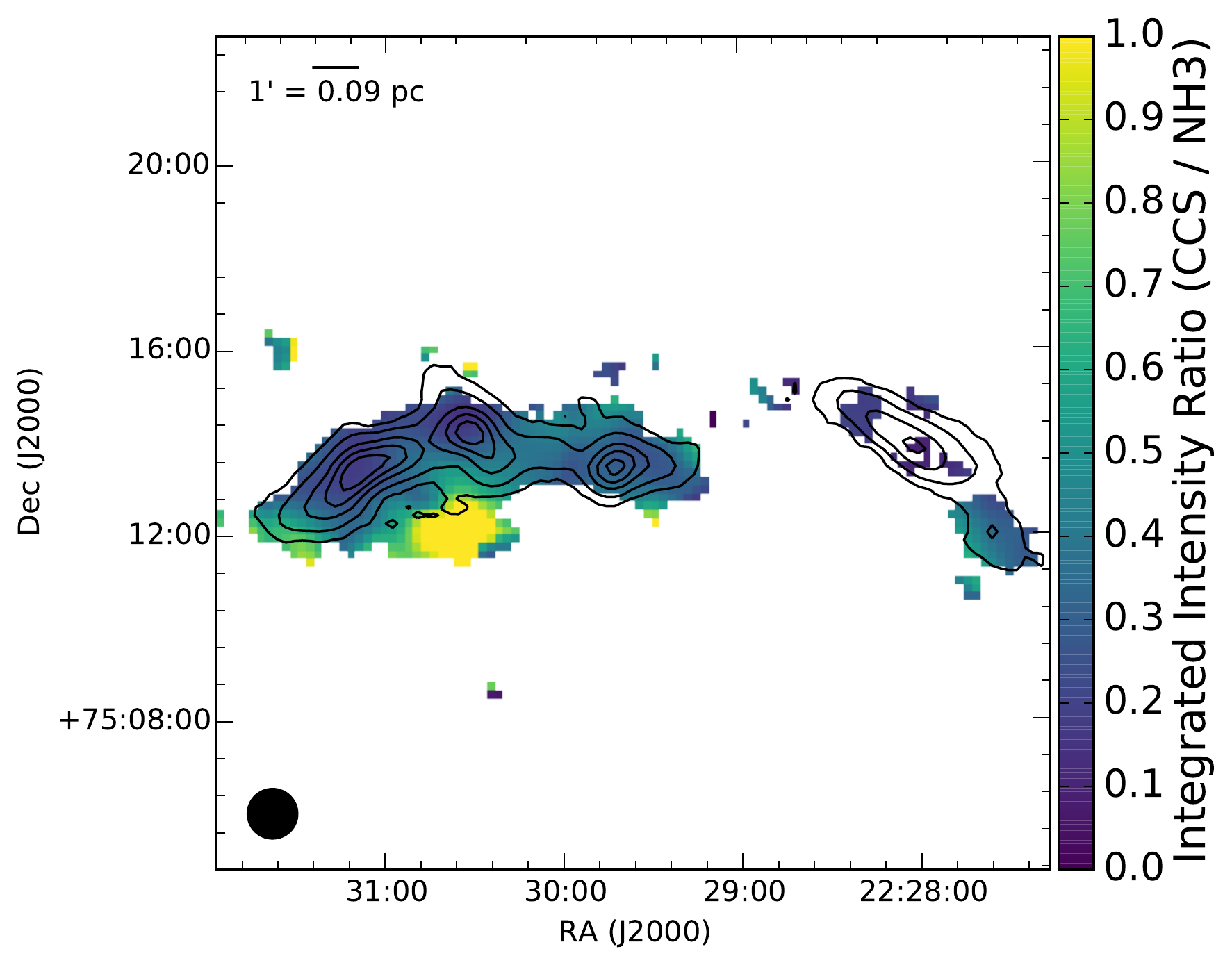}{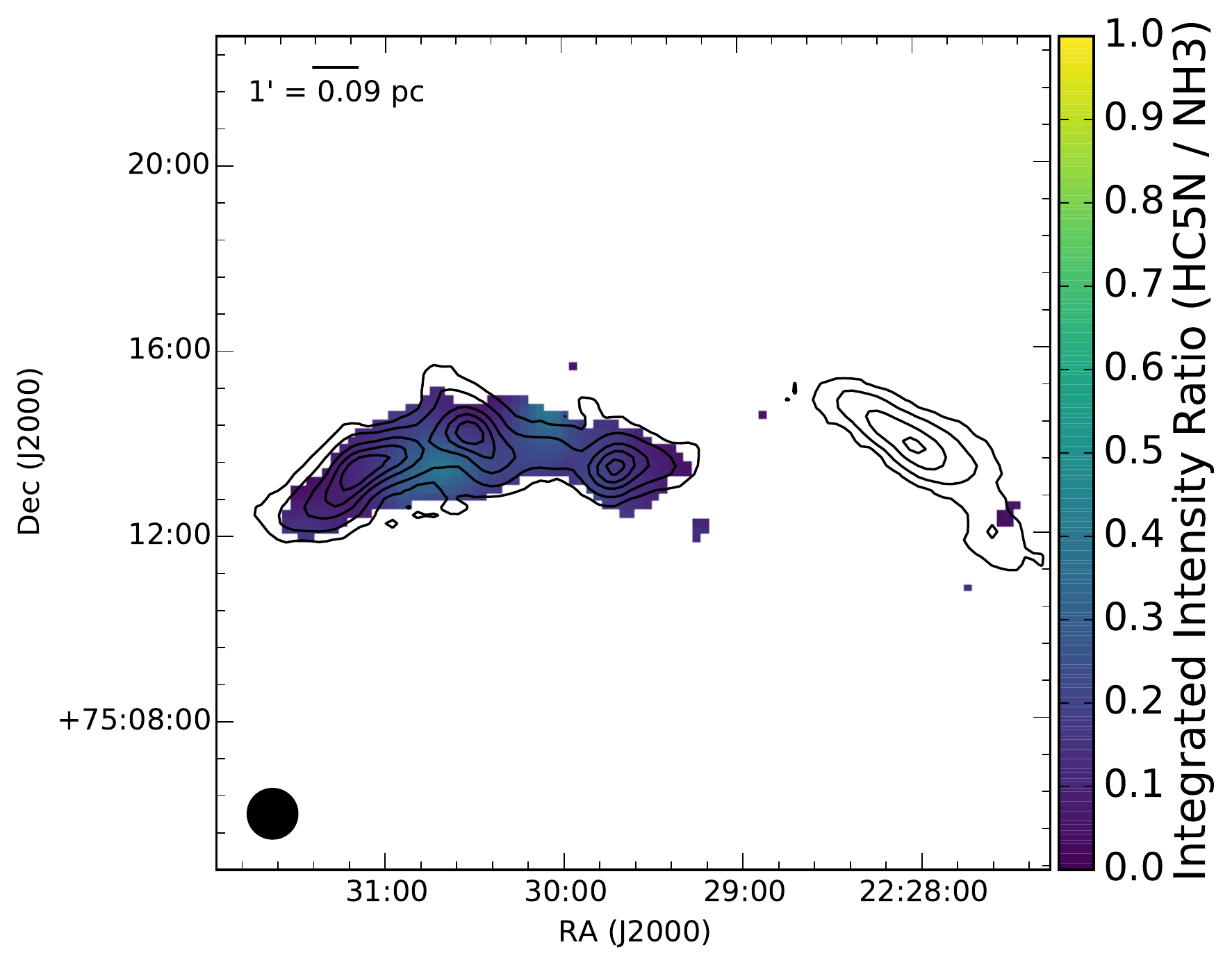} \\
\plottwo{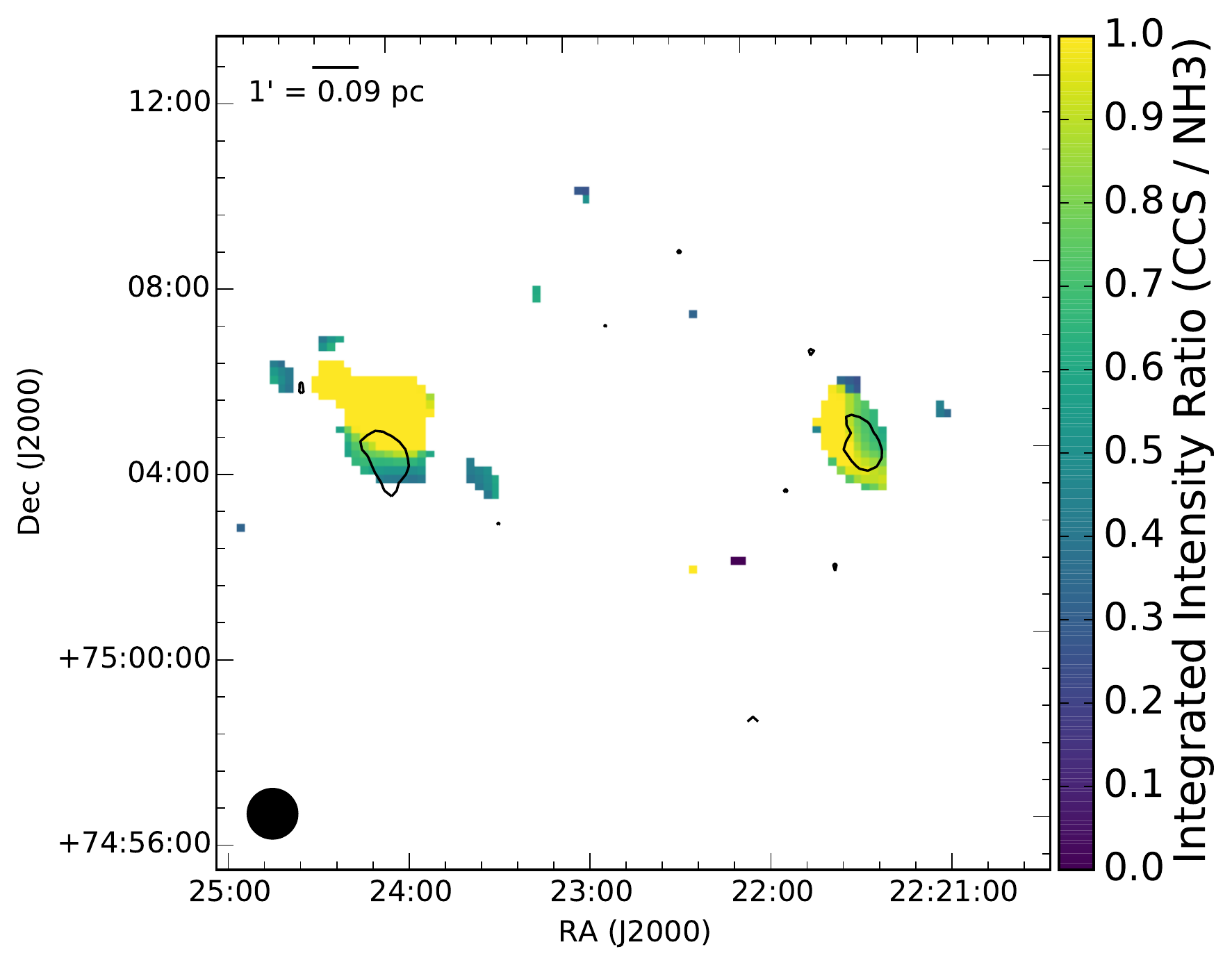}{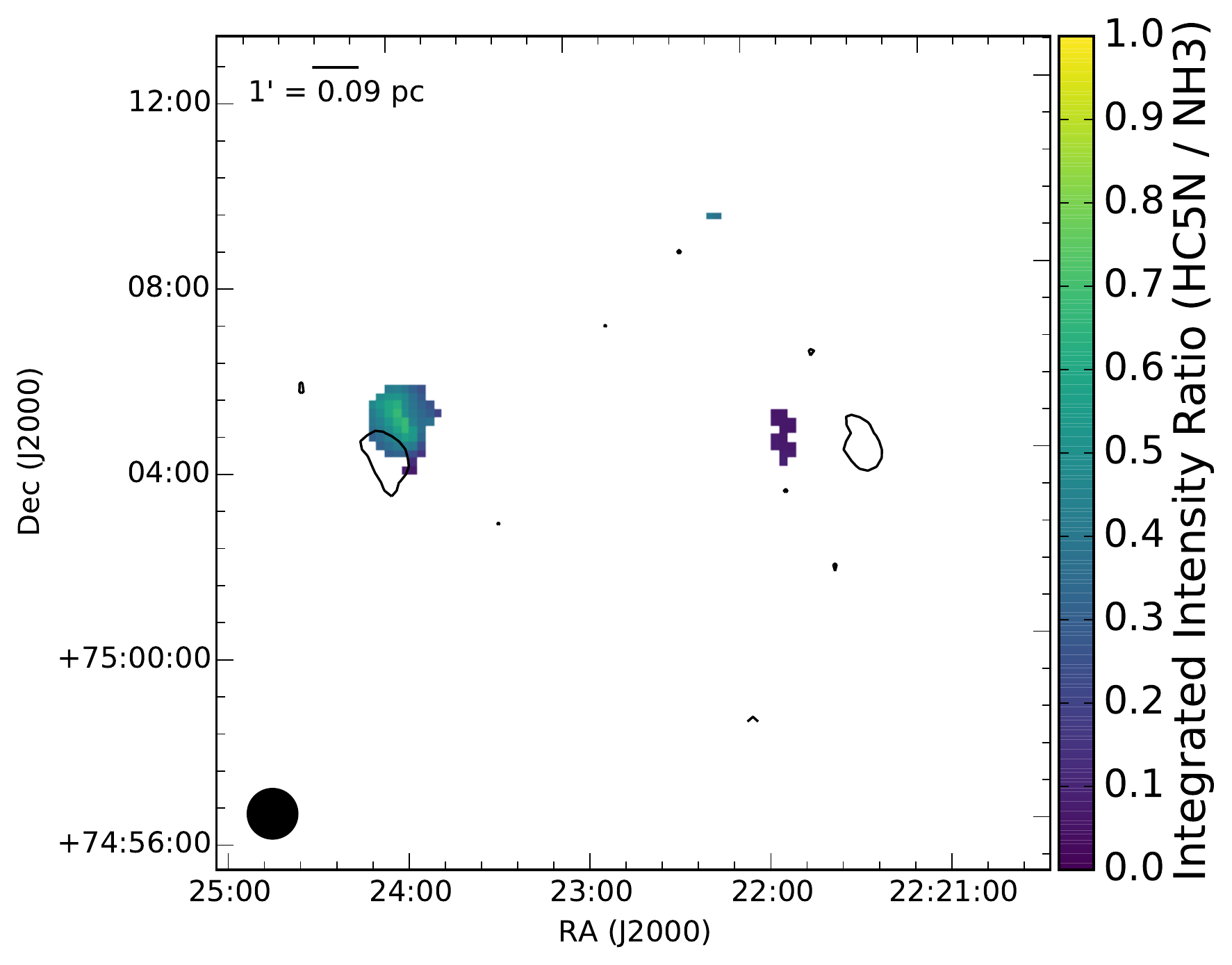}
\caption{Ratios of CCS $(2_0-1_0)$ integrated intensity (left column) and HC$_5$N $(9-8)$ integrated intensity (right column) to NH$_3$ (1,1) integrated intensity for the same fields shown in Figure \ref{CCS_HC5N}.  Black contours represent the NH$_3$ (1,1) integrated intensity at 0.5, 1.5, 3.5, 5.5, and 7.5 K km s$^{-1}$.}
\label{CCS_HC5N_ratios}
\end{figure}

\begin{figure}[ht]
\epsscale{0.6}
\plotone{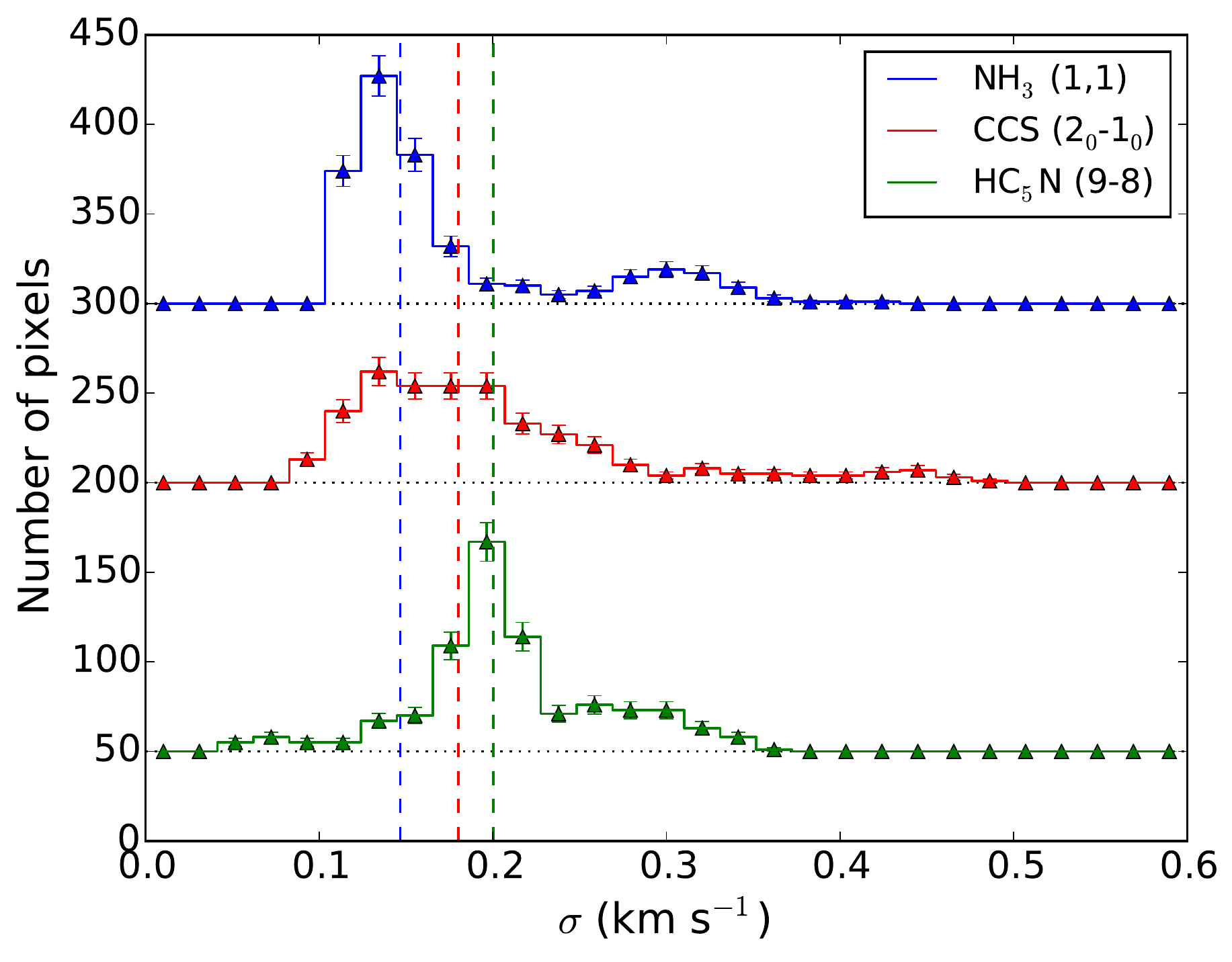}
\plotone{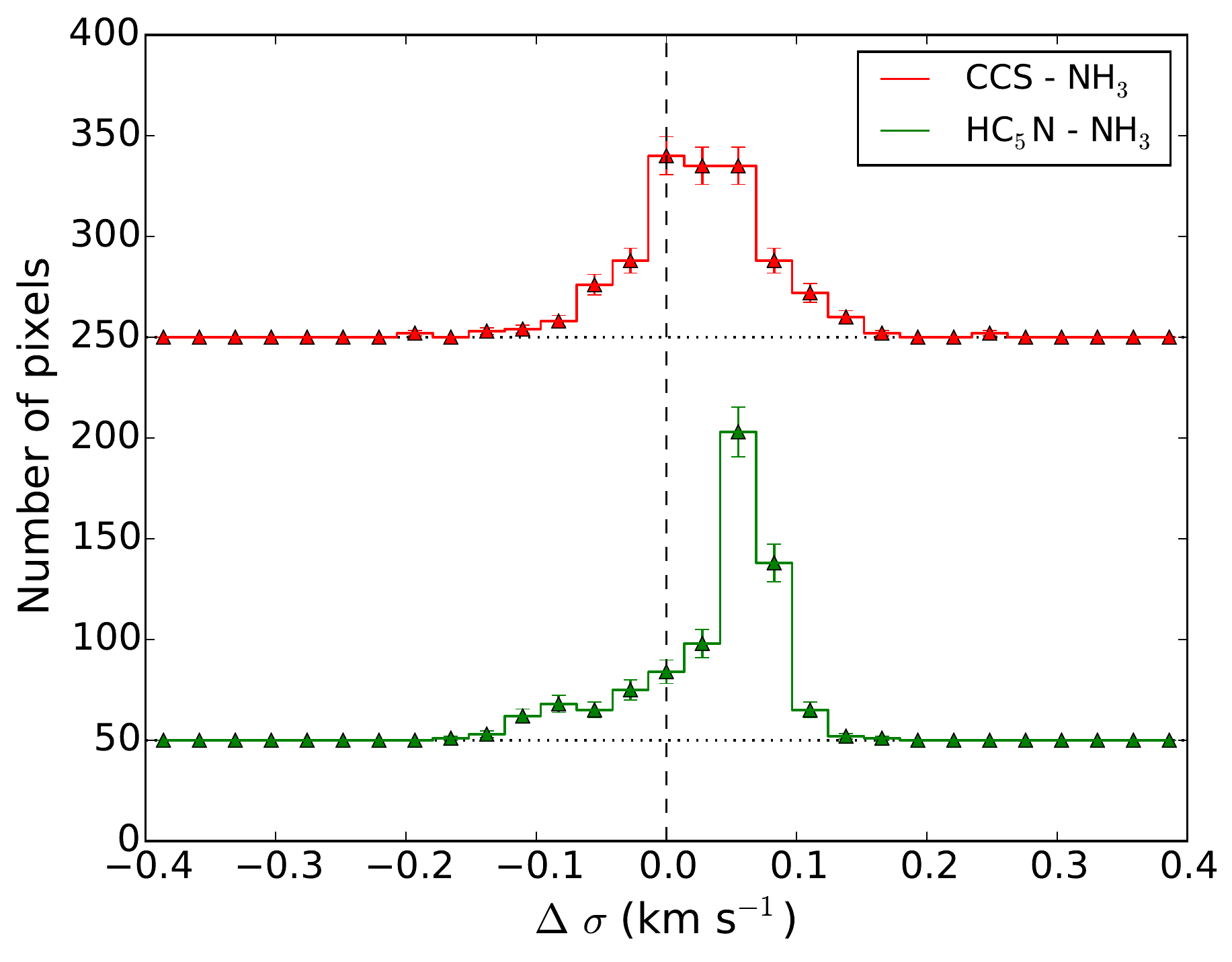}
\caption{Top: histograms of velocity dispersion for all pixels falling within an ammonia-identified leaf in the NH$_3$ (1,1) (blue), CCS $(2_0-1_0)$ (red) and HC$_5$N $(9-8)$ (green) emission maps.  All maps were convolved to a spatial resolution of 64$\arcsec$ before line-fitting and only pixels with reliable fits in all three transitions are included in the histograms.  The vertical dashed lines represent the median value of $\sigma$ for each distribution.  The HC$_5$N $(9-8)$, CCS $(2_0-1_0)$, and  NH$_3$ (1,1) histograms have been offset from the zero position of the $y$-axis by 50, 200, and 300, respectively (denoted by the horizontal dotted lines).  Bin widths are set to 0.021 km s$^{-1}$.  Bottom: $\Delta\sigma$ offset between HC$_5$N $(9-8)$ and NH$_3$ (1,1) (green) and between CCS $(2_0-1_0)$ and NH$_3$ (1,1) (red) for all pixels in the top panel.  Bin widths are set to 0.028 km s$^{-1}$.  The green and red histograms have been offset from the zero position of the $y$-axis by 50 and 250, respectively.  The vertical dotted black line denotes $\Delta\sigma$ = 0.}
\label{sigma_carbons}
\end{figure}

\begin{figure}[ht]
\epsscale{1.1}
\plottwo{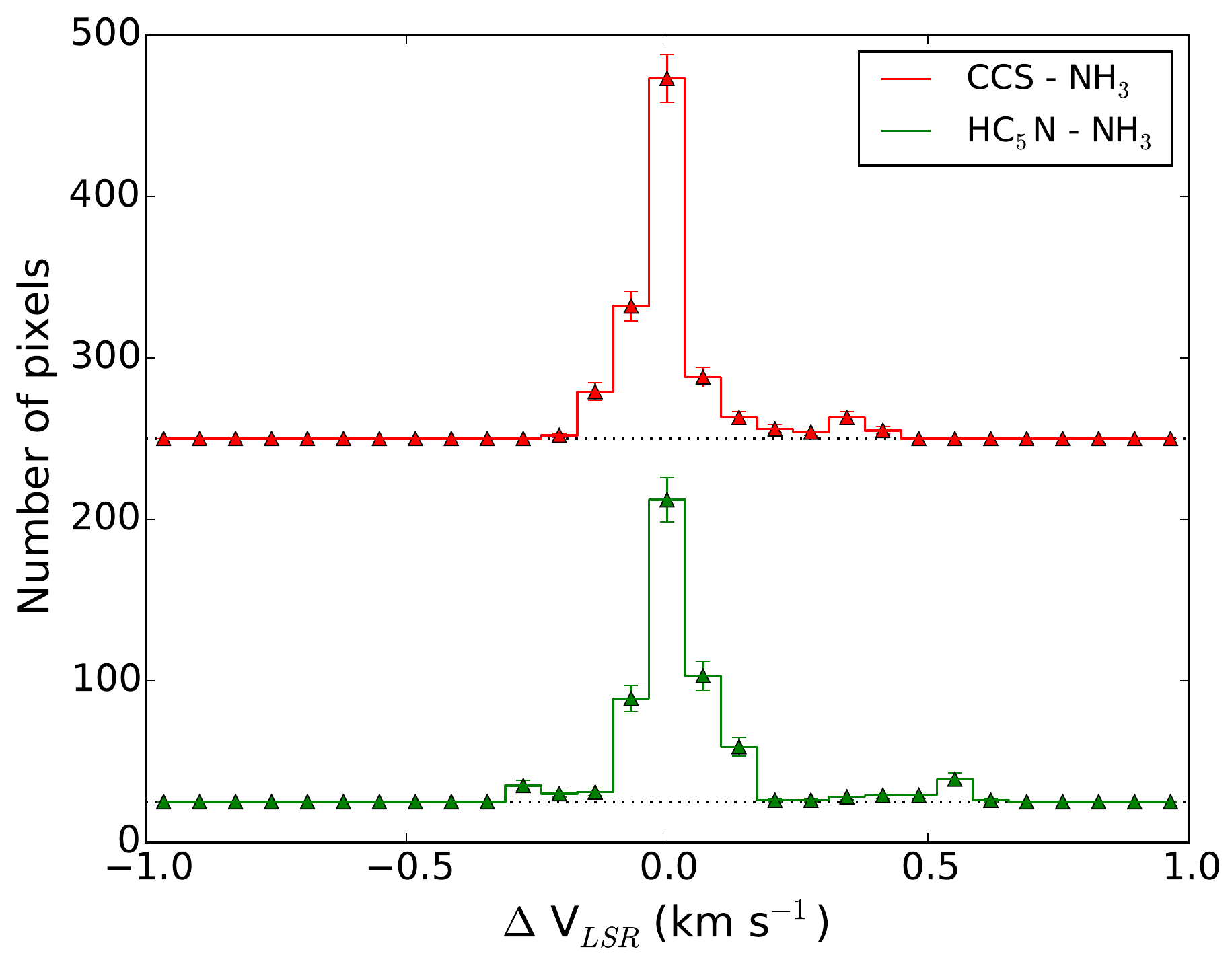}{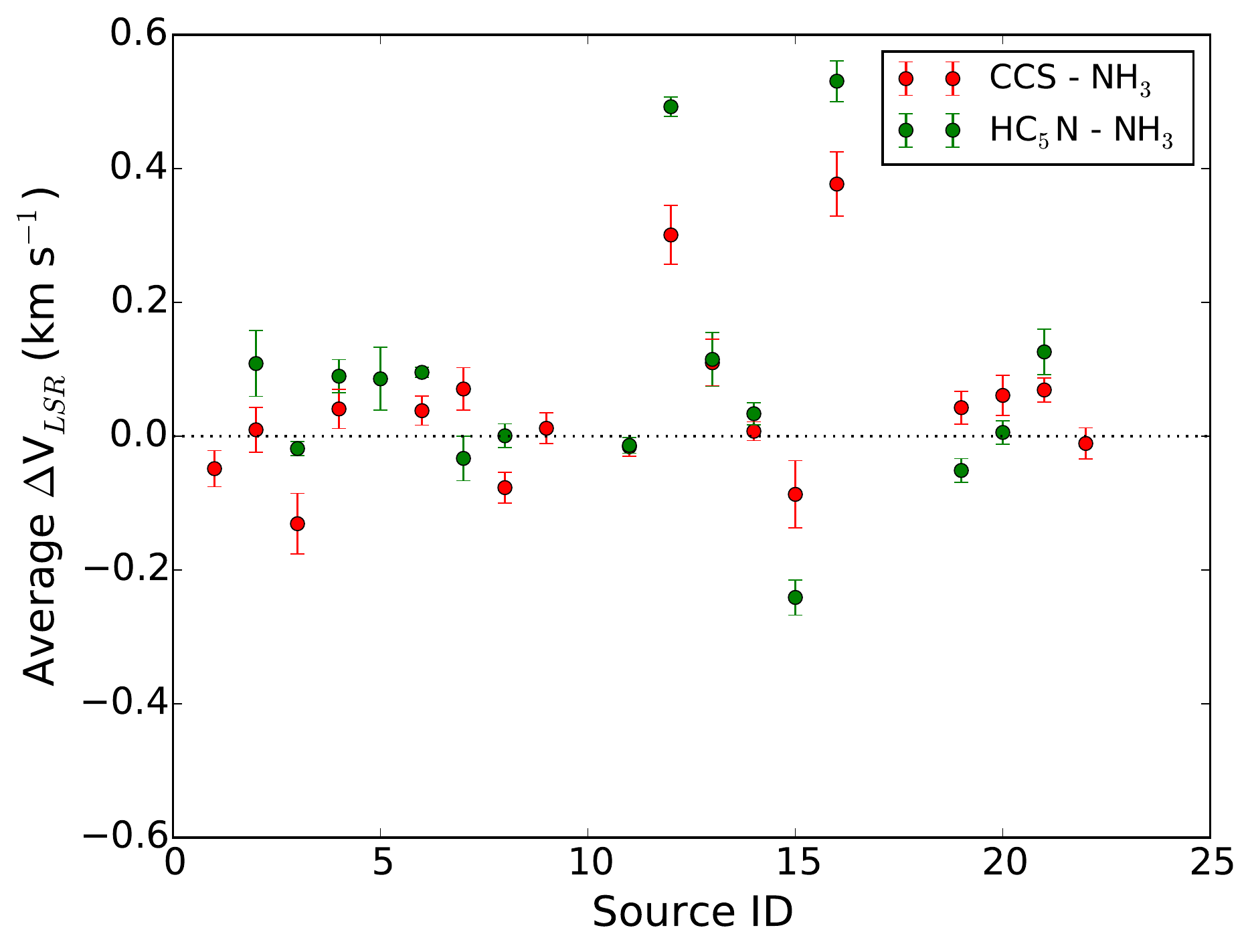} \\
\plottwo{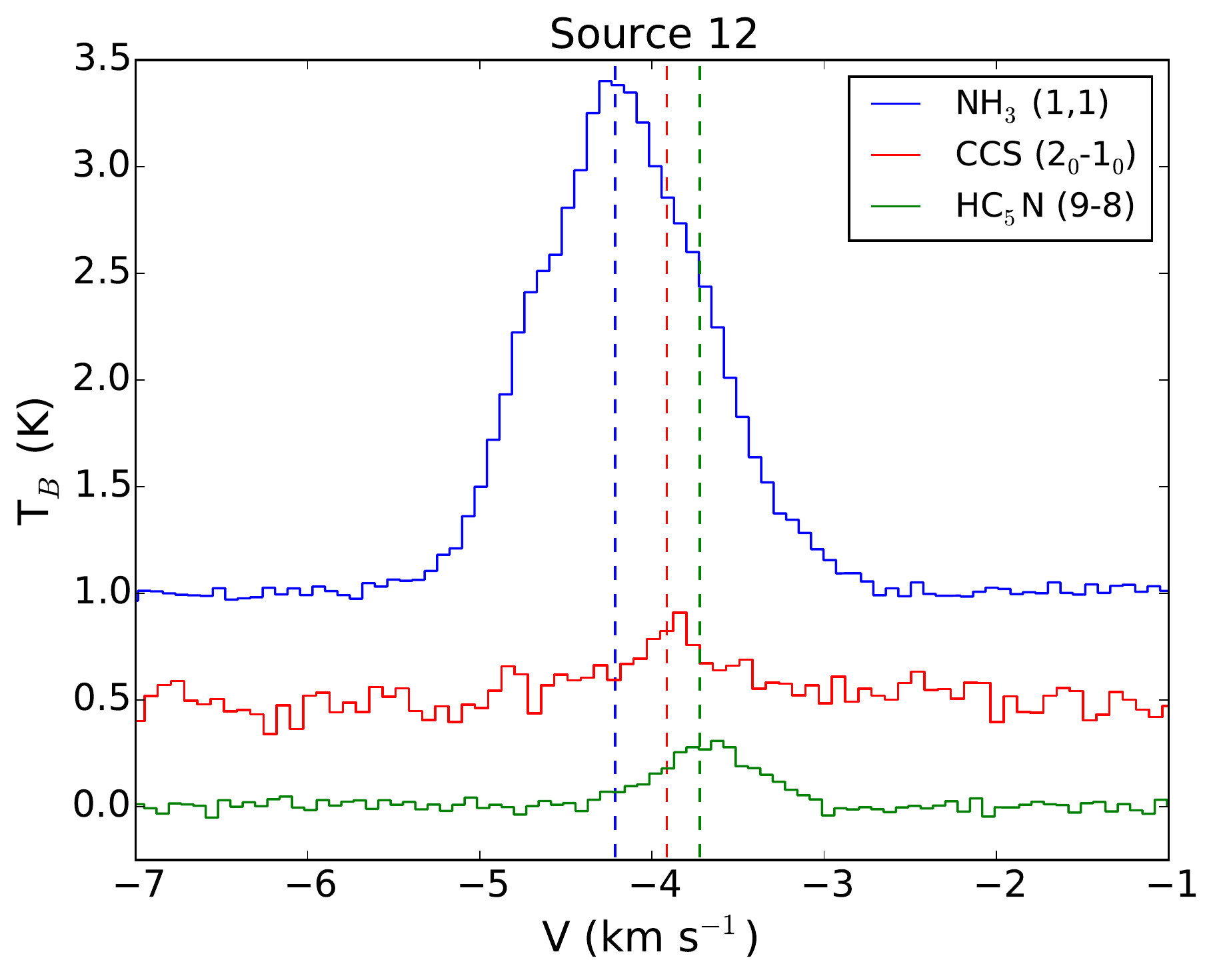}{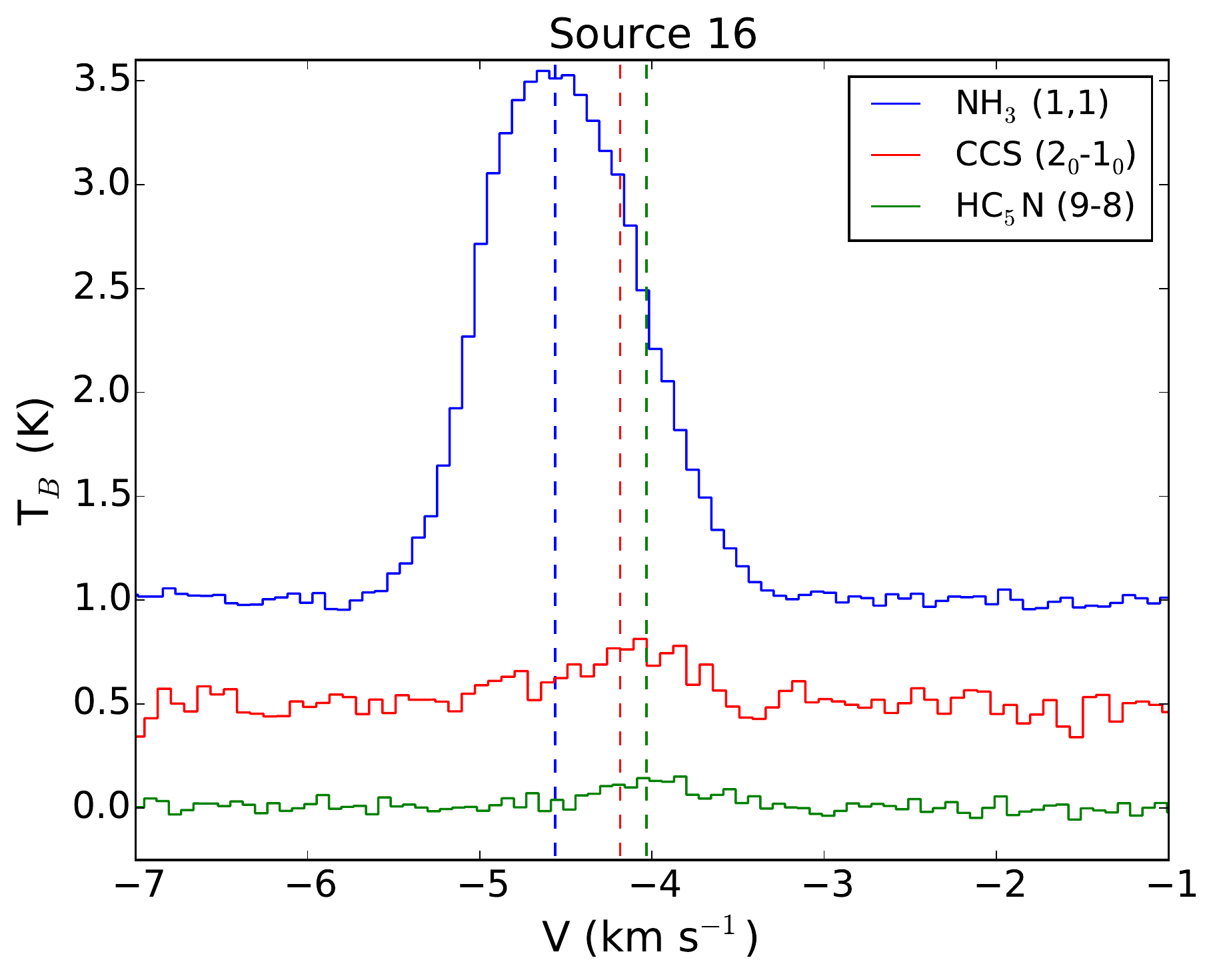}
\caption{Top left: histograms of the difference between the centroid velocity of HC$_5$N $(9-8)$ (green) or CCS $(2_0-1_0)$ (red) and the centroid velocity of NH$_3$ (1,1) for all pixels falling within an ammonia-identified leaf.  The green and red histograms have been offset from the zero position of the $y$-axis by 25 and 250, respectively (denoted by the horizontal dotted black lines).  Top right: Average $\Delta V_{LSR}$ for all pixels falling within each individual ammonia-identified source, weighted by the integrated intensity maps of each respective molecular transition.  The horizontal dotted line marks $\Delta V_{LSR}$ = 0. Bottom row: Average observed spectra for all pixels falling within source 12 (left) and source 16 (right) listed in Table \ref{Table_NH3}, weighted by the integrated intensity maps of each respective molecular transition.  Vertical lines represent the average $V_{LSR}$ for the source in each transition.  The CCS $(2_0-1_0)$ and NH$_3$ (1,1) spectra have been offset from the zero position on the $y$-axis by 0.5 and 1.0 K, respectively. Only the central hyperfine group of the NH$_3$ (1,1) spectrum is shown.}
\label{vlsr_carbons}
\end{figure}

\begin{figure}[ht]
\epsscale{0.5}
\plotone{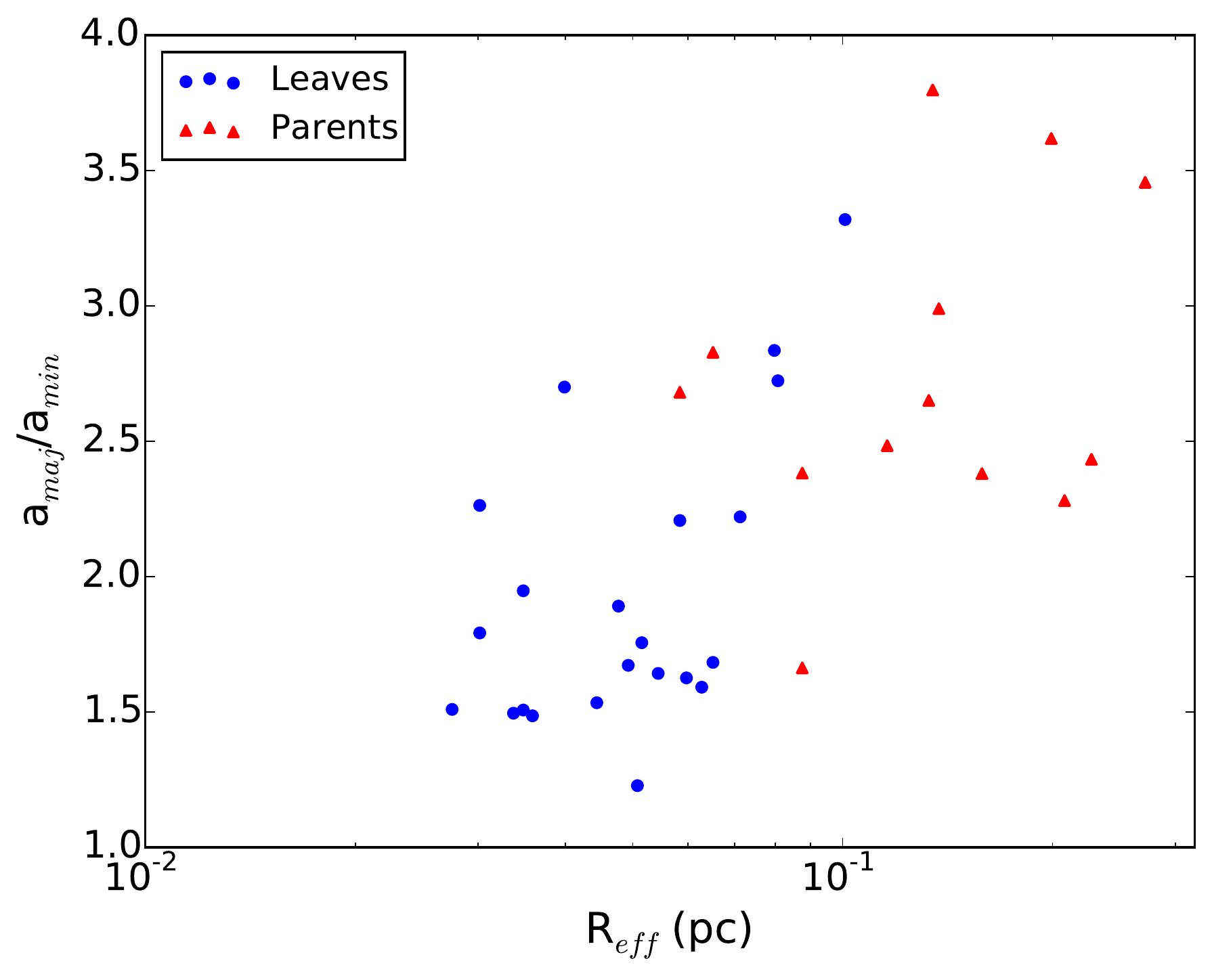}
\plotone{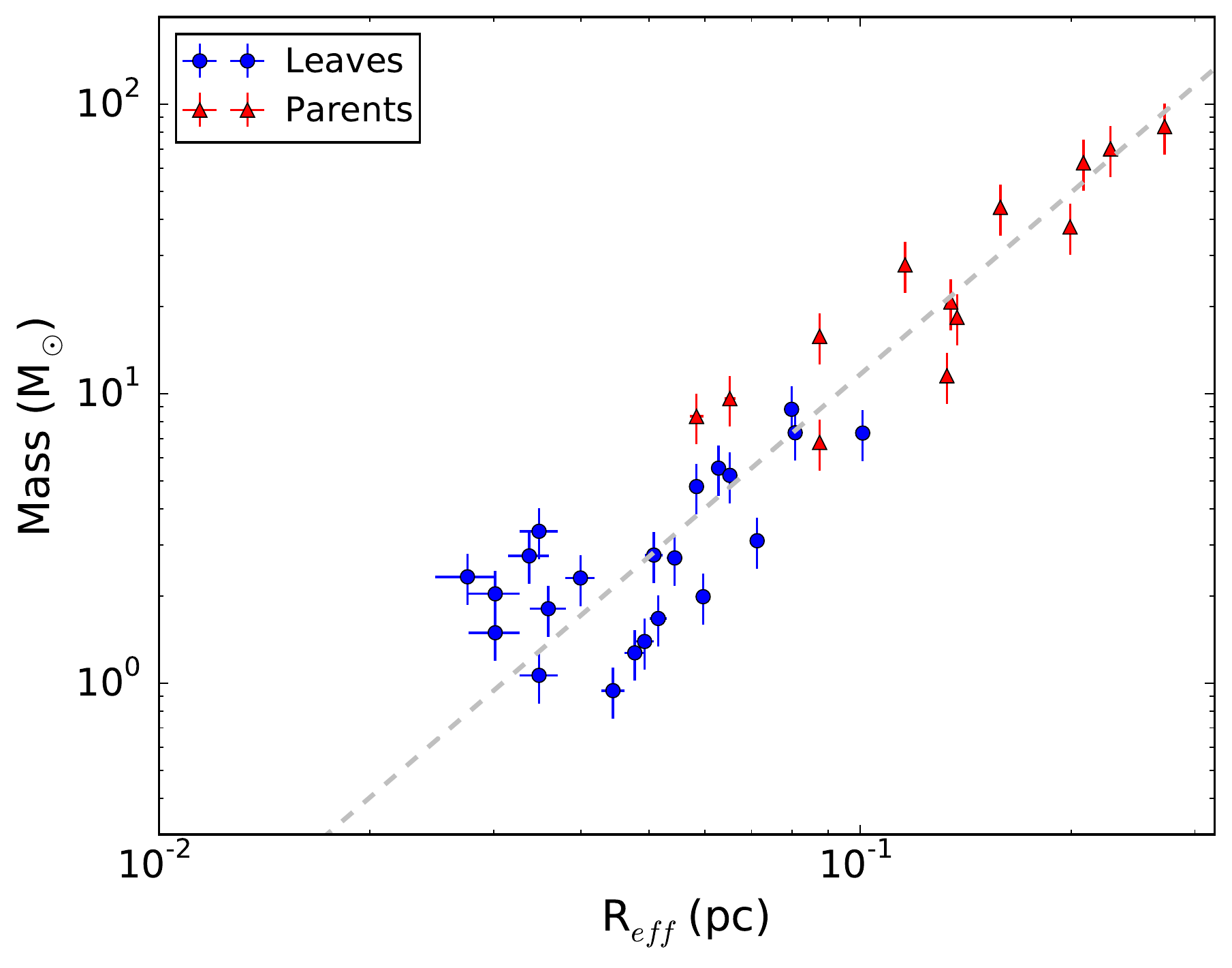}
\plotone{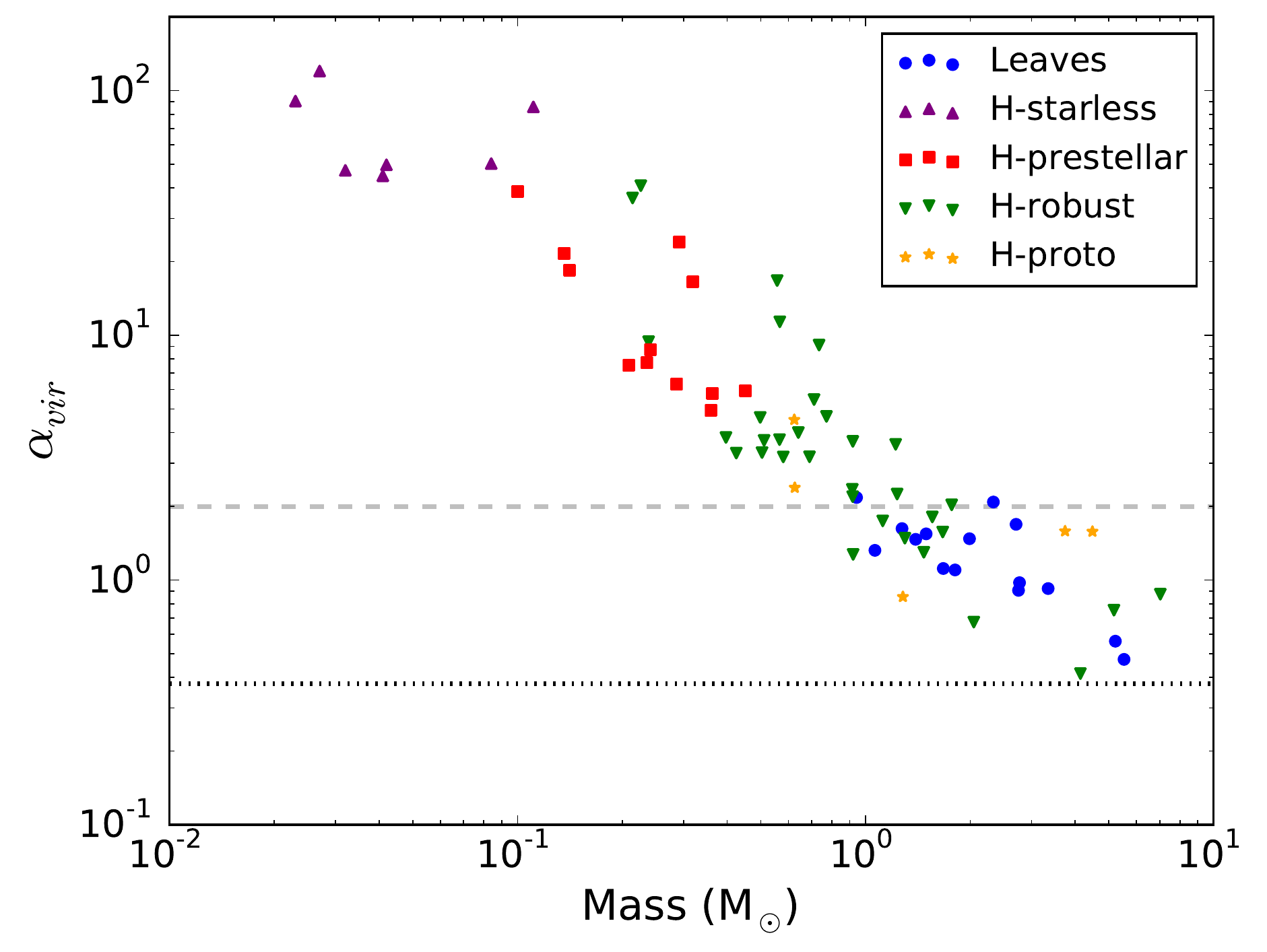}
\caption{Same as Figure \ref{R_vs_aspect} (top), Figure \ref{R_vs_mass} (middle), and Figure \ref{Mass_vs_virial} (bottom), but using $R = (A/\pi)^{1/2}$ as the effective radius for all structures.  The best-fit power law slope in the middle panel is $2.09 \pm 0.15$.}
\label{Reff_new}
\end{figure}

\begin{figure}[ht]
\epsscale{0.5}
\plotone{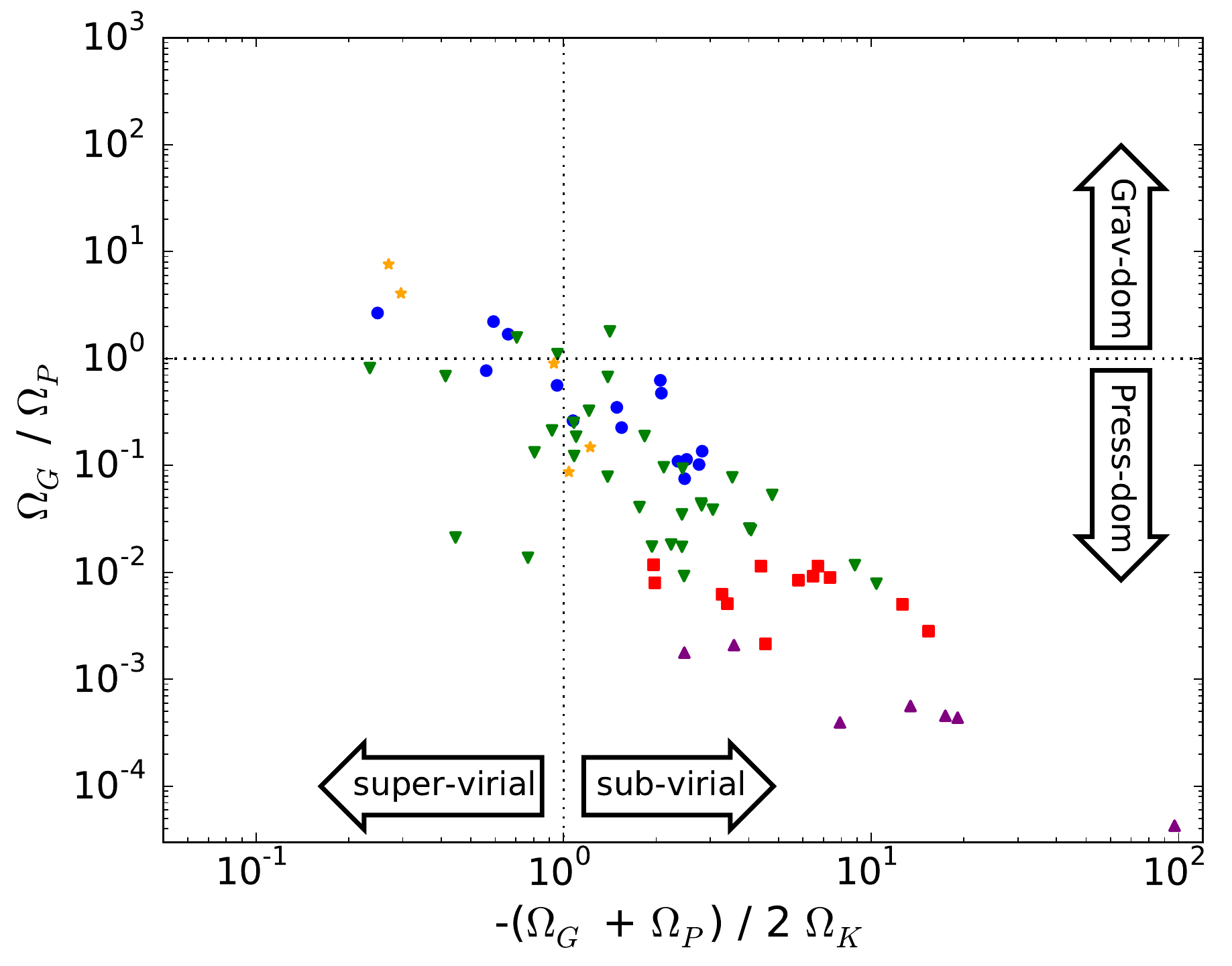}
\plotone{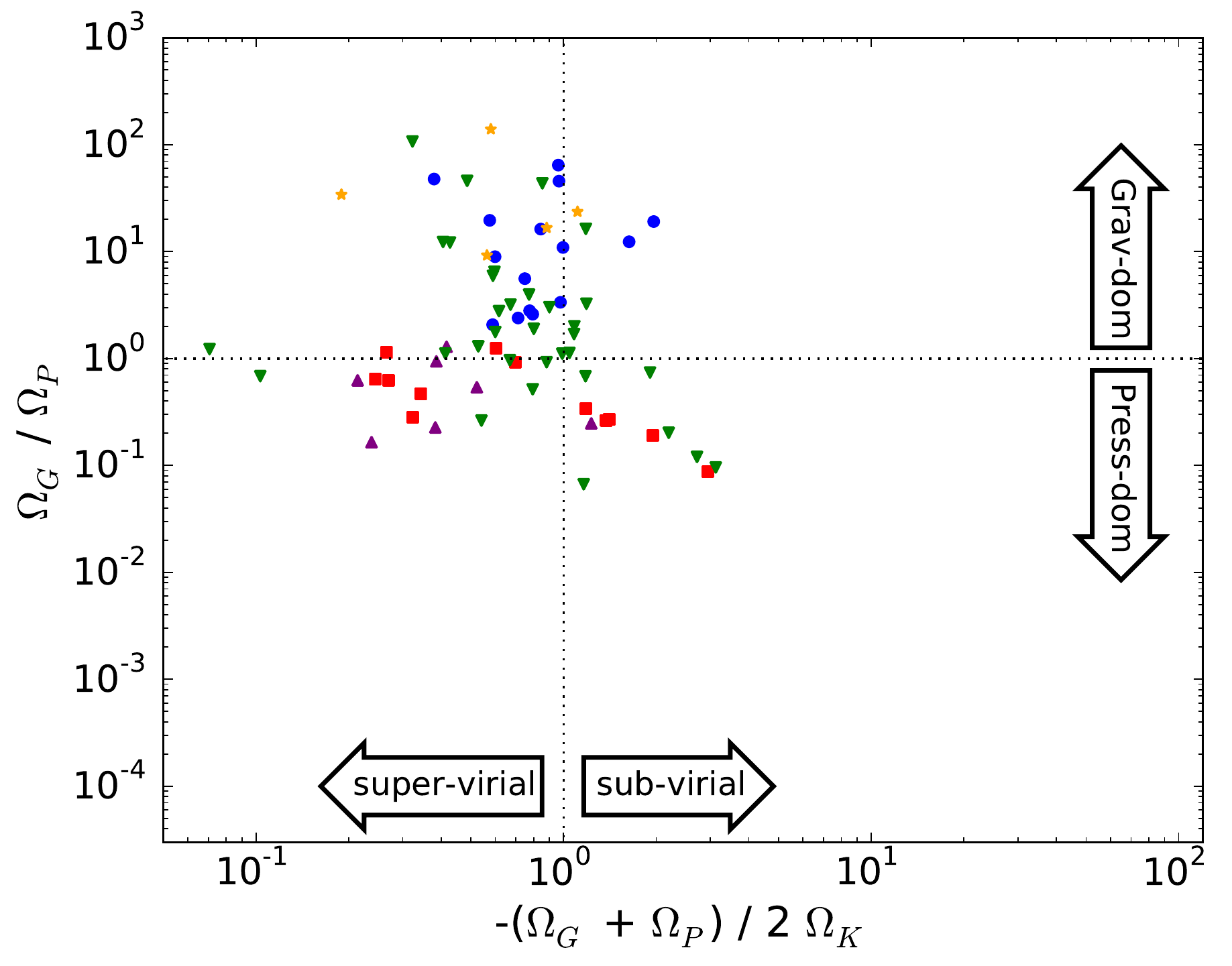}
\plotone{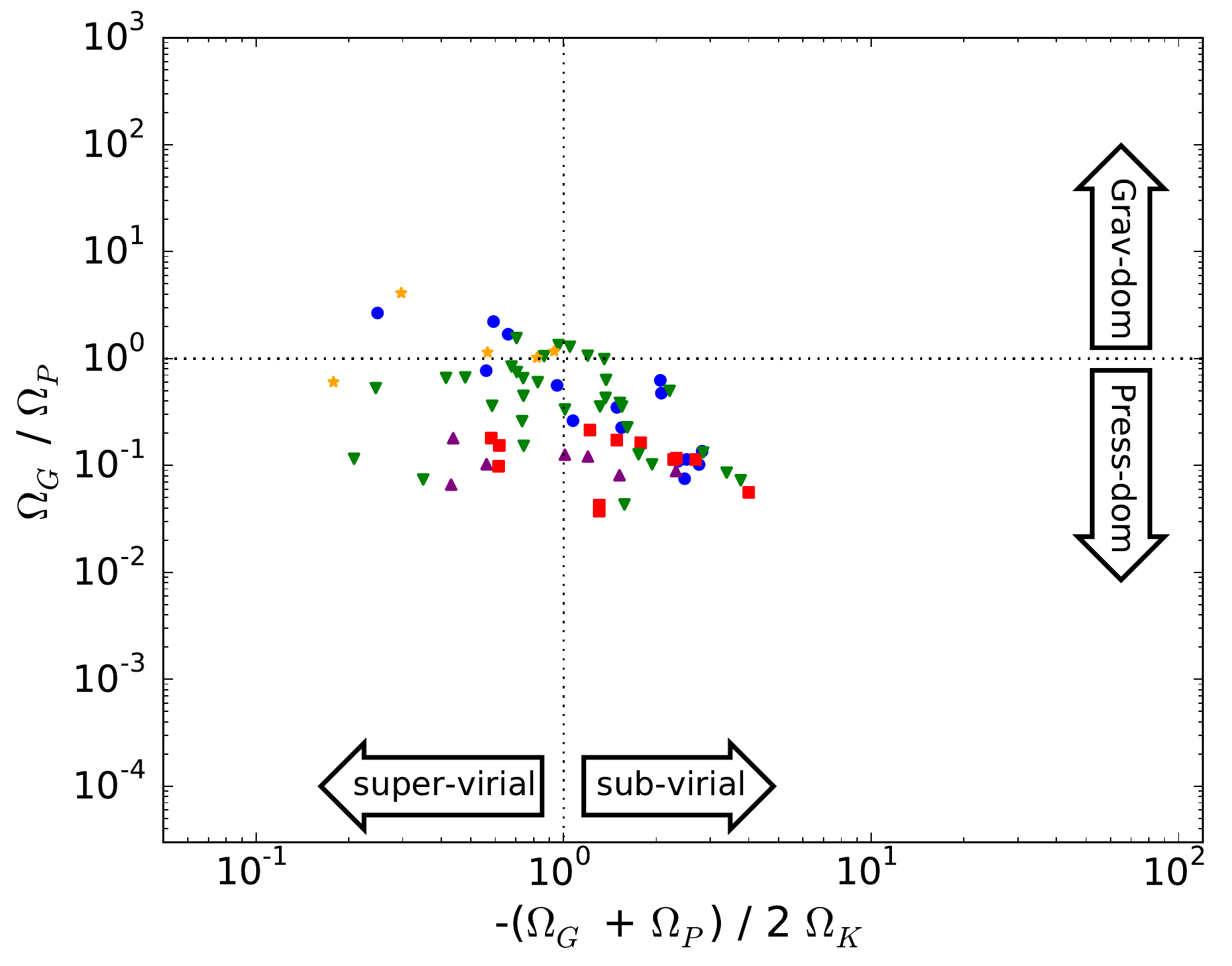}
\caption{Top: Virial plane for the same structures as in Figure \ref{virial_plane} when using $R_{eff} = (A/\pi)^{1/2}$ as the effective radius (see Appendix A for details). Middle: Virial plane when using the H$_2$ column density map to determine the masses of the \textit{Herschel}-identified dense cores (see Appendix B for details).  Bottom: Virial plane when using both $R_{eff} = (A/\pi)^{1/2}$ and the H$_2$ column density map to determine structure masses.  In all panels, a Gaussian density profile is assumed for the structures, which is consistent with the virial analysis presented in \cite{Kirk_submitted}.}
\label{virial_plane2}
\end{figure}

\begin{figure}[ht]
\epsscale{0.6}
\plotone{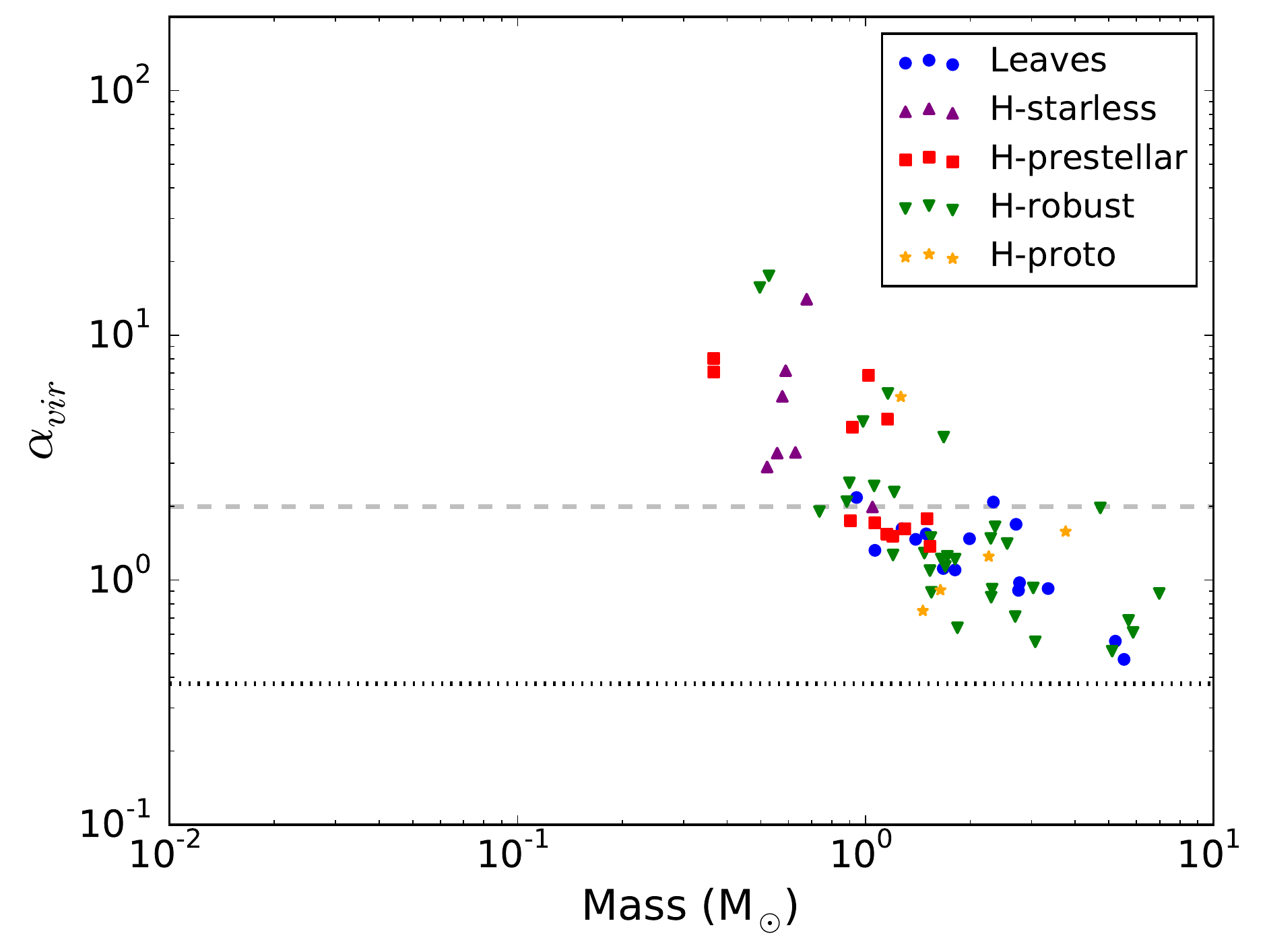}
\plotone{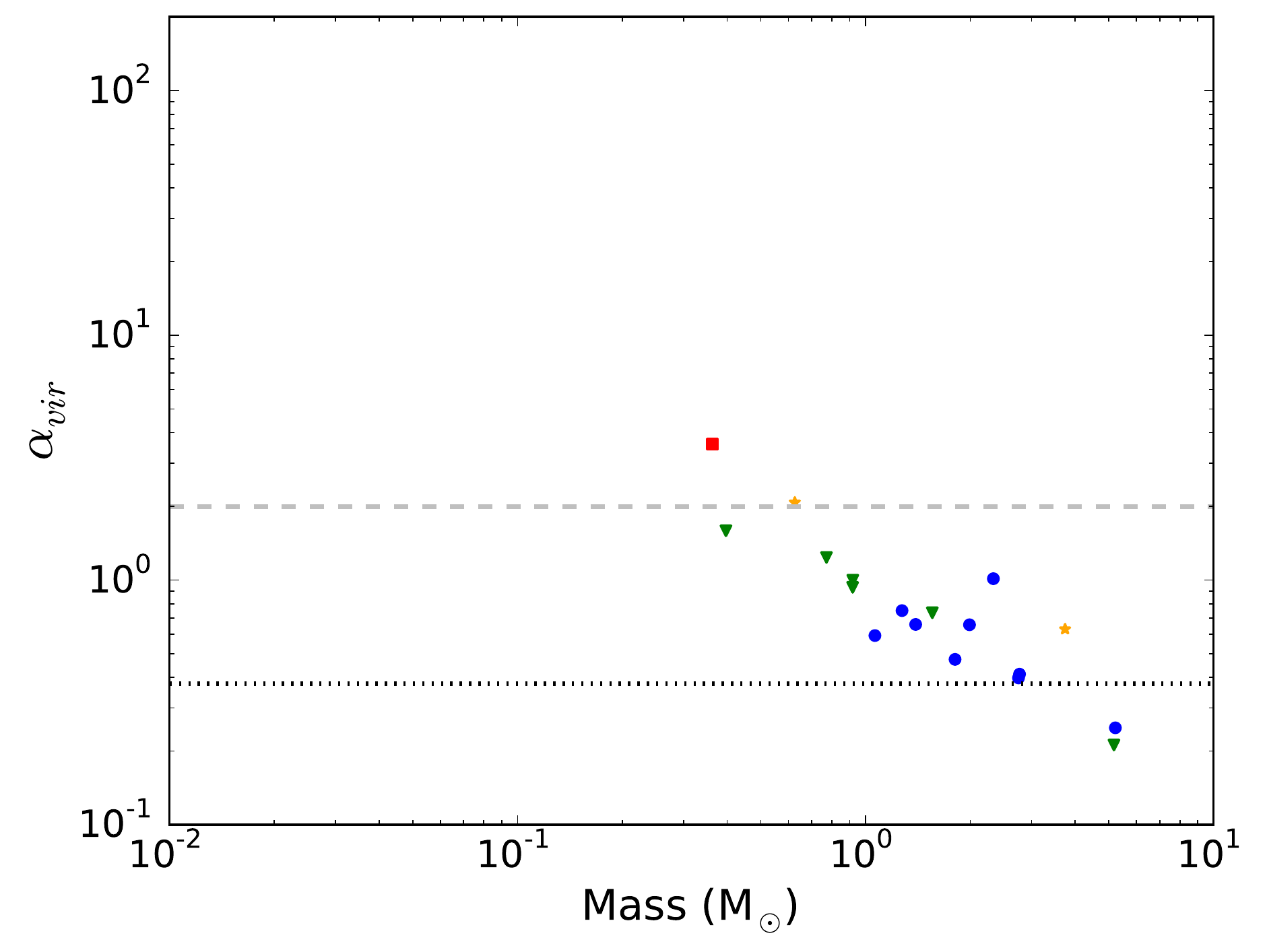}
\caption{Top panel: Same as Figure \ref{Mass_vs_virial}, but using an alternative method for obtaining the observed mass for the \textit{Herschel}-identified cores.  Namely, the observed mass for each \textit{Herschel}-identified core was obtained by summing all pixels in the \textit{Herschel}-derived H$_2$ column density map that fall within its elliptical mask.  Bottom panel: Where ammonia-identified leaves have a \textit{Herschel}-identified counterpart, the mass of the \textit{Herschel}-identified core is used to calculate the virial parameter of the ammonia-identified structure.  The blue points represent the original virial parameter calculations (see Section 3.4) for the leaves which have \textit{Herschel}-identified core counterparts. The non-blue points represent the virial parameters for the leaves when using their \textit{Herschel}-identified core counterpart's SED-derived mass.  The color and shape of the non-blue points represent the core type for the cross-matched \textit{Herschel}-identified source.}
\label{Mass_new4}
\end{figure}

To understand the impact large-scale filtering has upon the virial parameters derived in this paper, we re-derive masses for the \textit{Herschel}-identified dense cores by summing all the H$_2$ column density within the pixels that fall inside their respective elliptical 2D masks.  This method is consistent with the method described in Section 3.3, which we used to estimate the observed masses for the top-level ammonia-identified structures.  Figure \ref{Mass_new4} re-plots Figure \ref{Mass_vs_virial} using the re-derived masses for the \textit{Herschel}-identified dense cores.  The higher observed masses for the \textit{Herschel}-identified cores result in lower virial parameters and a higher fraction of the structures falling below $\alpha_{vir} = 2$.  Relatedly, the middle panel of Figure \ref{virial_plane2} shows that many of the cores move into the super-virial zone of the virial plane after altering the method by which their mass is calculated.  When combining both the altered mass calculation for the \textit{Herschel}-identified cores with the altered $R_{eff}$ calculation discussed in Appendix A, however, the structures remain split between the sub-virial and super-virial portions of the virial plane (as can be seen in the bottom panel of Figure \ref{virial_plane2}).  


An additional caveat arises when comparing the virial parameters derived for the \textit{Herschel}-identified dense cores and ammonia-identified structures when considering the former were identified in 2D while the latter were identified in 3D.  To circumvent any issues related to these projection effects and provide a direct comparison between the two types of structures, we first perform a cross-match to find the \textit{Herschel}-identified counterpart to each ammonia-identified leaf.  If the center of a \textit{Herschel}-identified dense core is within 32$\arcsec$ of an ammonia-identified leaf (i.e., the GBT beam size for our NH$_3$ (1,1) observations), the two structures are considered matched.  Nine of the 15 ammonia-identified leaves with aspect ratios less than two were found to have a \textit{Herschel}-identified counterpart using this criterion. For matched pairs, we re-calculate the virial parameter of the ammonia-identified structure using the SED-derived mass determined for its \textit{Herschel}-identified counterpart.  

The bottom panel of Figure \ref{Mass_new4} shows the originally calculated virial parameters for the ammonia-identified leaves (blue points) that contained a \textit{Herschel}-identified counterpart, and the re-calculated virial parameters using the \textit{Herschel} mass, which are shown in the color corresponding to the core type of the \textit{Herschel}-identified counterpart.  Although the virial parameters calculated using the (lower) \textit{Herschel} masses tend to be higher, nearly all remain below $\alpha_{vir}$ = 2.  This result shows that, at least for the ammonia-identified structures, filtering out large-scale structure prior to estimating masses does not significantly alter the conclusions of our virial analysis. 

\end{appendix}


\bibliographystyle{apj}
\bibliography{Cepheus_GAS_stability}

\begin{thebibliography}{}
\expandafter\ifx\csname natexlab\endcsname\relax\def\natexlab#1{#1}\fi

\bibitem[{{Andr{\'e}} {et~al.}(2010){Andr{\'e}}, {Men'shchikov}, {Bontemps},
  {K{\"o}nyves}, {Motte}, {Schneider}, {Didelon}, {Minier}, {Saraceno},
  {Ward-Thompson}, {di Francesco}, {White}, {Molinari}, {Testi}, {Abergel},
  {Griffin}, {Henning}, {Royer}, {Mer{\'{\i}}n}, {Vavrek}, {Attard},
  {Arzoumanian}, {Wilson}, {Ade}, {Aussel}, {Baluteau}, {Benedettini},
  {Bernard}, {Blommaert}, {Cambr{\'e}sy}, {Cox}, {di Giorgio}, {Hargrave},
  {Hennemann}, {Huang}, {Kirk}, {Krause}, {Launhardt}, {Leeks}, {Le Pennec},
  {Li}, {Martin}, {Maury}, {Olofsson}, {Omont}, {Peretto}, {Pezzuto}, {Prusti},
  {Roussel}, {Russeil}, {Sauvage}, {Sibthorpe}, {Sicilia-Aguilar}, {Spinoglio},
  {Waelkens}, {Woodcraft}, \& {Zavagno}}]{Andre_2010}
{Andr{\'e}}, P., {Men'shchikov}, A., {Bontemps}, S., {et~al.} 2010, \aap, 518,
  L102

\bibitem[{{Astropy Collaboration} {et~al.}(2013){Astropy Collaboration},
  {Robitaille}, {Tollerud}, {Greenfield}, {Droettboom}, {Bray}, {Aldcroft},
  {Davis}, {Ginsburg}, {Price-Whelan}, {Kerzendorf}, {Conley}, {Crighton},
  {Barbary}, {Muna}, {Ferguson}, {Grollier}, {Parikh}, {Nair}, {Unther},
  {Deil}, {Woillez}, {Conseil}, {Kramer}, {Turner}, {Singer}, {Fox}, {Weaver},
  {Zabalza}, {Edwards}, {Azalee Bostroem}, {Burke}, {Casey}, {Crawford},
  {Dencheva}, {Ely}, {Jenness}, {Labrie}, {Lim}, {Pierfederici}, {Pontzen},
  {Ptak}, {Refsdal}, {Servillat}, \& {Streicher}}]{Astropy_2013}
{Astropy Collaboration}, {Robitaille}, T.~P., {Tollerud}, E.~J., {et~al.} 2013,
  \aap, 558, A33

\bibitem[{{Bal{\'a}zs} {et~al.}(2004){Bal{\'a}zs}, {{\'A}brah{\'a}m}, {Kun},
  {Kelemen}, \& {T{\'o}th}}]{Balazs_2004}
{Bal{\'a}zs}, L.~G., {{\'A}brah{\'a}m}, P., {Kun}, M., {Kelemen}, J., \&
  {T{\'o}th}, L.~V. 2004, \aap, 425, 133

\bibitem[{{Battersby} {et~al.}(2014){Battersby}, {Bally}, {Dunham}, {Ginsburg},
  {Longmore}, \& {Darling}}]{Battersby_2014}
{Battersby}, C., {Bally}, J., {Dunham}, M., {et~al.} 2014, \apj, 786, 116

\bibitem[{{Benson} \& {Myers}(1983)}]{Benson_1983}
{Benson}, P.~J., \& {Myers}, P.~C. 1983, \apj, 270, 589

\bibitem[{{Bertoldi} \& {McKee}(1992)}]{Bertoldi_1992}
{Bertoldi}, F., \& {McKee}, C.~F. 1992, \apj, 395, 140

\bibitem[{{Bonnor}(1956)}]{Bonnor_1956}
{Bonnor}, W.~B. 1956, \mnras, 116, 351

\bibitem[{{Caselli} {et~al.}(1999){Caselli}, {Walmsley}, {Tafalla}, {Dore}, \&
  {Myers}}]{Caselli_1999}
{Caselli}, P., {Walmsley}, C.~M., {Tafalla}, M., {Dore}, L., \& {Myers}, P.~C.
  1999, \apjl, 523, L165

\bibitem[{{Crapsi} {et~al.}(2007){Crapsi}, {Caselli}, {Walmsley}, \&
  {Tafalla}}]{Crapsi_2007}
{Crapsi}, A., {Caselli}, P., {Walmsley}, M.~C., \& {Tafalla}, M. 2007, \aap,
  470, 221

\bibitem[{{Di Francesco} {et~al.}(2007){Di Francesco}, {Evans}, {Caselli},
  {Myers}, {Shirley}, {Aikawa}, \& {Tafalla}}]{DiFrancesco_2007}
{Di Francesco}, J., {Evans}, II, N.~J., {Caselli}, P., {et~al.} 2007,
  Protostars and Planets V, 17

\bibitem[{{Di Francesco} {et~al.}(2002){Di Francesco}, {Hogerheijde}, {Welch},
  \& {Bergin}}]{DiFrancesco_2002}
{Di Francesco}, J., {Hogerheijde}, M.~R., {Welch}, W.~J., \& {Bergin}, E.~A.
  2002, \aj, 124, 2749

\bibitem[{{Di Francesco} {et~al.}(2017, in prep){Di Francesco}, {Keown},
  {Ladjelate}, {Fallscheer}, {Andre}, \& {Vera}}]{DiFrancesco_prep}
{Di Francesco}, J., {Keown}, J., {Ladjelate}, B., {et~al.} 2017, in prep, \apj

\bibitem[{Dunham {et~al.}(2015)Dunham, Allen, II, Broekhoven-Fiene, Cieza,
  Francesco, Gutermuth, Harvey, Hatchell, Heiderman, Huard, Johnstone, Kirk,
  Matthews, Miller, Peterson, \& Young}]{Dunham_2015}
Dunham, M.~M., Allen, L.~E., II, N. J.~E., {et~al.} 2015, The Astrophysical
  Journal Supplement Series, 220, 11

\bibitem[{{Dzib} {et~al.}(2011){Dzib}, {Loinard}, {Mioduszewski}, {Boden},
  {Rodr{\'{\i}}guez}, \& {Torres}}]{Dzib_2011}
{Dzib}, S., {Loinard}, L., {Mioduszewski}, A.~J., {et~al.} 2011, in Revista
  Mexicana de Astronomia y Astrofisica Conference Series, Vol.~40, Revista
  Mexicana de Astronomia y Astrofisica Conference Series, 231--232

\bibitem[{{Ebert}(1955)}]{Ebert_1955}
{Ebert}, R. 1955, \zap, 37, 217

\bibitem[{{Fiege} \& {Pudritz}(2000)}]{Fiege_2000}
{Fiege}, J.~D., \& {Pudritz}, R.~E. 2000, \mnras, 311, 105

\bibitem[{{Friesen} {et~al.}(2016){Friesen}, {Bourke}, {Di Francesco},
  {Gutermuth}, \& {Myers}}]{Friesen_2016}
{Friesen}, R.~K., {Bourke}, T.~L., {Di Francesco}, J., {Gutermuth}, R., \&
  {Myers}, P.~C. 2016, \apj, 833, 204

\bibitem[{{Friesen} {et~al.}(2009){Friesen}, {Di Francesco}, {Shirley}, \&
  {Myers}}]{Friesen_2009}
{Friesen}, R.~K., {Di Francesco}, J., {Shirley}, Y.~L., \& {Myers}, P.~C. 2009,
  \apj, 697, 1457

\bibitem[{{Friesen} {et~al.}(2013){Friesen}, {Medeiros}, {Schnee}, {Bourke},
  {Francesco}, {Gutermuth}, \& {Myers}}]{Friesen_2013}
{Friesen}, R.~K., {Medeiros}, L., {Schnee}, S., {et~al.} 2013, \mnras, 436,
  1513

\bibitem[{{Friesen} {et~al.}(2017){Friesen}, {Pineda}, {co-PIs}, {Rosolowsky},
  {Alves}, {Chac{\'o}n-Tanarro}, {How-Huan Chen}, {Chun-Yuan Chen}, {Di
  Francesco}, {Keown}, {Kirk}, {Punanova}, {Seo}, {Shirley}, {Ginsburg},
  {Hall}, {Offner}, {Singh}, {Arce}, {Caselli}, {Goodman}, {Martin}, {Matzner},
  {Myers}, {Redaelli}, \& {The GAS Collaboration}}]{Friesen_submitted}
{Friesen}, R.~K., {Pineda}, J.~E., {co-PIs}, {et~al.} 2017, \apj, 843, 63

\bibitem[{{Goldsmith}(2001)}]{Goldsmith_2001}
{Goldsmith}, P.~F. 2001, \apj, 557, 736

\bibitem[{{Ho} \& {Townes}(1983)}]{Ho_1983}
{Ho}, P.~T.~P., \& {Townes}, C.~H. 1983, \araa, 21, 239

\bibitem[{{Hotzel} {et~al.}(2001){Hotzel}, {Harju}, {Lemke}, {Mattila}, \&
  {Walmsley}}]{Hotzel_2001}
{Hotzel}, S., {Harju}, J., {Lemke}, D., {Mattila}, K., \& {Walmsley}, C.~M.
  2001, \aap, 372, 302

\bibitem[{{Kauffmann} {et~al.}(2008){Kauffmann}, {Bertoldi}, {Bourke}, {Evans},
  \& {Lee}}]{Kauffmann_2008}
{Kauffmann}, J., {Bertoldi}, F., {Bourke}, T.~L., {Evans}, II, N.~J., \& {Lee},
  C.~W. 2008, \aap, 487, 993

\bibitem[{Kauffmann {et~al.}(2013)Kauffmann, Pillai, \&
  Goldsmith}]{Kauffmann_2013}
Kauffmann, J., Pillai, T., \& Goldsmith, P.~F. 2013, The Astrophysical Journal,
  779, 185

\bibitem[{{Kim} {et~al.}(2015){Kim}, {Lee}, {Choi}, {Bourke}, {Evans}, {Di
  Francesco}, {Cieza}, {Dunham}, \& {Kang}}]{Kim_2015}
{Kim}, J., {Lee}, J.-E., {Choi}, M., {et~al.} 2015, \apjs, 218, 5

\bibitem[{{Kirk} {et~al.}(2006){Kirk}, {Johnstone}, \& {Di
  Francesco}}]{Kirk_2006}
{Kirk}, H., {Johnstone}, D., \& {Di Francesco}, J. 2006, \apj, 646, 1009

\bibitem[{{Kirk} {et~al.}(2013){Kirk}, {Myers}, {Bourke}, {Gutermuth},
  {Hedden}, \& {Wilson}}]{Kirk_2013}
{Kirk}, H., {Myers}, P.~C., {Bourke}, T.~L., {et~al.} 2013, \apj, 766, 115

\bibitem[{{Kirk} {et~al.}(2017){Kirk}, {Friesen}, {Pineda}, {Rosolowsky},
  {Offner}, {Matzner}, {Myers}, {Di Francesco}, {Caselli}, {Alves},
  {Chac{\'o}n-Tanarro}, {Chen}, {Chun-Yuan Chen}, {Keown}, {Punanova}, {Seo},
  {Shirley}, {Ginsburg}, {Hall}, {Singh}, {Arce}, {Goodman}, {Martin}, \&
  {Redaelli}}]{Kirk_submitted}
{Kirk}, H., {Friesen}, R.~K., {Pineda}, J.~E., {et~al.} 2017, \apj, 846, 144

\bibitem[{{K{\"o}nyves} {et~al.}(2010){K{\"o}nyves}, {Andr{\'e}},
  {Men'shchikov}, {Schneider}, {Arzoumanian}, {Bontemps}, {Attard}, {Motte},
  {Didelon}, {Maury}, {Abergel}, {Ali}, {Baluteau}, {Bernard}, {Cambr{\'e}sy},
  {Cox}, {di Francesco}, {di Giorgio}, {Griffin}, {Hargrave}, {Huang}, {Kirk},
  {Li}, {Martin}, {Minier}, {Molinari}, {Olofsson}, {Pezzuto}, {Russeil},
  {Roussel}, {Saraceno}, {Sauvage}, {Sibthorpe}, {Spinoglio}, {Testi},
  {Ward-Thompson}, {White}, {Wilson}, {Woodcraft}, \& {Zavagno}}]{Konyves_2010}
{K{\"o}nyves}, V., {Andr{\'e}}, P., {Men'shchikov}, A., {et~al.} 2010, \aap,
  518, L106

\bibitem[{{K{\"o}nyves} {et~al.}(2015){K{\"o}nyves}, {Andr{\'e}},
  {Men'shchikov}, {Palmeirim}, {Arzoumanian}, {Schneider}, {Roy}, {Didelon},
  {Maury}, {Shimajiri}, {Di Francesco}, {Bontemps}, {Peretto}, {Benedettini},
  {Bernard}, {Elia}, {Griffin}, {Hill}, {Kirk}, {Ladjelate}, {Marsh}, {Martin},
  {Motte}, {Nguy{\^e}n Luong}, {Pezzuto}, {Roussel}, {Rygl}, {Sadavoy},
  {Schisano}, {Spinoglio}, {Ward-Thompson}, \& {White}}]{Konyves_2015}
---. 2015, \aap, 584, A91

\bibitem[{{Kun}(1998)}]{Kun_1998}
{Kun}, M. 1998, \apjs, 115, 59

\bibitem[{{Kun} {et~al.}(2009){Kun}, {Balog}, {Kenyon}, {Mamajek}, \&
  {Gutermuth}}]{Kun_2009}
{Kun}, M., {Balog}, Z., {Kenyon}, S.~J., {Mamajek}, E.~E., \& {Gutermuth},
  R.~A. 2009, \apjs, 185, 451

\bibitem[{{Kun} \& {Prusti}(1993)}]{Kun_1993}
{Kun}, M., \& {Prusti}, T. 1993, \aap, 272, 235

\bibitem[{{Larson}(1981)}]{Larson_1981}
{Larson}, R.~B. 1981, \mnras, 194, 809

\bibitem[{{Lee} {et~al.}(2007){Lee}, {Di Francesco}, {Bourke}, {Evans}, \&
  {Wu}}]{Lee_2007}
{Lee}, J.-E., {Di Francesco}, J., {Bourke}, T.~L., {Evans}, II, N.~J., \& {Wu},
  J. 2007, \apj, 671, 1748

\bibitem[{{Lee} {et~al.}(2010){Lee}, {Lee}, {Shinn}, {Dunham}, {Kim}, {Kim},
  {Bourke}, {Evans}, \& {Choi}}]{Lee_2010}
{Lee}, J.-E., {Lee}, H.-G., {Shinn}, J.-H., {et~al.} 2010, \apjl, 709, L74

\bibitem[{{Levshakov} {et~al.}(2013){Levshakov}, {Henkel}, {Reimers}, {Wang},
  {Mao}, {Wang}, \& {Xu}}]{Levshakov_2013}
{Levshakov}, S.~A., {Henkel}, C., {Reimers}, D., {et~al.} 2013, \aap, 553, A58

\bibitem[{{Marsh} {et~al.}(2016){Marsh}, {Kirk}, {Andr{\'e}}, {Griffin},
  {K{\"o}nyves}, {Palmeirim}, {Men'shchikov}, {Ward-Thompson}, {Benedettini},
  {Bresnahan}, {Francesco}, {Elia}, {Motte}, {Peretto}, {Pezzuto}, {Roy},
  {Sadavoy}, {Schneider}, {Spinoglio}, \& {White}}]{Marsh_2016}
{Marsh}, K.~A., {Kirk}, J.~M., {Andr{\'e}}, P., {et~al.} 2016, \mnras, 459, 342

\bibitem[{{McKee}(1989)}]{Mckee_1989}
{McKee}, C.~F. 1989, \apj, 345, 782

\bibitem[{{Men'shchikov} {et~al.}(2012){Men'shchikov}, {Andr{\'e}}, {Didelon},
  {Motte}, {Hennemann}, \& {Schneider}}]{Menshchikov_2012}
{Men'shchikov}, A., {Andr{\'e}}, P., {Didelon}, P., {et~al.} 2012, \aap, 542,
  A81

\bibitem[{{Men'shchikov} {et~al.}(2010){Men'shchikov}, {Andr{\'e}}, {Didelon},
  {K{\"o}nyves}, {Schneider}, {Motte}, {Bontemps}, {Arzoumanian}, {Attard},
  {Abergel}, {Baluteau}, {Bernard}, {Cambr{\'e}sy}, {Cox}, {di Francesco}, {di
  Giorgio}, {Griffin}, {Hargrave}, {Huang}, {Kirk}, {Li}, {Martin}, {Minier},
  {Miville-Desch{\^e}nes}, {Molinari}, {Olofsson}, {Pezzuto}, {Roussel},
  {Russeil}, {Saraceno}, {Sauvage}, {Sibthorpe}, {Spinoglio}, {Testi},
  {Ward-Thompson}, {White}, {Wilson}, {Woodcraft}, \&
  {Zavagno}}]{Menshchikov_2010}
---. 2010, \aap, 518, L103

\bibitem[{{Miville-Desch{\^e}nes} {et~al.}(2010){Miville-Desch{\^e}nes},
  {Martin}, {Abergel}, {Bernard}, {Boulanger}, {Lagache}, {Anderson},
  {Andr{\'e}}, {Arab}, {Baluteau}, {Blagrave}, {Bontemps}, {Cohen},
  {Compiegne}, {Cox}, {Dartois}, {Davis}, {Emery}, {Fulton}, {Gry}, {Habart},
  {Huang}, {Joblin}, {Jones}, {Kirk}, {Lim}, {Madden}, {Makiwa}, {Menshchikov},
  {Molinari}, {Moseley}, {Motte}, {Naylor}, {Okumura}, {Pinheiro Gon{\c
  c}alves}, {Polehampton}, {Rod{\'o}n}, {Russeil}, {Saraceno}, {Schneider},
  {Sidher}, {Spencer}, {Swinyard}, {Ward-Thompson}, {White}, \&
  {Zavagno}}]{Miville_2010}
{Miville-Desch{\^e}nes}, M.-A., {Martin}, P.~G., {Abergel}, A., {et~al.} 2010,
  \aap, 518, L104

\bibitem[{{Ohashi} {et~al.}(2016){Ohashi}, {Tatematsu}, {Sanhueza}, {Hirota},
  {Choi}, \& {Mizuno}}]{Ohashi_2016}
{Ohashi}, S., {Tatematsu}, K., {Sanhueza}, P., {et~al.} 2016, \mnras, 459, 4130

\bibitem[{{Ortiz-Le{\'o}n} {et~al.}(2017){Ortiz-Le{\'o}n}, {Dzib}, {Kounkel},
  {Loinard}, {Mioduszewski}, {Rodr{\'{\i}}guez}, {Torres}, {Pech}, {Rivera},
  {Hartmann}, {Boden}, {Evans}, {Brice{\~n}o}, {Tobin}, \&
  {Galli}}]{Ortiz_2017}
{Ortiz-Le{\'o}n}, G.~N., {Dzib}, S.~A., {Kounkel}, M.~A., {et~al.} 2017, \apj,
  834, 143

\bibitem[{{Palmeirim} {et~al.}(2013){Palmeirim}, {Andr{\'e}}, {Kirk},
  {Ward-Thompson}, {Arzoumanian}, {K{\"o}nyves}, {Didelon}, {Schneider},
  {Benedettini}, {Bontemps}, {Di Francesco}, {Elia}, {Griffin}, {Hennemann},
  {Hill}, {Martin}, {Men'shchikov}, {Molinari}, {Motte}, {Nguyen Luong},
  {Nutter}, {Peretto}, {Pezzuto}, {Roy}, {Rygl}, {Spinoglio}, \&
  {White}}]{Palmeirim_2013}
{Palmeirim}, P., {Andr{\'e}}, P., {Kirk}, J., {et~al.} 2013, \aap, 550, A38

\bibitem[{{Pattle}(2016)}]{Pattle_2016}
{Pattle}, K. 2016, \mnras, 459, 2651

\bibitem[{{Pattle} {et~al.}(2015){Pattle}, {Ward-Thompson}, {Kirk}, {White},
  {Drabek-Maunder}, {Buckle}, {Beaulieu}, {Berry}, {Broekhoven-Fiene},
  {Currie}, {Fich}, {Hatchell}, {Kirk}, {Jenness}, {Johnstone}, {Mottram},
  {Nutter}, {Pineda}, {Quinn}, {Salji}, {Tisi}, {Walker-Smith}, {Francesco},
  {Hogerheijde}, {Andr{\'e}}, {Bastien}, {Bresnahan}, {Butner}, {Chen},
  {Chrysostomou}, {Coude}, {Davis}, {Duarte-Cabral}, {Fiege}, {Friberg},
  {Friesen}, {Fuller}, {Graves}, {Greaves}, {Gregson}, {Griffin}, {Holland},
  {Joncas}, {Knee}, {K{\"o}nyves}, {Mairs}, {Marsh}, {Matthews},
  {Moriarty-Schieven}, {Rawlings}, {Richer}, {Robertson}, {Rosolowsky},
  {Rumble}, {Sadavoy}, {Spinoglio}, {Thomas}, {Tothill}, {Viti}, {Wouterloot},
  {Yates}, \& {Zhu}}]{Pattle_2015}
{Pattle}, K., {Ward-Thompson}, D., {Kirk}, J.~M., {et~al.} 2015, \mnras, 450,
  1094

\bibitem[{{Pattle} {et~al.}(2017){Pattle}, {Ward-Thompson}, {Kirk}, {Di
  Francesco}, {Kirk}, {Mottram}, {Keown}, {Buckle}, {Beaulieu}, {Berry},
  {Broekhoven-Fiene}, {Currie}, {Fich}, {Hatchell}, {Jenness}, {Johnstone},
  {Nutter}, {Pineda}, {Quinn}, {Salji}, {Tisi}, {Walker-Smith}, {Hogerheijde},
  {Bastien}, {Bresnahan}, {Butner}, {Chen}, {Chrysostomou}, {Coud{\'e}},
  {Davis}, {Drabek-Maunder}, {Duarte-Cabral}, {Fiege}, {Friberg}, {Friesen},
  {Fuller}, {Graves}, {Greaves}, {Gregson}, {Holland}, {Joncas}, {Knee},
  {Mairs}, {Marsh}, {Matthews}, {Moriarty-Schieven}, {Mowat}, {Rawlings},
  {Richer}, {Robertson}, {Rosolowsky}, {Rumble}, {Sadavoy}, {Thomas},
  {Tothill}, {Viti}, {White}, {Wouterloot}, {Yates}, \& {Zhu}}]{Pattle_2017}
---. 2017, \mnras, 464, 4255

\bibitem[{{Pineda} {et~al.}(2010){Pineda}, {Goodman}, {Arce}, {Caselli},
  {Foster}, {Myers}, \& {Rosolowsky}}]{Pineda_2010}
{Pineda}, J.~E., {Goodman}, A.~A., {Arce}, H.~G., {et~al.} 2010, \apjl, 712,
  L116

\bibitem[{{Ragan} {et~al.}(2011){Ragan}, {Bergin}, \& {Wilner}}]{Ragan_2011}
{Ragan}, S.~E., {Bergin}, E.~A., \& {Wilner}, D. 2011, \apj, 736, 163

\bibitem[{{Rosolowsky} {et~al.}(2008){Rosolowsky}, {Pineda}, {Kauffmann}, \&
  {Goodman}}]{Rosolowsky_2008}
{Rosolowsky}, E.~W., {Pineda}, J.~E., {Kauffmann}, J., \& {Goodman}, A.~A.
  2008, \apj, 679, 1338

\bibitem[{{Sato} {et~al.}(1994){Sato}, {Mizuno}, {Nagahama}, {Onishi},
  {Yonekura}, \& {Fukui}}]{Sato_1994}
{Sato}, F., {Mizuno}, A., {Nagahama}, T., {et~al.} 1994, \apj, 435, 279

\bibitem[{{Seo} \& {Youdin}(2016)}]{Seo_2016}
{Seo}, Y.~M., \& {Youdin}, A.~N. 2016, \mnras, 461, 1088

\bibitem[{{Seo} {et~al.}(2015){Seo}, {Shirley}, {Goldsmith}, {Ward-Thompson},
  {Kirk}, {Schmalzl}, {Lee}, {Friesen}, {Langston}, {Masters}, \&
  {Garwood}}]{Seo_2015}
{Seo}, Y.~M., {Shirley}, Y.~L., {Goldsmith}, P., {et~al.} 2015, \apj, 805, 185

\bibitem[{{Shirley}(2015)}]{Shirley_2015}
{Shirley}, Y.~L. 2015, \pasp, 127, 299

\bibitem[{{Spezzano} {et~al.}(2016){Spezzano}, {Bizzocchi}, {Caselli}, {Harju},
  \& {Br{\"u}nken}}]{Spezzano_2016}
{Spezzano}, S., {Bizzocchi}, L., {Caselli}, P., {Harju}, J., \& {Br{\"u}nken},
  S. 2016, \aap, 592, L11

\bibitem[{{Strai{\v z}ys} {et~al.}(2003){Strai{\v z}ys}, {{\v C}ernis}, \&
  {Barta{\v s}i{\= u}t{\.e}}}]{S_2003}
{Strai{\v z}ys}, V., {{\v C}ernis}, K., \& {Barta{\v s}i{\= u}t{\.e}}, S. 2003,
  \aap, 405, 585

\bibitem[{{Suzuki} {et~al.}(1992){Suzuki}, {Yamamoto}, {Ohishi}, {Kaifu},
  {Ishikawa}, {Hirahara}, \& {Takano}}]{Suzuki_1992}
{Suzuki}, H., {Yamamoto}, S., {Ohishi}, M., {et~al.} 1992, \apj, 392, 551

\bibitem[{{Swift} {et~al.}(2005){Swift}, {Welch}, \& {Di
  Francesco}}]{Swift_2005}
{Swift}, J.~J., {Welch}, W.~J., \& {Di Francesco}, J. 2005, \apj, 620, 823

\bibitem[{{Tafalla} {et~al.}(2006){Tafalla}, {Santiago-Garc{\'{\i}}a}, {Myers},
  {Caselli}, {Walmsley}, \& {Crapsi}}]{Tafalla_2006}
{Tafalla}, M., {Santiago-Garc{\'{\i}}a}, J., {Myers}, P.~C., {et~al.} 2006,
  \aap, 455, 577

\bibitem[{{Ward-Thompson} {et~al.}(2007){Ward-Thompson}, {Di Francesco},
  {Hatchell}, {Hogerheijde}, {Nutter}, {Bastien}, {Basu}, {Bonnell}, {Bowey},
  {Brunt}, {Buckle}, {Butner}, {Cavanagh}, {Chrysostomou}, {Curtis}, {Davis},
  {Dent}, {van Dishoeck}, {Edmunds}, {Fich}, {Fiege}, {Fissel}, {Friberg},
  {Friesen}, {Frieswijk}, {Fuller}, {Gosling}, {Graves}, {Greaves}, {Helmich},
  {Hills}, {Holland}, {Houde}, {Jayawardhana}, {Johnstone}, {Joncas}, {Kirk},
  {Kirk}, {Knee}, {Matthews}, {Matthews}, {Matzner}, {Moriarty-Schieven},
  {Naylor}, {Padman}, {Plume}, {Rawlings}, {Redman}, {Reid}, {Richer},
  {Shipman}, {Simpson}, {Spaans}, {Stamatellos}, {Tsamis}, {Viti}, {Weferling},
  {White}, {Whitworth}, {Wouterloot}, {Yates}, \& {Zhu}}]{WT_2007}
{Ward-Thompson}, D., {Di Francesco}, J., {Hatchell}, J., {et~al.} 2007, \pasp,
  119, 855

\bibitem[{{Ward-Thompson} {et~al.}(2010){Ward-Thompson}, {Kirk}, {Andr{\'e}},
  {Saraceno}, {Didelon}, {K{\"o}nyves}, {Schneider}, {Abergel}, {Baluteau},
  {Bernard}, {Bontemps}, {Cambr{\'e}sy}, {Cox}, {di Francesco}, {di Giorgio},
  {Griffin}, {Hargrave}, {Huang}, {Li}, {Martin}, {Men'shchikov}, {Minier},
  {Molinari}, {Motte}, {Olofsson}, {Pezzuto}, {Russeil}, {Sauvage},
  {Sibthorpe}, {Spinoglio}, {Testi}, {White}, {Wilson}, {Woodcraft}, \&
  {Zavagno}}]{Ward-Thompson_2010}
{Ward-Thompson}, D., {Kirk}, J.~M., {Andr{\'e}}, P., {et~al.} 2010, \aap, 518,
  L92

\bibitem[{{Yonekura} {et~al.}(1997){Yonekura}, {Dobashi}, {Mizuno}, {Ogawa}, \&
  {Fukui}}]{Yonekura_1997}
{Yonekura}, Y., {Dobashi}, K., {Mizuno}, A., {Ogawa}, H., \& {Fukui}, Y. 1997,
  \apjs, 110, 21

\end{thebibliography}

\clearpage

\begin{deluxetable}{ccccccccccccccc}
\tabletypesize{\tiny}
\rotate
\tablewidth{0pt}
\tablecolumns{15}
\tablecaption{NH$_3$ (1,1) Leaves Catalog}
\tablehead{\colhead{ID} & \colhead{$x_{center}$\tablenotemark{a}} & \colhead{$y_{center}$\tablenotemark{a}} & \colhead{$\sigma_{major}$\tablenotemark{b}} & \colhead{$\sigma_{minor}$\tablenotemark{b}} & \colhead{PA\tablenotemark{c}} & \colhead{$R_{eff}$\tablenotemark{d}} & \colhead{$M_{obs}$\tablenotemark{e}}  & \colhead{$\sigma_{obs}$\tablenotemark{f}}  & \colhead{$T_{K}$\tablenotemark{g}} & \colhead{$T_{dust}$\tablenotemark{h}} & \colhead{$N_{para-NH_3}$\tablenotemark{i}} & \colhead{$N_{H_2}$\tablenotemark{j}} & \colhead{$\alpha_{vir}$\tablenotemark{k}} & \colhead{$\alpha_{vir, fil}$\tablenotemark{l}} \\
   & ($\deg$) & ($\deg$) & ($\arcsec$) & ($\arcsec$) & ($\deg$) & (10$^{-2}$ pc) & (M$_\odot$)  & (km s$^{-1}$)  & (K) & (K) & (10$^{13}$ cm$^{-2}$) & (10$^{21}$ cm$^{-2}$) & & }
\startdata
1 & 339.173424 & 75.139485 & 23.3 & 14.3 & 162 & 2.7  $\pm$ 0.1 & 2.0  $\pm$ 0.4 & 0.14  $\pm$ 0.06 & 11.0  $\pm$ 1.7 & 12.7  $\pm$ 0.2 & 6.1  $\pm$ 1.9 & 7.9  $\pm$ 1.3 & 0.7 & \nodata \\ 
2 & 339.403121 & 75.195569 & 16.4 & 13.3 & 157 & 2.1  $\pm$ 0.1 & 2.8  $\pm$ 0.6 & 0.16  $\pm$ 0.01 & 10.5  $\pm$ 0.6 & 11.5  $\pm$ 0.2 & 15.1  $\pm$ 3.8 & 15.2  $\pm$ 3.0 & 0.4 & \nodata \\ 
3 & 339.762110 & 75.196272 & 13.9 & 6.1 & 139 & 1.3  $\pm$ 0.3 & 2.0  $\pm$ 0.4 & 0.18  $\pm$ 0.02 & 11.3  $\pm$ 0.5 & 11.5  $\pm$ 0.7 & 24.3  $\pm$ 2.2 & 31.7  $\pm$ 5.0 & 0.4 & 0.7 \\ 
4 & 339.422677 & 75.162841 & 32.3 & 14.5 & 163 & 3.1  $\pm$ 0.1 & 3.1  $\pm$ 0.6 & 0.12  $\pm$ 0.01 & 10.3  $\pm$ 1.6 & 12.0  $\pm$ 0.2 & 5.8  $\pm$ 1.5 & 8.7  $\pm$ 1.1 & 0.4 & 0.8 \\ 
5 & 339.697981 & 75.187238 & 11.3 & 7.5 & 99 & 1.3  $\pm$ 0.3 & 2.3  $\pm$ 0.5 & 0.39  $\pm$ 0.05 & 13.1  $\pm$ 0.7 & 14.3  $\pm$ 0.6 & 36.9  $\pm$ 6.8 & 43.6  $\pm$ 27.9 & 1.0 & \nodata \\ 
6 & 339.822184 & 75.176348 & 12.4 & 6.9 & -136 & 1.3  $\pm$ 0.3 & 1.5  $\pm$ 0.3 & 0.22  $\pm$ 0.05 & 10.7  $\pm$ 0.5 & 11.0  $\pm$ 0.1 & 22.1  $\pm$ 2.8 & 23.3  $\pm$ 2.5 & 0.7 & \nodata \\ 
7 & 339.503914 & 75.194370 & 15.0 & 7.7 & -177 & 1.6  $\pm$ 0.2 & 1.1  $\pm$ 0.2 & 0.12  $\pm$ 0.01 & 9.4  $\pm$ 0.7 & 12.0  $\pm$ 0.1 & 8.2  $\pm$ 1.6 & 12.4  $\pm$ 1.1 & 0.6 & \nodata \\ 
8 & 337.418972 & 75.225552 & 23.2 & 14.6 & -171 & 2.7  $\pm$ 0.1 & 5.5  $\pm$ 1.1 & 0.13  $\pm$ 0.01 & 9.5  $\pm$ 0.6 & 11.1  $\pm$ 0.3 & 24.7  $\pm$ 9.6 & 19.9  $\pm$ 6.3 & 0.2 & \nodata \\ 
9 & 336.896899 & 75.201494 & 21.1 & 12.0 & 115 & 2.3  $\pm$ 0.1 & 1.7  $\pm$ 0.3 & 0.11  $\pm$ 0.01 & 8.9  $\pm$ 1.3 & 11.8  $\pm$ 0.1 & 4.7  $\pm$ 1.2 & 8.9  $\pm$ 1.1 & 0.5 & \nodata \\ 
10 & 337.004425 & 75.233212 & 52.7 & 15.9 & 150 & 4.2  $\pm$ 0.1 & 7.3  $\pm$ 1.5 & 0.11  $\pm$ 0.01 & 9.1  $\pm$ 0.9 & 11.5  $\pm$ 0.3 & 14.4  $\pm$ 5.8 & 10.2  $\pm$ 3.1 & 0.2 & 0.5 \\ 
11 & 337.630159 & 75.235730 & 25.7 & 15.3 & 151 & 2.9  $\pm$ 0.1 & 5.2  $\pm$ 1.0 & 0.14  $\pm$ 0.02 & 9.8  $\pm$ 0.5 & 11.5  $\pm$ 0.3 & 22.0  $\pm$ 7.0 & 17.4  $\pm$ 5.7 & 0.2 & \nodata \\ 
12 & 339.875346 & 75.183526 & 12.4 & 8.3 & 172 & 1.5  $\pm$ 0.2 & 2.8  $\pm$ 0.6 & 0.22  $\pm$ 0.03 & 10.0  $\pm$ 0.6 & 10.3  $\pm$ 0.2 & 30.7  $\pm$ 2.8 & 34.3  $\pm$ 5.0 & 0.4 & \nodata \\ 
13 & 339.622874 & 75.190326 & 13.0 & 8.7 & 165 & 1.5  $\pm$ 0.2 & 1.8  $\pm$ 0.4 & 0.16  $\pm$ 0.05 & 11.6  $\pm$ 0.4 & 12.4  $\pm$ 0.4 & 21.9  $\pm$ 2.4 & 19.9  $\pm$ 2.9 & 0.5 & \nodata \\ 
14 & 337.789904 & 75.222149 & 41.0 & 14.5 & -145 & 3.5  $\pm$ 0.1 & 8.8  $\pm$ 1.8 & 0.11  $\pm$ 0.01 & 9.9  $\pm$ 0.4 & 11.1  $\pm$ 0.2 & 23.4  $\pm$ 7.8 & 19.7  $\pm$ 6.5 & 0.2 & 0.3 \\ 
15 & 338.852562 & 75.314655 & 19.8 & 12.1 & 70 & 2.2  $\pm$ 0.1 & 2.7  $\pm$ 0.5 & 0.23  $\pm$ 0.06 & 12.5  $\pm$ 1.6 & 12.4  $\pm$ 0.3 & 6.2  $\pm$ 1.7 & 13.0  $\pm$ 3.2 & 0.7 & \nodata \\ 
16 & 339.910415 & 75.196766 & 12.7 & 8.4 & 90 & 1.5  $\pm$ 0.2 & 3.3  $\pm$ 0.7 & 0.26  $\pm$ 0.04 & 9.5  $\pm$ 0.3 & 10.2  $\pm$ 0.1 & 31.9  $\pm$ 4.9 & 39.1  $\pm$ 6.9 & 0.4 & \nodata \\ 
17 & 338.764785 & 75.217186 & 16.5 & 10.8 & 173 & 1.9  $\pm$ 0.2 & 0.9  $\pm$ 0.2 & 0.14  $\pm$ 0.01 & 10.3  $\pm$ 1.7 & 12.2  $\pm$ 0.2 & 5.1  $\pm$ 1.8 & 6.8  $\pm$ 1.1 & 0.9 & \nodata \\ 
18 & 339.834372 & 75.224749 & 25.5 & 11.5 & 81 & 2.5  $\pm$ 0.1 & 4.8  $\pm$ 1.0 & 0.20  $\pm$ 0.04 & 9.4  $\pm$ 1.3 & 11.0  $\pm$ 0.3 & 11.7  $\pm$ 2.9 & 19.9  $\pm$ 4.9 & 0.3 & 0.6 \\ 
19 & 338.970673 & 75.312996 & 38.3 & 14.0 & 102 & 3.4  $\pm$ 0.1 & 7.3  $\pm$ 1.5 & 0.13  $\pm$ 0.02 & 10.5  $\pm$ 0.8 & 12.0  $\pm$ 0.5 & 12.9  $\pm$ 4.5 & 15.9  $\pm$ 3.9 & 0.2 & 0.4 \\ 
20 & 338.889854 & 75.282631 & 21.1 & 7.8 & 172 & 1.9  $\pm$ 0.2 & 2.3  $\pm$ 0.5 & 0.25  $\pm$ 0.09 & 12.9  $\pm$ 2.7 & 15.3  $\pm$ 1.7 & 5.0  $\pm$ 1.2 & 20.6  $\pm$ 2.6 & 0.8 & 1.5 \\ 
21 & 336.016977 & 75.072091 & 20.8 & 11.0 & 126 & 2.2  $\pm$ 0.2 & 1.3  $\pm$ 0.3 & 0.13  $\pm$ 0.02 & 10.0  $\pm$ 1.9 & 12.1  $\pm$ 0.2 & 4.2  $\pm$ 1.0 & 7.9  $\pm$ 1.1 & 0.7 & \nodata \\ 
22 & 335.354540 & 75.072309 & 19.7 & 11.8 & 115 & 2.2  $\pm$ 0.2 & 1.4  $\pm$ 0.3 & 0.11  $\pm$ 0.02 & 10.5  $\pm$ 1.7 & 12.4  $\pm$ 0.2 & 3.6  $\pm$ 1.6 & 8.1  $\pm$ 0.9 & 0.7 & \nodata \\ 
\enddata
\tablecomments{Top-level, ``leaves'' identified in our dendrogram analysis of the NH$_3$ (1,1) data.}
\tablenotetext{a}{Mean position of the structure in R.A. and Dec. (J2000) coordinates.}
\tablenotetext{b}{Major and minor axis of the structure, projected onto the position-position plane.} 
\tablenotetext{c}{Position angle of the major axis, measured in degrees counter-clockwise from the west direction on the sky.}
\tablenotetext{d}{Geometric mean of $\sigma_{major}$ and $\sigma_{minor}$.  Uncertainty represents $\sqrt{A_{pix}/\pi N_{pix}}$, where $A_{pix}$ is the area of a pixel in the NH$_3$ (1,1) emission map and $N_{pix}$ is the number of pixels falling within the structure.}
\tablenotetext{e}{Observed mass estimated from the \textit{Herschel}-derived H$_2$ column density map presented in \cite{DiFrancesco_prep}.  Uncertainties estimated assuming a 20$\%$ error on the H$_2$ column densities used to calculate mass. }
\tablenotetext{f}{Average and standard deviation of the velocity dispersion measured from the NH$_3$ (1,1) emission for all pixels falling within the structure, weighted by the NH$_3$ (1,1) integrated intensity.}
\tablenotetext{g}{Average and standard deviation of the kinetic gas temperature for all pixels falling within the structure, weighted by the NH$_3$ (1,1) integrated intensity.}
\tablenotetext{h}{Average and standard deviation of the dust temperature for all pixels falling within the structure in the \textit{Herschel}-derived $T_{dust}$ map presented in \cite{DiFrancesco_prep}.}
\tablenotetext{i}{Average and standard deviation of the para-NH$_3$ column density for all pixels falling within the structure, weighted by the NH$_3$ (1,1) integrated intensity.}
\tablenotetext{j}{Average and standard deviation of the H$_2$ column density for all pixels falling within the structure in the \textit{Herschel}-derived $N_{H_2}$ map presented in \cite{DiFrancesco_prep}.}
\tablenotetext{k}{Virial parameter ($\alpha_{vir} = M_{vir} / M_{obs}$) for the structure.}
\tablenotetext{l}{Filamentary virial parameter for the leaves with $a_{maj}/a_{min} \geq 2$. }

\label{Table_NH3}
\end{deluxetable}


\end{document}